\documentclass[aps,prb,10pt
]{revtex4-2}
\usepackage{graphicx,bm,amssymb,color,enumitem}

\def\be{\begin{equation}}
\def\ee{\end{equation}}
\def\ba{\begin{eqnarray}}
\def\ea{\end{eqnarray}}

\def\bc{\begin{center}}
\def\ec{\end{center}}
\def\p{\partial}

\begin{document}

\title{Toward a new theory of the fractional quantum Hall effect
}
\author{S. A. Mikhailov}
\email[Electronic mail: ]{sergey.mikhailov@physik.uni-augsburg.de} 
\affiliation{Institute of Physics, University of Augsburg, D-86135 Augsburg, Germany} 

\date{\today}

\begin{abstract}
The fractional quantum Hall effect was experimentally discovered in 1982. It was observed that the Hall conductivity $\sigma_{yx}$ of a two-dimensional electron system is quantized, $\sigma_{yx}=e^2/3h$, in the vicinity of the Landau level filling factor $\nu=1/3$. In 1983, Laughlin proposed a trial many-body wave function, which he claimed described a ``new state of matter'' -- a homogeneous incompressible liquid with fractionally charged quasiparticles. Here I develop an exact diagonalization theory that allows calculation of the energy and other physical properties of the ground and excited states of a system of $N$ two-dimensional Coulomb interacting electrons in a strong magnetic field. I analyze the energies, electron densities, and other physical properties of the systems with $N\le 7$ electrons, continuously as a function of magnetic field in the range $1/4\lesssim\nu<1$. The results show that both the ground and excited states of the system resemble a sliding Wigner crystal, whose parameters are influenced by the magnetic field. Energy gaps in the many-particle spectra appear and disappear as the magnetic field changes. I also calculate the physical properties of the $\nu=1/3$ Laughlin state for $N\le 8$ and show that neither this state nor its fractionally charged excitations describe the physical reality. The results obtained shed new light on the nature of the ground and excited states in the fractional quantum Hall effect.
\end{abstract}

\maketitle

\tableofcontents

\section{Introduction\label{sec:intro}}

\subsection{Historical background}

The quantum Hall effect was discovered by Klaus von Klitzing in 1980 \cite{Klitzing80}. He studied the longitudinal ($R_{xx}$) and Hall ($R_{H}=R_{xy}$) resistances of a degenerate two-dimensional (2D) electron gas (EG) in the inversion layer of a Si-MOSFET (metal-oxide-semiconductor field effect transistor). The sample was placed in a strong perpendicular magnetic field $B\approx 18$ T and cooled down to $T\approx 1.5$ K. The resistances $R_{xx}$ and $R_{xy}$ were measured as a function of the gate voltage $V_g$, applied between the metallic gate and the 2DEG, which changed the density $n_s$ of 2D electrons and the Landau level filling factor 
\be 
\nu=\pi n_s\lambda^2;\label{LLff}
\ee
here 
\be 
\lambda=\sqrt{\frac{2\hbar c}{|e|B}}=\sqrt{\frac{2\hbar}{m^\star\omega_c}},
\label{lambda}
\ee 
$\omega_c=|e|B/m^\star c$ is the cyclotron frequency, and $m^\star$ is the effective mass of electrons ($\lambda/\sqrt{2}\equiv l_B$ is the magnetic length). He found that, when $\nu$ is close to integer values $\nu\approx i$, $i=1,2,3,\dots$, the diagonal resistance $R_{xx}$ becomes negligibly small, while the Hall resistance takes on, with a very high accuracy, quantized values, corresponding to the Hall conductivity 
\be 
\sigma_{yx}=\frac{e^2}h \nu=\frac{e^2}h  i, \ \ i=1,2,3,\dots \label{iqhe}
\ee 
The origin of this fascinating physical phenomenon, which was called the \textit{integer quantum Hall effect}, was quickly understood \cite{Klitzing80} in terms of the single-particle picture. The Landau quantization of electron motion leads to the appearance of energy gaps in the electron spectrum when $\nu\approx i$; the classical formula for the Hall conductivity $\sigma_{yx}=n_sec/B$, together with the relation (\ref{LLff}) immediately gives the quantized values (\ref{iqhe}). The stabilization of $\sigma_{yx}$ at the levels  (\ref{iqhe}) and the vanishing of $\sigma_{xx}$ in finite intervals around $\nu=i$ was explained by the influence of disorder, see, e.g., Ref. \cite{Prange90}.

The time of mysteries came a little later. In 1982 Tsui, Stormer and Gossard published a paper \cite{Tsui82} where the same transport coefficients ($R_{xx}$ and $R_{xy}$) were measured in another material system, GaAs/AlGaAs heterojunction. The main difference between the new experiment and the one of von Klitzing was that the mobility of 2D electrons was higher ($\mu\sim 10^5$ cm$^2$/Vs) and the temperature was lower ($T$ down to $\sim 0.48$ K). In the experiment \cite{Tsui82} the density of electrons was fixed while the magnetic field varied from zero up to $\sim 22$ T. Like in Ref. \cite{Klitzing80}, the already familiar integer quantization of $R_{xy}$ was observed around $\nu=1,2,3,\dots$, but -- very surprisingly -- a very similar plateau was found around $\nu\approx 1/3$, where the measured $R_{H}$ corresponded to the Hall conductivity 
\be 
\sigma_{yx}= \frac{e^2}h\nu, \ \ \nu=\frac 13.
\ee
Subsequent experimental studies showed that such a \textit{fractional} quantization of $\sigma_{yx}$ and the corresponding suppression of $\sigma_{xx}$ is the case around many fractions of the form $\nu=p/q$ where $p$ and $q$ are integers and $q$ is odd, as well as around some fractions with an even denominator, see, e.g., Ref. \cite{Willett87}. 

If $\nu<1$, all electrons occupy the highly degenerate lowest Landau level, and there are no energy gaps in the single-particle electron spectrum. Therefore the mysterious feature at $\nu=1/3$ could be explained only within a many-body approach, taking into account electron-electron interactions. As known, when considered as classical point particles, Coulomb-interacting electrons form the Wigner crystal \cite{Wigner34}, and Tsui et al. \cite{Tsui82} put forward a hypothesis that the observed $1/3$ feature in the Hall conductivity is related to the formation of the Wigner crystal (or a charge density wave) with a triangular symmetry. However, in 1983 Laughlin \cite{Laughlin83} proposed the following trial wave function for the ground state of the $N$-particle system at $\nu=1/m$: 
\be 
\Psi_{\rm LS}^{(m)}(\bm r_1,\bm r_2,\dots,\bm r_N)\propto \left(\prod_{1\le j<k\le N} (z_j-z_k)^m \right)\exp\left(-\frac 12\sum_{j=1}^N|z_j|^2\right),
\label{LaughlinWF}
\ee
where LS means the ``Laughlin state'', $m$ is odd integer, $\bm r_j=(x_j,y_j)$, and $z_j=(x_j-iy_j)/\lambda$ are the normalized complex coordinates of 2D electrons. The function $\Psi_{\rm LS}^{(m)}$ is an eigenfunction of the total angular momentum operator with the eigenvalue ${\cal L}=mN(N-1)/2$ (in units of $\hbar$). If $m=1$, it coincides with the wave function of the so called maximum density droplet (MDD) state proposed earlier in Ref. \cite{Bychkov81} for the ground state of the system at $\nu=1$. The MDD state is characterized by a uniform electron density at $r\lesssim R=\sqrt{N/\pi n_s}$, see Section \ref{sec:MDD} for further details.

For $m=3$ and $5$ the energy of the states (\ref{LaughlinWF}) in the thermodynamic limit was evaluated in Ref. \cite{Laughlin83}, and it was found that it is lower than the energy of the charge density wave calculated in Refs. \cite{Yoshioka79,Yoshioka83b} using the Hartree-Fock \cite{Yoshioka79} and second-order perturbation theory \cite{Yoshioka83b}. The projections of $\Psi_{\rm LS}^{(m=3)}$ and $\Psi_{\rm LS}^{(m=5)}$ onto the numerically calculated exact ground states for three and four particles were also calculated and found to be close to 1. Apart from the function (\ref{LaughlinWF}) Laughlin also ``generated'' many-body wave functions for elementary excitations of the system and stated that they describe quasiparticles with fractional charge $e/m$. Finally he concluded that the wave function (\ref{LaughlinWF}) describes the ground state of the system at $\nu=1/m$ and is ``an incompressible quantum fluid with fractionally charged excitations'' \cite{Laughlin83}. A few critical comments on the wave function (\ref{LaughlinWF}) followed \cite{Tao84}, and several more attempts to find an alternative ground state of the fractional quantum Hall effect (FQHE) system \cite{Maki83,Yoshioka83a} have been made, but finally the LS (\ref{LaughlinWF}) was accepted by the community \cite{NP98} as the closest approximation to the ground-state wave function at $\nu=1/m$ (with $m=3$ and 5). Laughlin's ideas have been developed in a very large number of subsequent publications, see, e.g., Refs. \cite{Girvin83,Haldane83,Levesque84,Jain89,Kasner94,Tsiper01,Morf02,Wan02,Wan03,Ciftja03,Ciftja04,Ciftja11} and review articles \cite{Eisenstein90,Girvin04,Jain12,Hansson17,Murthy03}. In order to explain the fractions $\nu=p/q$ different from $1/3$, various theoretical approaches have been proposed, for example, hierarchical schemes \cite{Haldane83} or the composite fermions theory \cite{Jain89}. According to the currently accepted version of the FQHE theory, based on the theory \cite{Laughlin83}, a 2D electron system placed in a strong magnetic field undergoes a sequence of phase transitions at various fractional values of $\nu$ into highly idealized dissipationless states \cite{Girvin04}. Reports on the experimental observations of fractionally charged quasiparticles were published in Refs. \cite{Goldman95,Saminadayar97,Picciotto97}. 

\subsection{Brief overview of results of this work}

In this paper, I develop an exact diagonalization  theory that enables calculation of the energy and other physical properties of the ground and excited states of $N$ two-dimensional Coulomb interacting electrons placed in a strong magnetic field. It is assumed that all electrons are spin-polarized and occupy only the lowest Landau level states. It is also assumed that the electrons are in the field of a neutralizing positively charged background, which has the shape of a disk of radius $R=\sqrt{N/\pi n_s}$ and constant surface density $n_s$. I present exact results for the ground and excited states of the systems of $N\le 7$ electrons, both for $\nu=1/3$ and for arbitrary $\nu$ varying from $\nu=1$ to $\nu\simeq 1/4$, in dependence of the magnetic field $B$. The results show that electron-electron and background-electron interactions lift the degeneracy of the Landau levels and lead to the appearance of energy gaps in the many-body spectra of the FQHE system. As the magnetic field $B$ changes, the width of the energy gaps oscillates, remaining on order of $e^2/l_B$ in finite intervals of the magnetic field and disappearing at separate $B$-points. The oscillations of the gap width are caused by the interplay of the Coulomb repulsive forces and the compressive action of the $B$ field. Both the ground and excited states of the FQHE system have a shape reminiscent of a Wigner crystal (or a Wigner molecule) in the sense that the electron density maxima are in the same places where one would expect to find Coulomb interacting point charges. All these results shed new light on the true nature of the ground and excited states of the 2D electron systems in strong magnetic fields and lead to a better understanding of the FQHE effect.

A large part of this paper is devoted to a detailed analysis of the currently accepted theory of this phenomenon. I investigate physical properties of the state (\ref{LaughlinWF}) for $N\le 8$, as well as of its ``fractionally charged'' excitations, and show that they have no relation to the true ground and excited states of the FQHE system. The results obtained in this work force the conclusion that the Laughlin liquid with its fractionally charged excitations does not exist.

It should be noted that, although the currently accepted FQHE theory claims that the Laughlin function well describes the properties of the system in the thermodynamic limit, no precise evidence for such statements has been presented. Moreover, it is obvious that such evidence cannot exist, since in order to obtain it, it would be necessary to solve the many-body Schr\"odinger equation for a very large number of ($N\ggg 1$) strongly interacting particles. On the other hand, some exact results have been obtained in the literature for systems with a small number of particles, see, e.g., exact diagonalization calculations of Ref. \cite{Kasner94} (for $\nu=1/3$ and $N\le 9$) and of Ref. \cite{Tsiper01} (for $\nu=1/3$ and $N\le 12$). Although the results of these \textit{exact} calculations have been found in clear contradiction with the \textit{variational} theory of Ref. \cite{Laughlin83}, the Laughlin theory was not questioned. In Ref. \cite{Kasner94} the discrepancies were ignored, while in Ref. \cite{Tsiper01} a complicated and questionable interpretation in terms of an edge reconstruction of the LS into a ``chiral striped phase'' was put forward. I discuss these and some other statements of the currently accepted FQHE theory in Sections \ref{sec:remark7} and \ref{sec:discussionAcceptedTheory} below.

The rest of the paper is organized as follows. In Section \ref{sec:formulation} I formulate the problem and discuss all the technical issues needed for the remaining part of the paper; in particular, the many-particle matrix elements of the Hamiltonian and other physical quantities are calculated there. In Section \ref{sec:Wigner} the classical solution of the problem, the Wigner crystal, is briefly discussed. In Section \ref{sec:ExactSolution} results of the exact solution of the problem for $\nu=1/3$ and up to $N=7$ particles are presented. Then I switch to a discussion of the LS. In Section \ref{sec:MDD} I overview the physical properties of the  MDD state ($\nu=1$) which are used in the subsequent discussion of the FQHE problem. In Section \ref{sec:Laughlin13} the energy and other physical properties of the trial state (\ref{LaughlinWF}) for $m=1/\nu=3$ and $N\le 8$ are calculated and compared with the exact results from Section \ref{sec:ExactSolution}. 

After the complete analysis of the case $\nu=1/3$, I present in Section \ref{sec:ExactSolutionNu<1} the results of the exact solution of the problem for $\nu\le 1$. In Section \ref{sec:discussionAcceptedTheory} a number of statements of the currently accepted FQHE theory are analyzed and discussed, and finally in Section \ref{sec:conclusions} all results of this work are summarized and conclusions are formulated. Mathematical details are given in the Appendices. 

\section{Theory \label{sec:formulation}}

\subsection{Single-particle problem\label{sec:single-particle}}

Let us consider a single 2D electron moving in the plane $z=0$ in the presence of a uniform external magnetic field $\bm B=(0,0,B)$. Its quantum-mechanical motion is described by the single-particle Schr\"odinger equation 
\be
\frac{1}{2m^\star}\left(\hat {\bm p}+\frac {|e|}{2c}{\bm B}\times{\bm
r}\right)^2\phi({\bm r})=\epsilon\phi({\bm r}).
\label{1particle}
\ee 
Its solution,
\be
\epsilon\equiv\epsilon_{n,l}=\hbar\omega_c\left(n+\frac{l+|l|+1}2\right),\ \ 0\le n<\infty , \ \ -\infty<l<+\infty,
\label{sp-energy}
\ee
\be
\phi({\bf r})\equiv\phi_{n,l}({\bm
r})=\frac{e^{il\theta}}{\sqrt{\pi}\lambda}\left(\frac{n!}{(n+|l|)!}\right)^{1/2} \exp\left(-\frac{r^2}{2\lambda^2}\right)\left(\frac
r\lambda\right)^{|l|}L_n^{|l|}\left(\frac{r^2}{\lambda^2}\right),
\label{sp-wavefun}
\ee
is characterized by the radial quantum number $n$ and the azimuthal (angular momentum) quantum number $l$. The functions (\ref{sp-wavefun}) represent a complete basis set in a 2D space.

The states with $n=0$ and non-positive $l$, $l\le 0$, belong to the lowest Landau level. The corresponding energy equals $\hbar\omega_c/2$ and the corresponding wave functions are
\be
|L\rangle\equiv \psi_L({\bm r})\equiv \phi_{n=0,l\le 0}({\bm r})= \frac{1}{\lambda\sqrt{\pi L!}} 
\left(\frac r\lambda e^{-i\theta}\right)^{L}
\exp\left(-\frac{r^2}{2\lambda^2}\right)= \frac{z^{L} e^{-|z|^2/2}}{\lambda\sqrt{\pi L!}} ,
\label{spwf_psiL}
\ee
where $L=-l=0,1,2,\dots $, and $z=(x-iy)/\lambda$. The states (\ref{spwf_psiL}) are normalized, $\langle L|L'\rangle=\delta_{LL'}$, and represent a complete subset of functions in two dimensions belonging to the lowest Landau level with $\epsilon_{0,l}=\hbar\omega_c/2$. The matrix elements of the exponential function $e^{i\bm q\cdot\bm r}$ between the single-particle states (\ref{spwf_psiL}) are
\be 
\langle L |  e^{i\bm q\cdot\bm r} | L'\rangle=
i^{|L-L'|}
e^{i(L-L')\alpha} 
\sqrt{\frac{(\min\{L,L'\})!}{(\max\{L,L'\})!}}
\left(\frac{q\lambda}{2}\right)^{|L-L'|}\exp\left(-\frac{(q\lambda)^2}{4}\right) L_{\min\{L,L'\}}^{|L-L'|}\left(\frac{(q\lambda)^2}{4}\right),
\label{Exp_iqr_matrel}
\ee
where $\alpha$ is the polar angle of the vector $\bm q$, $\bm q=q(\cos\alpha,\sin\alpha)$, and $L_n^l(x)$ are the Laguerre polynomials.

\subsection{Positive background}

$N$ electrons repel each other by Coulomb forces, therefore a compensating positive background is required to keep them together and to ensure the system electroneutrality. I will use two physically reasonable models of the positive background density. The first model, also used, for example, in Refs. \cite{Laughlin83,Ciftja03,Ciftja04,Ciftja11}, assumes that the positive background has the shape of a disk with a uniform charge density $n_s$ and a radius $R=\sqrt{N/\pi n_s}$,
\be 
n_b^{\rm st}(\bm r)=n_s\Theta(R-r)=n_s\Theta(\sqrt{N}-x).
\label{dens-step}
\ee 
Here $x=r/a_0$, $\Theta(x)$ is the Heaviside step function, and the length $a_0$ is defined as 
\be 
\pi n_sa_0^2=1. \label{def-a0}
\ee
In the second model, the density profile is smoothed near the disk edge at the length $\sim a_0$, 
\be
n_b^{\rm sm}({\bm r})= n_se^{-x^2}\sum_{k=0}^{N-1}\frac{(x^2)^k}{k!}=n_s\frac{\Gamma(N,x^2)}{\Gamma(N)}=n_s Q(N,x^2),
\label{dens-smooth}
\ee
where $\Gamma(N)$ is the Euler Gamma function, $\Gamma(N,z)$ is the incomplete Gamma function, and
\be 
Q(N,x)=\frac{\Gamma(N,x)}{\Gamma(N)} \label{Qfunc}
\ee 
is the regularized incomplete Gamma function. The smooth density profile (\ref{dens-smooth}) actually gives a more adequate description of the real density distribution, since in real systems the edge is always smeared over a certain length, for example, over the distance between the 2D gas and the donor layer in GaAs/AlGaAs heterostructures or over the average distance between electrons $a_0$. The length $a_0$, defined in (\ref{def-a0}), will be used as the length unit throughout the paper. In contrast to other possible options for choosing the length unit (e.g., $l_B$ or $\lambda$, as was done in many publications), $a_0$ does not depend on $B$, which is more convenient for the system behavior analysis at varying magnetic fields, see Section \ref{sec:ExactSolutionNu<1}. 

\begin{figure}[ht!]
\includegraphics[width=0.5\columnwidth]{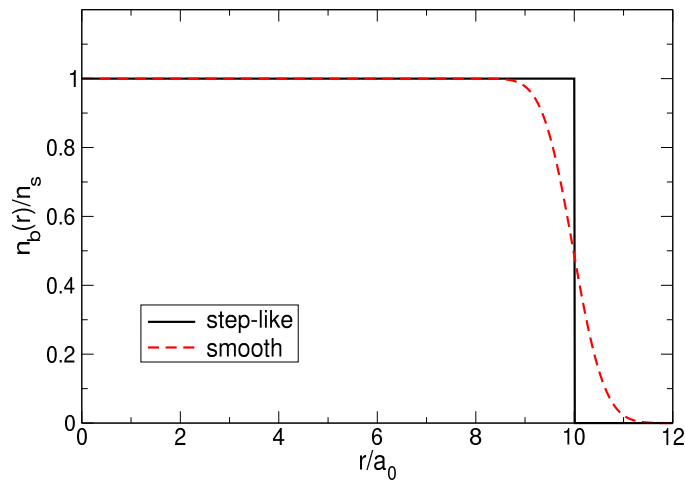}
\caption{\label{fig:bcgrdens}The positive background density $n_b (r)$ for $N = 100$ in the step-like and smooth density profile models.
}
\end{figure}

The both density profiles, (\ref{dens-step}) and (\ref{dens-smooth}), satisfy the condition 
\be 
\int n_b({\bm r}) d\bm r =N\label{fullNcondition}
\ee
and are shown in Figure \ref{fig:bcgrdens} for $N=100$. The Fourier transforms of the density profiles (\ref{dens-step}) and (\ref{dens-smooth}) are determined by the formulas
\be
n_{\bm q}^{b,\rm st}\equiv\int d\bm r n_b({\bm r})e^{i\bm q\cdot\bm r}=N \frac {J_1(qR)}{qR/2}=2N \frac {J_1(qa_0\sqrt{N})}{qa_0\sqrt{N}},
\label{BGdensStep_Fourier}
\ee
and
\be 
n_{\bm q}^{b,\rm sm}=\exp\left(-\frac{(qa_0)^2}4\right)
 L_{N-1}^1\left(\frac{(qa_0)^2}4\right) ,\label{BGdensSmooth_Fourier}
\ee
respectively, where $J_1$ is the Bessel function, and $L_n^k$ are the Laguerre polynomials. 

\begin{figure}[ht!]
\includegraphics[width=0.49\columnwidth]{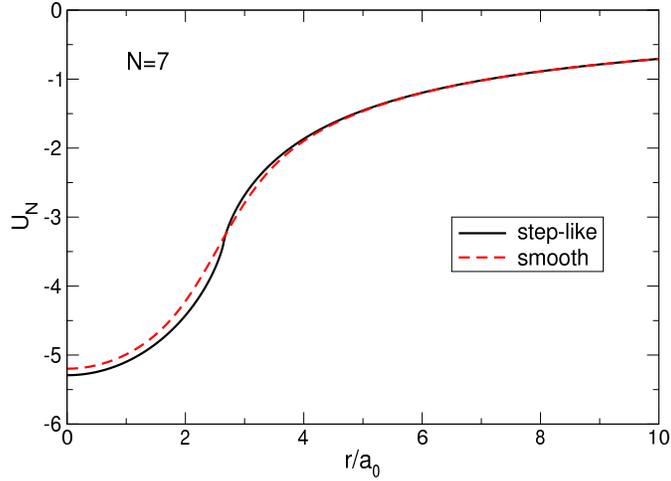}
\caption{\label{fig:Ubpotential}The potential energies (\ref{Upot-step}) and (\ref{Upot-smooth}) of the positive background with the density profiles (\ref{dens-step}) and (\ref{dens-smooth}) for $N=7$.}
\end{figure}

The potential well created by the positively charged background disk with the density (\ref{dens-step}) or (\ref{dens-smooth}) is described by the formula
\be 
V_b(\bm r)=\frac{e^2}{a_0}U_N\left(\frac{r}{a_0}\right),\label{Upot}
\ee
where
\be 
U_N^{\rm st}\left(x\right)=
-\frac 4\pi \left\{
\begin{array}{ll}
\sqrt{N}\textrm{E}\left(\frac{x^2}{N}\right), & x^2\le N \\
x\textrm{E}\left(\frac{N}{x^2}\right)-\frac{x^2-N}{x}\textrm{K}\left(\frac{N}{x^2}\right), & x^2\ge N \\
\end{array}
\right. ,
\label{Upot-step}
\ee
for the step-like density profile (\ref{dens-step}), and 
\be 
U_N^{\rm sm}(x)=-\sum_{m=0}^{N-1}
\left(
\begin{array}{c}
N \\ m+1 \\
\end{array}
\right)
\frac{(-1)^m }{m!} \Gamma\left(m+\frac 12\right){_1F_1}\left(m+\frac 12,1;-x^2\right) \label{Upot-smooth}
\ee
for the smooth density profile (\ref{dens-smooth}). Here the functions $\textrm{K}(m)$ and $\textrm{E}(m)$, defined as 
\be  
\textrm{K}(m)=\int_0^{\pi/2}\frac{d\theta}{\sqrt{1-m\sin^2\theta}},\ \ \ \textrm{E}(m)=\int_0^{\pi/2}d\theta\sqrt{1-m\sin^2\theta},
\ee 
are complete elliptic integrals of the first and second kind, respectively, $\left(\begin{array}{c}
n \\ m \\
\end{array}\right)$ are the binomial coefficients, $\Gamma(x)$ is the Gamma function, and ${_1F_1}\left(a,b;z\right)$ is the degenerate (confluent) hypergeometric function, Eq. (\ref{DegenHyperGfun}). Figure \ref{fig:Ubpotential} shows the functions (\ref{Upot-step}) and (\ref{Upot-smooth}) for $N=7$. In the case of the step-like density profile, the potential well is slightly deeper, while in the case of the smooth density profile, it is slightly wider. The depths of both potential wells grow with $N$ as $\sqrt{N}$, $U_N^{\rm sm}(0)=-2\Gamma(N+1/2)/\Gamma(N)\approx U_N^{\rm st}(0)=-2\sqrt{N}$. In real, macroscopically large samples ($N\sim 10^{11}-10^{12}$) this depth is on the keV scale.

\subsection{Many-body Hamiltonian}

The Hamiltonian of $N$ interacting 2D electrons, placed in the magnetic field $\bm B=(0,0,B)$ and in the attractive potential (\ref{Upot}) of the positively charged background with the density (\ref{dens-step}) or (\ref{dens-smooth}), has the form
\be 
\hat {\cal H}=\hat K+\hat V_C =\frac{1}{2m}\sum_{j=1}^N\left(\hat {\bm p}_j+\frac {|e|}{2c}{\bm B}\times{\bm
r}_j\right)^2 +\hat V_C .
\label{MBhamilt}
\ee
Here $\hat K$ is the total kinetic energy operator and the Coulomb interaction energy $\hat V_C=\hat V_{bb}+\hat V_{eb}+\hat V_{ee}$ consists of the sum of background-background, background-electron and electron-electron interaction energies,
\be 
\hat V_C
=\frac{e^2}{2}\int \frac{n_b(\bm r)n_b(\bm r')d\bm r d\bm r'}{|\bm r-\bm r'|}
-e^2 \int n_b(\bm r) d\bm r\sum_{j=1}^N \frac 1{|\bm r-\bm r_j|}
+\frac{e^2}{2}\sum_{j\neq k=1}^N \frac 1{|\bm r_j-\bm r_k|}.
\label{CoulombEnergy}
\ee
In order to calculate the energy $\hat V_C$ it is convenient to rewrite (\ref{CoulombEnergy}) in terms of Fourier transforms of the electron and background charge densities. This gives
\ba 
\hat V_C=
\frac{e^2}{2\pi} \int \frac{d\bm q}{q}
\left(
\frac{1}{2}\left(n_{\bm q}^b\right)^2 
- n_{\bm q}^b \sum_{j=1}^N  e^{-i\bm q\cdot \bm r_j}
+\frac{1}{2}\sum_{j\neq k=1}^N e^{i\bm q\cdot \bm r_j} e^{-i\bm q\cdot \bm r_k}
\right),
\label{CoulombEnergyFourier}
\ea
where $n_{\bm q}^b$ are given by Eqs. (\ref{BGdensStep_Fourier}) and (\ref{BGdensSmooth_Fourier}), for the step-like and smooth density profiles respectively. The Hamiltonian (\ref{MBhamilt}) commutes with the total angular momentum operator 
\be 
\hat {\cal L}_z=\sum_{j=1}^N({\bm r}_j \times \hat {\bm p}_j)_z.
\ee
The total angular momentum quantum number ${\cal L}\equiv {\cal L}_z$ can be used to classify the many-body basis wave functions.

\subsection{Basis many-body wave functions\label{sec:basisMBfunc}}

Let us consider $N$ spin-polarized electrons at the lowest Landau level. Each of the particles can occupy one of the states (\ref{spwf_psiL}). If the $j$-th particle is in the single-particle state $|L_j\rangle=\psi_{L_j}({\bm r})$, the corresponding many-body wave function can be written as a Slater deteminant 
\be
|L_1,L_2,\dots,L_N\rangle=
\frac{1}{\sqrt{N!}}
\left |\begin{array}{cccc}
\psi_{L_1}(\bm r_1) & 
\psi_{L_1}(\bm r_2) & 
\dots &
\psi_{L_1}(\bm r_N) \\
\psi_{L_2}(\bm r_1) & 
\psi_{L_2}(\bm r_2) &
\dots & 
\psi_{L_2}(\bm r_N) \\
\dots &\dots &\dots &\dots \\
\psi_{L_N}(\bm r_1) & 
\psi_{L_N}(\bm r_2) & 
\dots & 
\psi_{L_N}(\bm r_N) \\
\end{array}
\right |.
\label{manybodyWF}
\ee
The functions (\ref{manybodyWF}) are orthogonal and normalized, 
\be 
\langle L'_1,L'_2,\dots,L'_N|L_1,L_2,\dots,L_N\rangle=\delta_{L'_1L_1}\delta_{L'_2L_2}\dots\delta_{L'_NL_N}.
\ee 
They are eigenfunctions of the kinetic energy operator
\be 
 \hat K|L_1,L_2,\dots,L_N\rangle= N\frac{\hbar\omega_c}2 |L_1,L_2,\dots,L_N\rangle,
\ee
 and of the total angular momentum operator $\hat {\cal L}$
\be 
\hat {\cal L}_z|L_1,L_2,\dots,L_N\rangle= \left(\sum_{i=1}^{N} L_i\right) |L_1,L_2,\dots,L_N\rangle,
\ee
where the latter is measured in units of $\hbar$. The many-body wave functions (\ref{manybodyWF}) represent the orthonormal basis set of functions belonging to the lowest Landau level. 

If $N$ electrons occupy the single-particle states with the lowest possible angular momenta $L$ from $L=0$ up to $L=N-1$, one gets the many-body configuration $\Psi_{\rm mdd}=|0,1,2,\dots,N-1\rangle$, Ref. \cite{Bychkov81}, which is often referred to as the maximum density droplet (MDD) state. This MDD configuration has the lowest possible total angular momentum 
\be 
{\cal L}= {\cal L}_{\min}= \sum_{L=0}^{N-1} L = \frac {N(N-1)}{2}.
\label{Lmin}
\ee
If ${\cal L}>{\cal L}_{\min}$, there exist, in general, more than one many-body electronic configurations corresponding to the given $N$ and ${\cal L}$. For example, Tables \ref{tab:N2configs} and \ref{tab:N3configs} show possible many-body configurations for $N=2$ and $N=3$ and several ${\cal L}$'s. The number $N_{mbs}(N,{\cal L})$ of many-body configurations grows with ${\cal L}$ for a given $N$.

\begin{table}[!ht]
\caption{Possible many-body configurations in a system of $N=2$ electrons. $N_{mbs}(N,{\cal L})$ is the total number of all many-particle configurations with given $N$ and ${\cal L}$. \label{tab:N2configs}}
\begin{tabular}{clc}
${\cal L}$ & \hspace{4mm} Configurations  & $N_{mbs}$\\
\hline
1 & $|0,1\rangle $ & 1 \\
2 & $|0,2\rangle $ & 1 \\
3 & $|0,3\rangle $ $|1,2\rangle $ & 2 \\
4 & $|0,4\rangle $ $|1,3\rangle $ & 2 \\
5 & $|0,5\rangle $ $|1,4\rangle $ $|2,3\rangle $ & 3 \\
6 & $|0,6\rangle $ $|1,5\rangle $ $|2,4\rangle $ & 3 \\
7 & $|0,7\rangle $ $|1,6\rangle $ $|2,5\rangle $ $|3,4\rangle $ & 4 \\
\end{tabular}
\end{table}

\begin{table}[ht!]
\caption{Possible many-body configurations in a system of $N=3$ particles. $N_{mbs}(N,{\cal L})$ is the total number of all many-particle configurations with given $N$ and ${\cal L}$. \label{tab:N3configs}}
\begin{tabular}{clc}
${\cal L}$ & \hspace{5.5cm} Configurations  & $N_{mbs}$\\
\hline
3 & $|0,1,2\rangle $ & 1\\
4 & $|0,1,3\rangle $ & 1\\
5 & $|0,1,4\rangle $ $|0,2,3\rangle $ & 2 \\
6 & $|0,1,5\rangle $ $|0,2,4\rangle $ $|1,2,3\rangle $ & 3 \\
7 & $|0,1,6\rangle $ $|0,2,5\rangle $ $|0,3,4\rangle $ $|1,2,4\rangle $ & 4 \\
8 & $|0,1,7\rangle $ $|0,2,6\rangle $ $|0,3,5\rangle $ $|1,2,5\rangle $ $|1,3,4\rangle $ & 5 \\
9 & $|0,1,8\rangle $ $|0,2,7\rangle $ $|0,3,6\rangle $ $|0,4,5\rangle $ $|1,2,6\rangle $ $|1,3,5\rangle $ $|2,3,4\rangle $ & 7 \\
10 & $|0,1,9\rangle $ $|0,2,8\rangle $ $|0,3,7\rangle $ $|0,4,6\rangle $ $|1,2,7\rangle $ $|1,3,6\rangle $ $|1,4,5\rangle $ $|2,3,5\rangle $ & 8 \\
11 & $|0,1,10\rangle $ $|0,2,9\rangle $ $|0,3,8\rangle $ $|0,4,7\rangle $  $|0,5,6\rangle $ $|1,2,8\rangle $ $|1,3,7\rangle $ $|1,4,6\rangle $ $|2,3,6\rangle $ $|2,4,5\rangle $ & 10 \\
12 & $|0,1,11\rangle $ $|0,2,10\rangle $ $|0,3,9\rangle $ $|0,4,8\rangle $  $|0,5,7\rangle $ $|1,2,9\rangle $ $|1,3,8\rangle $ $|1,4,7\rangle $ $|1,5,6\rangle $ $|2,3,7\rangle $ $|2,4,6\rangle $ $|3,4,5\rangle $ & 12 \\
\end{tabular}
\end{table}

\subsection{Many-body matrix elements\label{sec:MBmatrixelements}}

To calculate various physical properties of an $N$-electron system one needs matrix elements of one-particle or two-particle operators,
\be 
\langle L'_1,L'_2,\dots,L'_N|\sum_{j=1}^N\hat F_1(\bm r_j)|L_1,L_2,\dots,L_N\rangle,\label{1partOperator}
\ee 
\be 
\langle L'_1,L'_2,\dots,L'_N|
\sum_{j=1}^N\sum_{k=1,k\neq j}^N\hat F_2(\bm r_j,\bm r_k)
|L_1,L_2,\dots,L_N\rangle,\label{2partOperator}
\ee 
with the many-body states (\ref{manybodyWF}). In this paper, only the matrix elements (\ref{1partOperator})--(\ref{2partOperator}) between the many-body states $\langle L'_1,L'_2,\dots,L'_N|$ and $|L_1,L_2,\dots,L_N\rangle$ that belong to the same total angular momentum ${\cal L}$ will be needed, i.e., $\sum_{j=1}^N L_j=\sum_{j=1}^N L'_j={\cal L}$. This means that, if the bra and ket configurations are different, they differ by two or more single-particle states. For example, the configurations $\langle 0,1,8|$ and $|0,2,7\rangle$,  $\langle 0,3,6| $ and $|1,3,5\rangle$ differ by two single-particle states, while the configurations $\langle 0,1,8|$ and $|2,3,4\rangle$ differ by three single-particle states, see Table \ref{tab:N3configs}. The matrix elements between the bra and ket configurations which differ by only one single-particle state, e.g., between the configurations $\langle 0,1,8|$ and $|0,1,9\rangle $, will not be considered since they correspond to different values of the total angular momentum ${\cal L}$.

Now I calculate matrix elements of several one-particle and two-particle operators of the type (\ref{1partOperator})--(\ref{2partOperator}). For brevity,  the short notations $|s\rangle\equiv|\Psi_s\rangle\equiv|L_1^{(s)},L_2^{(s)},\dots,L_N^{(s)}\rangle$ for the functions (\ref{manybodyWF}) will be used.

\subsubsection{Electron density\label{sec:densoperator}}

The operator of the electron density has the form
\be 
\hat n_e(\bm r)=\sum_{j=1}^N\delta(\bm r-\bm r_j). \label{densityoperator}
\ee
The off-diagonal matrix elements of (\ref{densityoperator}) are evidently zero. Then one gets
\be 
\langle \Psi_s|\hat n_e(\bm r)|\Psi_{s'}\rangle =\delta_{ss'} 
\sum_{j=1}^N \langle L_j^{(s)}|\delta(\bm r-\bm r_j)|L_j^{(s)}\rangle =
\delta_{ss'} 
\sum_{j=1}^N |\psi_{L_j^{(s)}}(\bm r)|^2,
\label{electrondensity}
\ee
where $\psi_{L_j^{(s)}}(\bm r)$ is the single-particle wave function (\ref{spwf_psiL}) of the $j$-th particle in the $s$-th many-body configuration.

\subsubsection{Fourier transform of the electron density}

The Fourier transform of the density operator (\ref{densityoperator}) is
\be 
\hat n_{\bm q}^e=\int d\bm r \hat n_e(\bm r)e^{i\bm q\cdot \bm r}=
\sum_{j=1}^N e^{i\bm q\cdot \bm r_j}. 
\ee
Using Eq. (\ref{Exp_iqr_matrel}) I get 
\be 
\langle \Psi_s|\hat n_{\bm q}^e|\Psi_{s'}\rangle =\delta_{ss'} 
\sum_{j=1}^N \langle L_j^{(s)}|e^{i\bm q\cdot \bm r_j}|L_j^{(s)}\rangle =
\delta_{ss'} 
\exp\left(-\frac{(q\lambda)^2}{4}\right) \sum_{j=1}^N 
L_{L_j^{(s)}}^{0}\left(\frac{(q\lambda)^2}{4}\right).
\label{ELdens_Fourier}
\ee

\subsubsection{Background-background interaction energy\label{sec:BBmatrixelements}}

The background-background interaction energy $V_{bb}$ is given by the first term in Eq. (\ref{CoulombEnergyFourier}). Since $V_{bb}$ does not depend on the coordinates of electrons, the matrix $\langle \Psi_s|\hat V_{bb}|\Psi_{s'}\rangle$ is diagonal and all matrix elements are the same. For the step-like (\ref{dens-step}) and smooth (\ref{dens-smooth}) density profiles they are given by the following formulas
\be 
\langle \Psi_s|\hat V_{bb}^{\rm st}|\Psi_{s'}\rangle =\delta_{ss'}
\frac{e^2}{a_0} \frac 8{3\pi}N^{3/2},
\label{BBmbStep-matrixelements}
\ee
\be 
\langle \Psi_s|\hat V_{bb}^{\rm sm}|\Psi_{s'}\rangle =\delta_{ss'}
\frac{e^2}{a_0} \sqrt{\frac\pi 8}{\cal J}(N-1,N-1,1,1,0;1,1),
\label{BB_smooth_matrixelements}
\ee
where the integrals ${\cal J}(n_1,n_2,l_1,l_2,k;\alpha,\beta)$ are defined and calculated in Appendix \ref{app:intJ}, see Eqs. (\ref{intJ}) and (\ref{intJsolution}).

\subsubsection{Background-electron interaction energy\label{sec:BEmatrixelements}}

The background-electron interaction energy is given by the second term in Eq. (\ref{CoulombEnergyFourier}). Its many-body matrix elements are  
\be 
\langle \Psi_{s}| \hat V_{be}|\Psi_{s'}\rangle=
-\frac{e^2}{2\pi} \int \frac{d\bm q}{q}n_{\bm q}^b
\langle \Psi_{s}|
 \left(\hat n_{\bm q}^e\right)^\star|\Psi_{s'}\rangle.
\label{EBmatr_el}
\ee 
Substituting the Fourier transforms of the background and electron densities from Eqs. (\ref{BGdensStep_Fourier}), (\ref{BGdensSmooth_Fourier}), and (\ref{ELdens_Fourier}) into (\ref{EBmatr_el}) I obtain the following results. In the case of the step-like density profile calculations give (for details see Appendix \ref{app:mddBEgen}) 
\be 
\langle \Psi_s|\hat V_{be}^{\rm st}|\Psi_{s'}\rangle =
-\delta_{ss'} \frac{e^2}{a_0} N \sqrt{\beta}\sum_{j=1}^N \sum_{m=0}^{L_j^{(s)}}\left(\begin{array}{c}
L_j^{(s)} \\ m \\
\end{array}\right)\frac{(-1)^m }{m!}\Gamma\left(m+\frac{1}{2}\right)
{_1F_1}\left(m+\frac{1}{2},2;-N\beta\right),
\label{BEmb-matrixelementsStep}
\ee
where 
\be 
\beta=\frac{a_0^2}{\lambda^2}=\frac 1\nu
\ee 
is the inverse Landau level filling factor. In the case of the smooth density profile the results are expressed it terms of the integral ${\cal J}$ (Appendix \ref{app:intJ})
\be 
\langle \Psi_{s}| \hat V_{be}^{\rm sm}|\Psi_{s'}\rangle=
-\delta_{ss'} 
\frac{e^2}{a_0} \sqrt{\frac{\pi\beta} 2}
\sum_{j=1}^N {\cal J}(L_j^{(s)},N-1,0,1,0;1,\beta).
\label{BE_smooth_matrixelements}
\ee 
The matrix elements (\ref{BEmb-matrixelementsStep}) and (\ref{BE_smooth_matrixelements}) depend on the magnetic field $B$.

\subsubsection{Electron-electron interaction energy\label{sec:EEmatrixelements}}

The electron-electron interaction energy is given by the third term in Eq. (\ref{CoulombEnergyFourier}). Calculating its many-body matrix elements I get the following result
\be 
\langle \Psi_{s}| \hat V_{ee}|\Psi_{s'}\rangle=
\delta_{ss'} \left(V^H_{ss}-V^F_{ss}\right) + (1-\delta_{ss'})V^{\rm off}_{ss'},
\label{EEenergy-MatrixElement}
\ee 
where the diagonal matrix elements are given by the difference of Hartree and Fock contributions, 
\be 
V^H_{ss}=\frac{e^2}{2a_0}\int_0^\infty dq a_0 
\sum_{i=1}^N 
\langle L_i^{(s)} |e^{i\bm q\cdot \bm r}|L_i^{(s)}\rangle 
\sum_{j=1}^N 
\langle L_j^{(s)} |e^{-i\bm q\cdot \bm r}|L_j^{(s)}\rangle,
\ee
\be 
V^F_{ss}=\frac{e^2}{2a_0}
\int_0^\infty dq a_0 
\sum_{i=1}^N \sum_{j=1}^N 
\left|
\langle L_i^{(s)} |e^{i\bm q\cdot \bm r}|L_j^{(s)}\rangle 
\right|^2.
\ee
Substituting here the matrix elements of the exponential functions from Eq. (\ref{Exp_iqr_matrel}) I get
\be 
V^H_{ss}=\frac{e^2}{a_0} \sqrt{\frac{\pi\beta}8}
\sum_{j=1}^N 
\sum_{k=1}^N 
{\cal K}(L_j^{(s)},L_k^{(s)},0),
\label{EE-Hartree}
\ee
\be 
V^F_{ss}=\frac{e^2}{a_0} \sqrt{\frac{\pi\beta}8}
\sum_{j=1}^N \sum_{k=1}^N 
{\cal K}(L_{\min},L_{\min},\delta L),
\label{EE-Fock}
\ee
where the integrals ${\cal K}$ are related to the integrals ${\cal J}$ defined above, see Appendix \ref{app:intJ}, and in the last formula
\be 
L_{\min}=\min\{L_j^{(s)},L_k^{(s)}\},\ \ L_{\max}=\max\{L_j^{(s)},L_k^{(s)}\},\ \ \delta L=|L_j^{(s)}-L_k^{(s)}|=L_{\max}-L_{\min}.
\ee 

The formulation of results for the off-diagonal matrix elements ($s\neq s'$) requires a bit longer discussion. First, since $s\neq s'$, the sets of numbers $L_j^s$ and $L_j^{s'}$, $j=1,2,\dots,N$, differ from each other. In general, these sets may differ by one, two or more numbers. The case when they differ by only one number is excluded, as explained above. If they differ by more than two numbers, the corresponding matrix elements equal zero, 
\be 
\langle \Psi_{s}| \hat V_{ee}|\Psi_{s'}\rangle=0\textrm{, if the sets }L_j^s\textrm{ and }L_j^{s'}\textrm{ differ by more than two numbers}.
\ee
Thus, the matrix elements $\langle \Psi_{s}| \hat V_{ee}|\Psi_{s'}\rangle$ are nonzero if and only if the states $|s\rangle$ and $|s'\rangle$ differ from each other by the single-particle states of exactly two particles. For example, for three particles with the total angular momentum ${\cal L}=9$, Table \ref{tab:N3configs}, the matrix elements
\be 
\langle 0,2,7| \hat V_{ee}|0,4,5\rangle \textrm{ and }
\langle 0,1,8| \hat V_{ee}|1,2,6\rangle \label{states1}
\ee
are finite (only two numbers in the bra and ket configurations are different), but the matrix element
\be 
\langle 0,2,7| \hat V_{ee}|1,3,5\rangle 
\ee
is zero (all three numbers are different). 

Let the configurations 
\be 
|s\rangle=|\dots,\underbrace{ L_1^s}_{p_1^s},\dots,\underbrace{ L_2^s}_{p_2^s},\dots\rangle\textrm{ and }|s'\rangle=|\dots,\underbrace{ L_1^{s'}}_{p_1^{s'}},\dots,\underbrace{ L_2^{s'}}_{p_1^{s'}},\dots\rangle
\label{ss'states}
\ee 
differ from each other by the $L$-states of exactly two particles. I designate them as $L_1^s$, $L_2^s$ and $L_1^{s'}$, $L_2^{s'}$, and their serial numbers in the sets $|s\rangle$ and $|s'\rangle$ as $p_1^s$, $p_2^s$, $p_1^{s'}$, $p_2^{s'}$; all other states in (\ref{ss'states}), designated by dots, are identical. For example, for the first of the matrix elements in (\ref{states1}), $\langle 0,2,7| \hat V_{ee}|0,4,5\rangle$, these numbers are $L_1^s=2$, $L_2^s=7$,  $p_1^s=2$, $p_2^s=3$, and $L_1^{s'}=4$, $L_2^{s'}=5$, $p_1^{s'}=2$, $p_2^{s'}=3$. For the second matrix element in (\ref{states1}), $\langle 0,1,8| \hat V_{ee}|1,2,6\rangle$, they are $L_1^s=0$, $L_2^s=8$, $p_1^s=1$, $p_2^s=3$, and $L_1^{s'}=2$, $L_2^{s'}=6$, $p_1^{s'}=2$, $p_2^{s'}=3$. 

Now one can formulate the results for the off-diagonal matrix elements of the electron-electron interaction energy. Calculations show that for $s\neq s'$
\ba 
V^{\rm off}_{ss'}&=&\frac{e^2}{a_0} \sqrt{\frac{\pi\beta}2}(-1)^{p_1^s+p_2^s+p_1^{s'}+p_2^{s'}}
\nonumber \\ &\times&
\left[
{\cal K}\Big(\min\{L_1^s,L_1^{s'}\},\min\{L_2^s,L_2^{s'}\},|L_1^s-L_1^{s'}|\Big)-{\cal K}\Big(\min\{L_1^s,L_2^{s'}\},\min\{L_2^s,L_1^{s'}\},|L_1^s-L_2^{s'}|\Big)
\right],
\label{EE-off}
\ea
where the integrals ${\cal K}$ are defined in Eq. (\ref{intK}).

Equations (\ref{EEenergy-MatrixElement}), (\ref{EE-Hartree}), (\ref{EE-Fock}) and (\ref{EE-off}) give the matrix elements of the electron-electron interaction between the basis many-body configurations (\ref{manybodyWF}). Note that the magnetic field enters these formulas only via the common prefactor $\sqrt{\beta}$, which means that for all values of $B$ the $ee$-interaction  matrix needs to be calculated only once.

\subsubsection{Pair correlation function\label{sec:PCFmatrixelements}}

The operator of the pair correlation function is defined as 
\be 
\hat P(\bm r,\bm r')=\sum_{j=1}^N \sum_{k=1,k\neq j }^N \delta(\bm r-\bm r_j)\delta(\bm r'-\bm r_k).
\label{PairCorrFun}
\ee
Its diagonal and off-diagonal matrix elements are determined by the following formulas,
\be 
\langle \Psi_{s}| \hat P(\bm r,\bm r')|\Psi_{s}\rangle=
\sum_{j=1}^N\sum_{k=1}^N
\left(|\psi_{L_j^{(s)}}(\bm r)|^2 |\psi_{L_k^{(s)}}(\bm r')|^2 -
\psi_{L_j^{(s)}}^\star(\bm r)  
\psi_{L_k^{(s)}}(\bm r)  
\psi_{L_j^{(s)}}(\bm r') 
\psi_{L_k^{(s)}}^\star(\bm r')
\right) ,\label{Prron}
\ee
\be 
\langle\Psi_s|\hat P(\bm r,\bm r')|\Psi_{s'}\rangle 
=
(-1)^{p_1^s+p_2^s+p_1^{s'}+p_2^{s'}}
\left(
\psi_{L_1^s}^\star(\bm r)
\psi_{L_2^s}^\star(\bm r')
\det\left|
\begin{array}{cc}
\psi_{L_1^{s'}}(\bm r)  & 
\psi_{L_2^{s'}}(\bm r) \\
\psi_{L_1^{s'}}(\bm r') & 
\psi_{L_2^{s'}}(\bm r') \\
\end{array}
\right|
+ (\bm r\leftrightarrow\bm r')
\right),\ \ s\neq s',
\label{Prroff}
\ee
where the numbers $L_1^s$, $L_2^s$, $L_1^{s'}$, $L_2^{s'}$, as well as $p_1^s$, $p_2^s$, $p_1^{s'}$, $p_2^{s'}$ have the same meaning as in the previous Section.

\subsection{General solution of the many-body Schr\"odinger problem}

Let us consider the $N$-particle Schr\"odinger equation
\be 
\hat {\cal H}\Psi(\bm r_1,\bm r_2,\dots,\bm r_N)=E\Psi(\bm r_1,\bm r_2,\dots,\bm r_N).
\label{NbodySchrEq}
\ee
In order to solve it for a given total angular momentum ${\cal L}$, the function $\Psi$ should be searched in the form of a linear combination of all $N_{mbs}(N,{\cal L})$ many-body configurations corresponding to given values of $N$ and ${\cal L}$, 
\be 
|\Psi\rangle =\sum_{s'=1}^{N_{mbs}} A_{s'} |\Psi_{s'}\rangle.
\label{PsiExpansion}
\ee
In this formula $A_s$ are unknown numbers. Substituting (\ref{PsiExpansion}) into (\ref{NbodySchrEq}) and multiplying the resulting equation by $\langle \Psi_{s}|$ one reduces the Schr\"odinger problem to the matrix equation 
\be 
\sum_{s'=1}^{N_{mbs}} \langle \Psi_{s}|\hat {\cal H}|\Psi_{s'}\rangle A_{s'}=EA_{s}.
\label{matrix-equation}
\ee
The size of the Hamiltonian matrix ${\cal H}_{ss'}\equiv \langle \Psi_{s}|\hat {\cal H}|\Psi_{s'}\rangle$ here is $N_{mbs}\times N_{mbs}$. Solving the eigenvalue problem (\ref{matrix-equation}) one can find $N_{mbs}$ solutions for given $N$ and ${\cal L}$: the energies $E_{N,{\cal L},k}$ and the sets of numbers $A_{s}^{N,{\cal L},k}$, $k=1,\dots,N_{mbs}$, which give the corresponding many-body wave functions according to the expansion (\ref{PsiExpansion}). After the numbers $A_{s}^{N,{\cal L},k}$ are found one can also calculate all physical properties of the ground or excited many-body states, for example, the electron density and the pair correlation function, using the corresponding matrix elements found in Section \ref{sec:MBmatrixelements}.

All matrix elements of the Hamiltonian ${\cal H}_{ss'}$ are calculated analytically, see Section \ref{sec:MBmatrixelements}. As a result, the energies and the wave functions of the $N$-electron system can be calculated, in principle, with a very high accuracy for any $\nu\le 1$ and any ${\cal L}$ and $N$. In practice, the computation time becomes too large if $N$ or ${\cal L}$ are much greater than one, but the ground state physics of a FQHE system can be well understood even if the number of particles is less than or on the order of ten. Especially valuable are results for $N=7$, because seven spin-polarized electrons form a highly symmetric piece of a macroscopic Wigner crystal, see Figure \ref{fig:WCconfigs}(g) below. 

In Section \ref{sec:ExactSolution} I present results of the theory for $2\le N\le 7$ and the Landau level filling factor $\nu=1/3$. Before doing so, however, it is useful to discuss the purely classical solution of the problem.

\section{Wigner crystal\label{sec:Wigner}}

\begin{figure}[ht!]
\includegraphics[height=4cm]{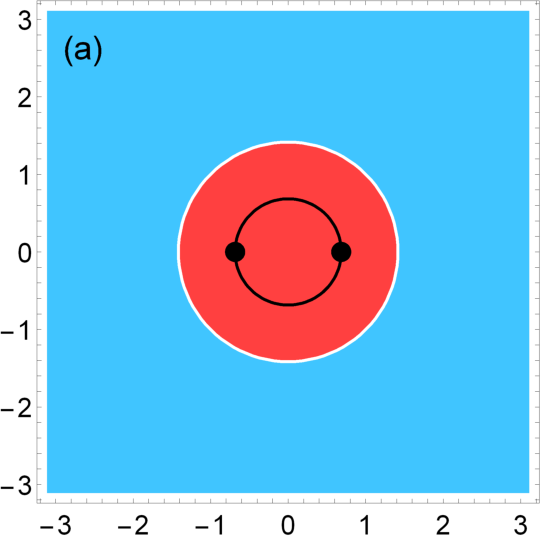}
\includegraphics[height=4cm]{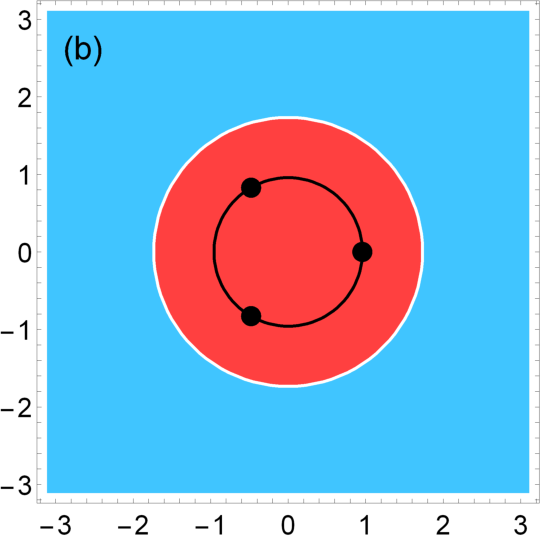}
\includegraphics[height=4cm]{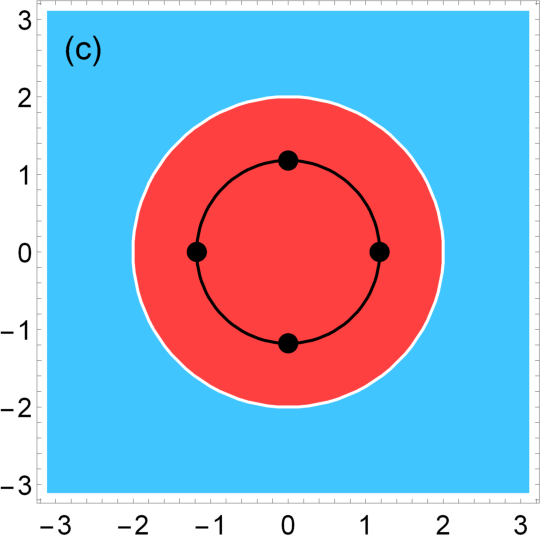}
\includegraphics[height=4cm]{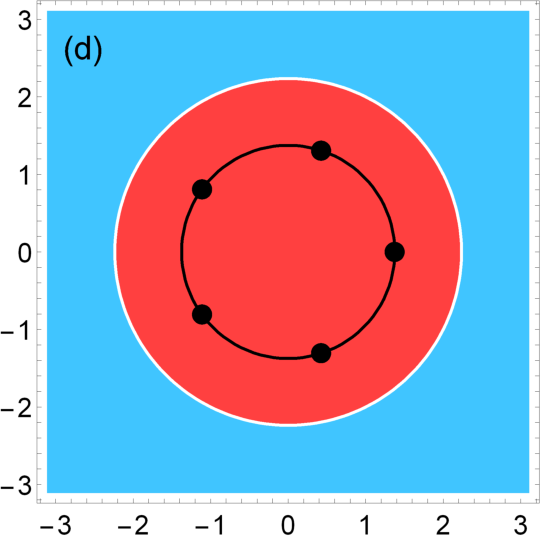}
\includegraphics[height=4cm]{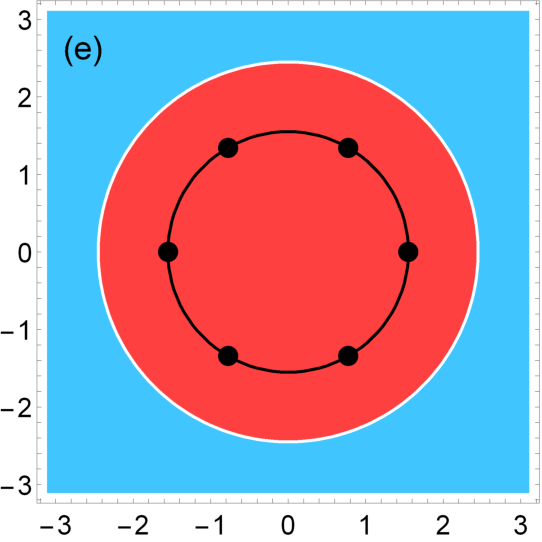}
\includegraphics[height=4cm]{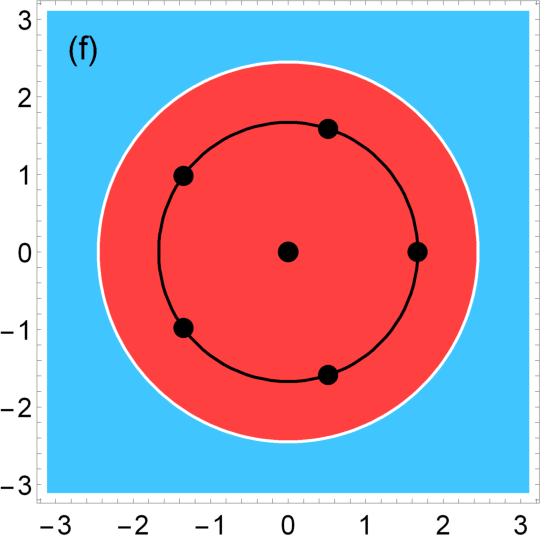}
\includegraphics[height=4cm]{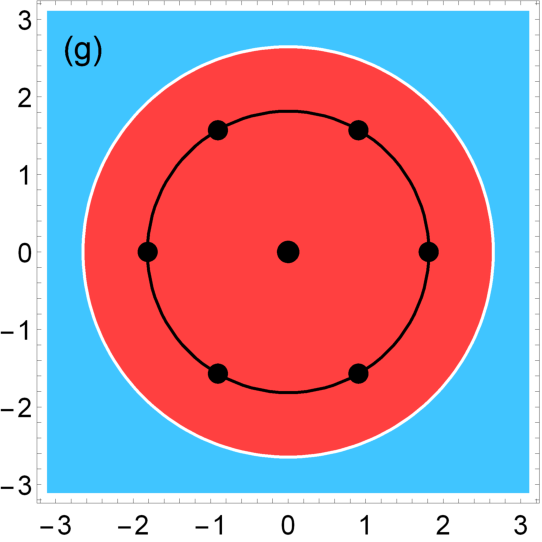}
\includegraphics[height=4cm]{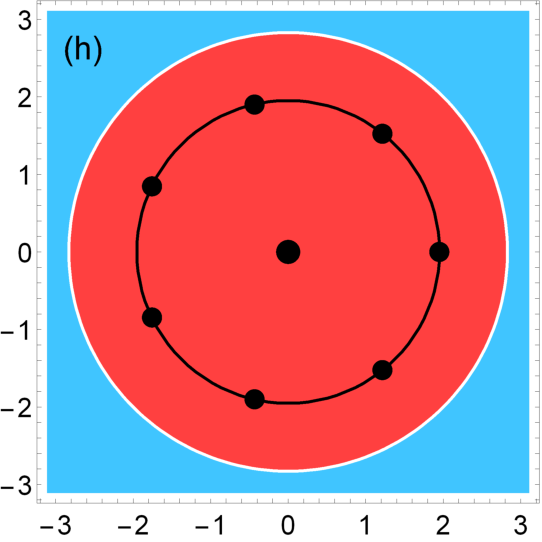}
\caption{\label{fig:WCconfigs}Configurations of the Wigner crystal molecules with $N=2,\dots,8$ in the field of the positively charged background with the step-like density (\ref{dens-step}). The white circle shows the boundary of the positively charged disk; the length unit is $a_0$. For $N\le 5$ the single-shell configurations (a)-(d) have a lower energy. For $N\ge 7$ the two-shell configurations (g)-(h) have a lower energy. For $N=6$ the energies of the single-shell (e) and the two-shell (f) configurations are very close.}
\end{figure}

The physics of the considered system is determined by the interplay of attractive forces acting on electrons by the positively charged background, and the inter-electron Coulomb forces repelling them from each other. If $N$ ($\le 8$) classical point charges are placed in the attractive potential of the positive background they may form two types of Wigner molecules (at small $N$), Fig. \ref{fig:WCconfigs}: with a single shell, when all $N$ particles are located on a ring of a finite radius $R_s$, Figs. \ref{fig:WCconfigs}(a)--\ref{fig:WCconfigs}(e), and with two shells, when one particle is at the center of the positively charged disk, and $N-1$ particles are located on a ring around the center, Figs. \ref{fig:WCconfigs}(f)--\ref{fig:WCconfigs}(h). I will denote these two configurations as $(0,N)$ and $(1,N-1)$, respectively. To understand which of the two possibilities is actually the case, one should calculate the total energy of the Wigner molecules in both situations. 

The potential energy of the positively charged background disk is described by Eqs. (\ref{Upot})--(\ref{Upot-step}) for the step-like density profile and by Eqs. (\ref{Upot}), (\ref{Upot-smooth}) for the smooth density profile. The both potential energies are shown in Figure \ref{fig:Ubpotential}. In the single-shell configuration $(0,N)$ the (dimensionless) complex coordinates of electrons can be written as
\be 
Z_j=R_se^{i2\pi (j-1)/N}, \ \ j=1,2,\dots,N,
\ee
with $R_s$ being the shell radius. Then the total energy of the system is 
\be
E_{(0,N)}(R_s)=\sum_{j=1}^N U_N(|Z_j|)+\sum_{j=1}^{N-1}\sum_{k=j+1}^N\frac {1}{|Z_j-Z_k|}
=
N U_N(R_s)+\frac {1}{R_s}S_N,
\label{WCenergy0N}
\ee
where $U_N(R)$ are given either by (\ref{Upot-step}) or by (\ref{Upot-smooth}), and 
\be 
S_N
=\sum_{j=1}^{N-1}\sum_{k=j+1}^N\frac {1}{|1-e^{i2\pi (k-j)/N}|}.
\ee
In the two-shell Wigner molecule $(1,N-1)$, the particle coordinates are
\be 
Z_j=R_se^{i2\pi (j-1)/(N-1)}, \ \ j=1,2,\dots,N-1,\ \ Z_N=0
\ee
and the total energy has the form
\be
E_{(1,N-1)}(R_s)=
(N-1) U_N(R_s)+U_N(0)+\frac 1{R_s}\Big( S_{N-1}+(N-1)\Big).
\label{WCenergy1N-1}
\ee
Minimizing the energies (\ref{WCenergy0N}) and (\ref{WCenergy1N-1}) with respect to $R_s$ one can find the radii of the shells $R_s(N)$, for both types of molecules and both types of the density profiles, and their total energies. The results of such calculations are shown in Tables \ref{tab:WCmoleculesStep} and \ref{tab:WCmoleculesSmooth} for the step-like and smooth density profiles. One sees that for both density profiles the energy of the $(0,N)$ configuration is smaller than that of the $(1,N-1)$ configuration, if $N\le 5$. The opposite inequality, $E_{(0,N)}>E_{(1,N-1)}$, is valid at $N= 7$ and $8$. If $N=6$, the energies of both configurations are very close to each other: in the case of the step-like density profile (\ref{dens-step}) the two-shell configuration has a slightly lower energy, with a difference of $0.051$\%. In the case of the smooth density profile (\ref{dens-smooth}) the one-shell configuration has a slightly lower energy, with a difference of $0.871$\%. Thus the classical solution of the problem has rotational symmetry $C_n$ of order $n$, where 
\be 
n=n^{\rm st}(N)=\left\{
\begin{array}{ll}
N &\textrm{ if }N\le 5 \\
N-1 &\textrm{ if }6\le N\le 8 \\
\end{array} 
\right. \label{NsymStep}
\ee
in the case of the step-like density profile, and 
\be 
n=n^{\rm sm}(N)=\left\{
\begin{array}{ll}
N &\textrm{ if }N\le 6 \\
N-1 &\textrm{ if }7\le N\le 8 \\
\end{array} 
\right. \label{NsymSmoo}
\ee
in the case of the smooth density profile. At $N=6$, the $C_6$ and $C_5$ configurations have very close energies, so that both of them can be realized in a real system, depending, for example, on tiny details of the confinement potential. With a further increase of the number of particles ($N\gtrsim 10$), one should expect that configurations with $n=n^{\rm st}(N)=n^{\rm sm}(N)=6$, will have the lowest energy, like in the macroscopic Wigner crystal.

\begin{table}[th!]
\caption{Parameters of the Wigner molecules in one-shell $(0,N)$ and two-shell $(1,N-1)$ configurations for the step-like density profile (\ref{dens-step}) and $2\le N\le 8$. $R$ is the radius of the positively charged background disk, and $R_s(N)$ are the radii of the shells obtained by the minimization of the energies (\ref{WCenergy0N}) and (\ref{WCenergy1N-1}). $E_{(0,N)}/N$ and $E_{(1,N-1)}/N$ are the Wigner molecule energies per particle for two configurations. The lengths are in units $a_0$, the energies are in units $e^2/a_0$. \label{tab:WCmoleculesStep}}
\begin{tabular}{||c|c||c|c|c||c|c|c||}
\hline
$N$ & $R$ & Configuration & $R_s(N)$ & $E_{(0,N)}/N$ & Configuration & $R_s(N)$ & $E_{(1,N-1)}/N$\\
\hline
2 & $1.41421$ & (0,2) & 0.684302  & $-2.289449$  &       &          & \\
3 & $1.73205$ & (0,3) & 0.956542  & $-2.578969$  & (1,2) & 1.196868 & $-2.460699$ \\
4 & $2.00000$  & (0,4) & 1.178838 & $-2.813969$  & (1,3) & 1.361107 & $-2.745380$ \\
5 & $ 2.23607$ & (0,5) & 1.373422 & $-3.012388$  & (1,4) & 1.518623 & $-2.983587$ \\
6 & $2.44949$ & (0,6) & 1.549288  & $-3.184857$  & (1,5) & 1.669034 & $-3.186483$ \\
7 & $2.64575$ & (0,7) & 1.711287  & $-3.337907$  & (1,6) & 1.812600 & $-3.362913$ \\
8 & $2.82843$ & (0,8) & 1.862429  & $-3.475824$  & (1,7) & 1.949843 & $-3.519063$ \\
\hline
\end{tabular}
\end{table}
\begin{table}[h!]
\caption{The same as in Table \ref{tab:WCmoleculesStep}, but for the smooth density profile (\ref{dens-smooth}).\label{tab:WCmoleculesSmooth}}
\begin{tabular}{||c|c||c|c|c||c|c|c||}
\hline
$N$ & $R$ & Configuration & $R_s(N)$ & $E_{(0,N)}/N$ & Configuration & $R_s(N)$ & $E_{(1,N-1)}/N$\\
\hline
2 & $1.41421$ & (0,2) & 0.621826  & $-2.067436$  &       &          & \\
3 & $1.73205$ & (0,3) & 0.888015  & $-2.379006$  & (1,2) & 1.127002 & $-2.250789$ \\
4 & $2.00000$  & (0,4) & 1.107700 & $-2.630306$  & (1,3) & 1.282076 & $-2.553625$ \\
5 & $ 2.23607$ & (0,5) & 1.301195 & $-2.841207$  & (1,4) & 1.437241 & $-2.782655$ \\
6 & $2.44949$ & (0,6) & 1.476787  & $-3.023620$  & (1,5) & 1.587616 & $-3.020984$ \\
7 & $2.64575$ & (0,7) & 1.638979  & $-3.184841$  & (1,6) & 1.732047 & $-3.206801$ \\
8 & $2.82843$ & (0,8) & 1.790593  & $-3.329631$  & (1,7) & 1.870522 & $-3.370707$ \\
\hline
\end{tabular}
\end{table}

\begin{figure}[ht!]
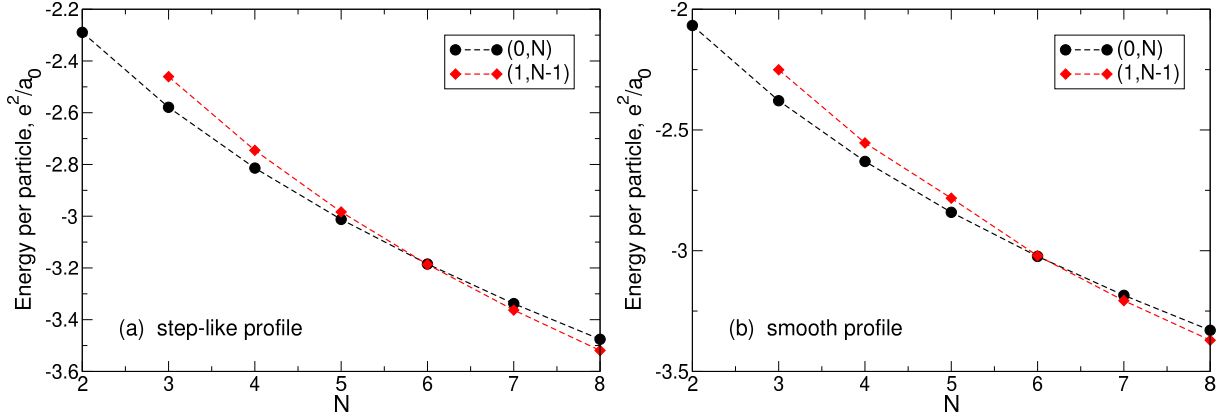

\includegraphics[width=8cm]{WCenergyStep.eps}
\includegraphics[width=8cm]{WCenergySmooth.eps}
\caption{\label{fig:WCenergy}The energy per particle of the Wigner crystal molecules in the $N$-electron systems with the $(0,N)$ and $(1,N-1)$ configurations for (a) the step-like and (b) smooth density profiles.
}
\end{figure}

The $N$-dependencies of the Wigner molecule energies for both configurations are shown in Figure \ref{fig:WCenergy}. In general, the energy of the system with the step-like density is lower than that of the system with the smooth density by $9.7$\%$-1.2$\% depending on $N$. This is due to the fact that in the case of a step-like density the potential well is deeper and narrower than in the case of a smooth density, Figure \ref{fig:Ubpotential}. The spatial distribution of the positive background density and the positions of charged point particles in the Wigner molecules are additionally illustrated in Figure \ref{fig:WCconfigs}. 

Thus, if $N\le 8$, electrons, considered as classical point particles, form single- or double-shell Wigner molecules shown in Figure \ref{fig:WCconfigs}. It should therefore be expected that in the quantum-mechanical solution one will have, instead of point particles, broadened (e.g., Gaussian type) wave functions near each points, these wave functions will overlap and their positions will be averaged over the angular coordinate. The electron density should therefore have the shape of a ring of radius $R_s(N)$ (Tables \ref{tab:WCmoleculesStep} and \ref{tab:WCmoleculesSmooth}), with an additional density maximum at $r=0$, if $N\gtrsim 6$. The exact quantum-mechanical solution obtained in Sections \ref{sec:ExactSolution} and \ref{sec:ExactSolutionNu<1} confirms these expectations. The features related to the symmetries of the Wigner molecules (\ref{NsymStep}) and (\ref{NsymSmoo}), including the ambiguity of the order of the symmetry axis at $N=6$, also manifest themselves in the quantum mechanical solution, see Section \ref{6el-energy}.

\section{Exact solution at $\nu=1/3$\label{sec:ExactSolution}}

Now, I present results of the exact quantum-mechanical solution of the FQHE problem for $N\le 7$. Since a detailed comparison of this solution with the LS will have to be made, a special attention will first be given, in this Section, to the case $\nu=1/3$. The positive background density is assumed to be step-like, Eq. (\ref{dens-step}), in this Section. 
 
\subsection{Ground state energy and wave function\label{sec:ExactEnergy}}

Laughlin assumed \cite{Laughlin83} that the total angular momentum in the ground state at $\nu=1/3$ equals ${\cal L}=3{\cal L}_{\min}=3N(N-1)/2$. I have calculated a few lowest energy levels $E_s^{3{\cal L}_{\min}}$, $s=1,2,3,\dots$, for $\nu=1/3$ and ${\cal L}=3{\cal L}_{\min}=3N(N-1)/2$, in the system with $N=2,3,\dots,7$ electrons. Results are shown in Table \ref{tab:ExactGSEnergy}. The third and fourth columns there give the energy of the lowest ($s=1$) and the next ($s=2$) energy levels; the second column gives the value of the corresponding total angular momentum ${\cal L}=3{\cal L}_{\min}$. 

\begin{table}[ht!]
\caption{Exact and Laughlin energies, as well as relevant angular momenta, of the lowest-energy states for different numbers of particles $N$. Second column: The total angular momentum $3{\cal L}_{\min}$ corresponding to the true ground and Laughlin states. Third and fourth columns: The exact energies of the two lowest levels of the states with ${\cal L}=3{\cal L}_{\min}$. Fifth and sixth columns: The total angular momentum $3{\cal L}_{\min}-\delta{\cal L}$ and the lowest energy $E_1^{3{\cal L}_{\min}-\delta{\cal L}}$ corresponding to the true first excited state; here $\delta{\cal L}=n^{\rm st}(N)$, Eq. (\ref{NsymStep}). Seventh column: The energy difference $E_{\rm 1st}-E_{\rm GS}$ between the true first excited and the true ground state, Eq. (\ref{GS1st}). Eighth column: The exactly calculated energy of the Laughlin state. Ninth column: The energy difference $E_{\rm LS}-E_{\rm GS}$ between the LS and the true ground state. All energies are in units $e^2/a_0$. The step-like background density profile is assumed. \label{tab:ExactGSEnergy}}
 \begin{tabular}{c||c|c|c|c|c|c||c|c}
  $N$ &\  $3{\cal L}_{\min}$\ & $E_1^{3{\cal L}_{\min}}$ & $E_2^{3{\cal L}_{\min}}$ &\ $3{\cal L}_{\min}-\delta{\cal L}$\ & $E_1^{3{\cal L}_{\min}-\delta{\cal L}}$\ & \ $E_{\rm 1st}-E_{\rm GS}$\ & \ $E_{\rm LS}$\ & \ $E_{\rm LS}-E_{\rm GS}$\ \\
 \hline
 2\ \ & \ \   3\ \ & \ \ $-1.872042$\ \ & \ \ $-1.447705$\ \ & \ \   1\ \ & \ \ $-1.773044$\ \ & \ \ $ 0.098998$\ \ & \ \ $-1.871568$\ \ & \ \ $ 0.000474$\ \ \\
 3\ \ & \ \   9\ \ & \ \ $-2.852911$\ \ & \ \ $-2.479623$\ \ & \ \   6\ \ & \ \ $-2.811722$\ \ & \ \ $ 0.041189$\ \ & \ \ $-2.840219$\ \ & \ \ $ 0.012692$\ \ \\
 4\ \ & \ \  18\ \ & \ \ $-3.837655$\ \ & \ \ $-3.534888$\ \ & \ \  14\ \ & \ \ $-3.813058$\ \ & \ \ $ 0.024597$\ \ & \ \ $-3.809984$\ \ & \ \ $ 0.027671$\ \ \\
 5\ \ & \ \  30\ \ & \ \ $-4.810806$\ \ & \ \ $-4.610344$\ \ & \ \  25\ \ & \ \ $-4.786288$\ \ & \ \ $ 0.024518$\ \ & \ \ $-4.779627$\ \ & \ \ $ 0.031179$\ \ \\
 6\ \ & \ \  45\ \ & \ \ $-5.783597$\ \ & \ \ $-5.645735$\ \ & \ \  40\ \ & \ \ $-5.745154$\ \ & \ \ $ 0.038443$\ \ & \ \ $-5.754097$\ \ & \ \ $ 0.029500$\ \ \\
 7\ \ & \ \  63\ \ & \ \ $-6.775065$\ \ & \ \ $-6.626187$\ \ & \ \  57\ \ & \ \ $-6.767761$\ \ & \ \ $ 0.007304$\ \ & \ \ $-6.732099$\ \ & \ \ $ 0.042965$\ \ \\
 \end{tabular}
 \end{table}

In addition, I have calculated the energies of the states with a few neighboring angular momenta. I have found that, for the total angular momentum ${\cal L}=3{\cal L}_{\min}-\delta{\cal L}$, the lowest energy level $E_1^{3{\cal L}_{\min}-\delta{\cal L}}$ lies between $E_1^{3{\cal L}_{\min}}$ and $E_2^{3{\cal L}_{\min}}$, 
\be 
E_1^{3{\cal L}_{\min}}<E_1^{3{\cal L}_{\min}-\delta{\cal L}}<E_2^{3{\cal L}_{\min}}.
\ee 
The value of $\delta{\cal L}$ here coincides with the order of the rotational symmetry axis (\ref{NsymStep}),
\be 
\delta{\cal L}=n^{\rm st}(N) \label{deltaL}.
\ee 
Thus, while the ground state of the system is the lowest energy state with the angular momentum ${\cal L}=3{\cal L}_{\min}$, the first excited state of the system is the lowest energy state with ${\cal L}=3{\cal L}_{\min}-\delta{\cal L}$, 
\be 
E_{\rm GS}=E_1^{3{\cal L}_{\min}},\ \ E_{\rm 1st}=E_1^{3{\cal L}_{\min}-\delta{\cal L}}.\label{GS1st}
\ee
This result is valid for the step-like density profile.

The fifth and sixth columns of Table \ref{tab:ExactGSEnergy} show the angular momentum $3{\cal L}_{\min}-\delta{\cal L}$ and the energy $E_1^{3{\cal L}_{\min}-\delta{\cal L}}$. The energy difference $E_{\rm 1st}-E_{\rm GS}$ between the first excited and the ground state is shown in the seventh column in Table \ref{tab:ExactGSEnergy}. This difference is always smaller than $0.1e^2/a_0$ and mainly decreases as $N$ grows. The case of $N=6$ particles is an exception related to the rearrangement of the shell structure from $(0,N)$ into $(1,N-1)$ at $N=6$, see Figures \ref{fig:WCconfigs}(e,f). 

\begin{figure}[ht!]
\includegraphics[width=0.49\columnwidth]{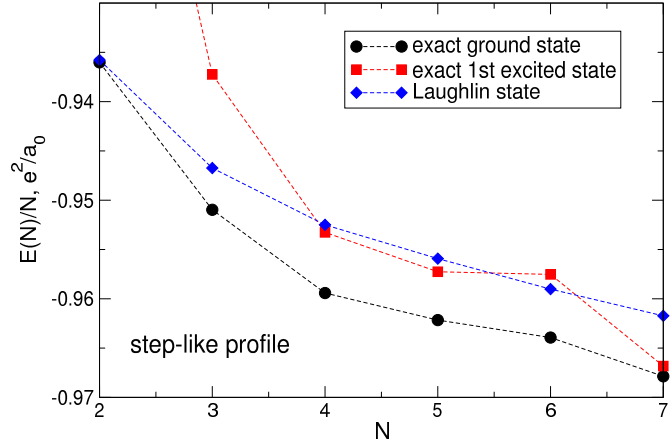}
\caption{\label{fig:ExactGSEnergy}The energy per particle of the exact ground (black circles) and the first excited (red squares) states, Eq. (\ref{GS1st}), as well as of the Laughlin state (\ref{LaughlinWF}), at $\nu=1/3$, as a function of the number of particles $N$. The energies are in units $e^2/a_0$. }
\end{figure}

Figure \ref{fig:ExactGSEnergy} shows the energies (per particle) of the ground and the first excited states, Eq. (\ref{GS1st}), as a function of the electron number $N$ (black circles and red squares, respectively). 

The expansion coefficients $A_s^{\rm GS}$ of the ground state many-body wave functions $|\Psi_{\rm GS}\rangle$ over the basis states $|\Psi_s\rangle$, see Eq. (\ref{PsiExpansion}), are shown in Tables \ref{tab:As2part} and \ref{tab:As3part} for two and three particles. The largest contribution to the ground state wave function for $N=2$ and $3$ is given by the states $|1,2\rangle$ and $|2,3,4\rangle$, respectively. The vectors $A_s^{\rm GS}$ for $N\ge 4$ particles can be found in Ref. \cite{datafiles}, see also Appendix \ref{app:suppl}.

\begin{table}[ht!]
 \caption{The expansion coefficients $A_s^{\rm GS}$ for the exact ground state of $N=2$ particles.
 \label{tab:As2part}}
 \begin{tabular}{c|c|r}
 $s$&\ \   state\ \ \  &\ \   $A_s^{\rm GS}$\ \ \ \\
 \hline
 1  & $|0,3\rangle$ & $-0.47078$ \\
 2  & $|1,2\rangle$ & $0.88225$ \\
 \end{tabular}
 \end{table}

\begin{table}[ht!]
 \caption{The expansion coefficients $A_s^{\rm GS}$ for the exact ground state of $N=3$ particles.
 \label{tab:As3part}}
 \begin{tabular}{c|c|r}
 $s$&\ \   state\ \ \  &\ \   $A_s^{\rm GS}$\ \ \ \\
 \hline
 1  & $|0,1,8\rangle$ & $0.01613$ \\
 2  & $|0,2,7\rangle$ & $-0.03971$ \\
 3  & $|0,3,6\rangle$ & $-0.08983$ \\
 4  & $|0,4,5\rangle$ & $0.32071$ \\
 5  & $|1,2,6\rangle$ & $0.29234$ \\
 6  & $|1,3,5\rangle$ & $-0.44922$ \\
 7  & $|2,3,4\rangle$ & $0.77458$ \\
 \end{tabular}
 \end{table}

\subsection{Density of electrons in the ground state\label{sec:ExactDensity}}

\begin{figure}[ht!]
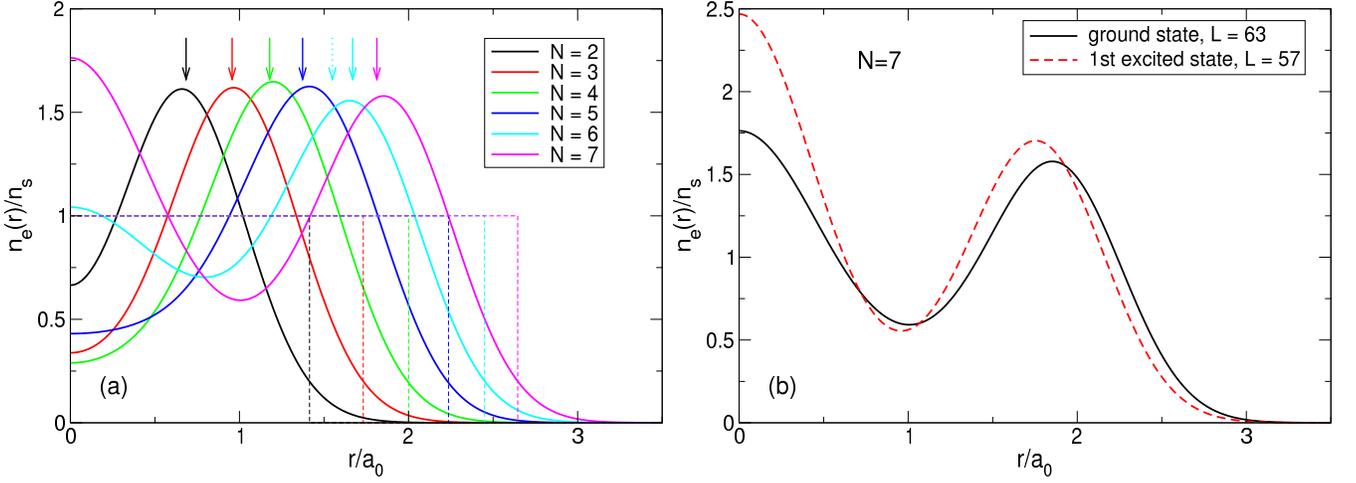

\includegraphics[width=0.49\columnwidth]{DensityExactN27step.eps}
\includegraphics[width=0.49\columnwidth]{DensityN7step.GS1st.eps}
\caption{\label{fig:DensityExact} (a) The density of electrons in the exact ground state (thick solid curves), together with the density of the positive background (thin dashed curves) for $N=2\dots 7$ for the step-like background density profile (\ref{dens-step}). Arrows above the curves show the radii of the outer shells $R_s(N)$ of the Wigner molecules, see Table \ref{tab:WCmoleculesStep}; for $N=6$ the shell radii of both $(0,N)$ (dashed arrow) and $(1,N-1)$ (solid arrow) configurations are shown. (b) The density of electrons in the exact ground state (${\cal L}=63$) and the first excited state (${\cal L}=57$) for $N=7$.}
\end{figure}

The density of electrons in the exact ground states in systems with different $N$ can be calculated using the matrix elements (\ref{electrondensity}) and the expansion coefficients of the exact ground state wave function $A_s^{\rm GS}$,
\be 
n_e^{\rm GS}(r)=\sum_{s=1}^{N_{mbs}} \left(A_s^{\rm GS}\right)^2\sum_{j=1}^N |\psi_{L_j^{(s)}}(\bm r)|^2.
\label{DensExactGS}
\ee
The results of calculations are shown in Figure \ref{fig:DensityExact}(a) for $N=2,\dots, 7$. The electron density calculated by quantum mechanics is in complete agreement with the expectations arising from the classical considerations in Section \ref{sec:Wigner}. When $N$ grows from 2 to 4, the density curves have maxima at the points which are very close to the radii of the classical Wigner molecules for these $N$, and the densities in the disk center $n_e(r=0)$ decrease, since as $N$ increases, the electrons repel stronger from the disk center. Starting from $N=5$, the probability of finding one electron in the disk center starts to grow. At $N=5$, the density at $r=0$ increases, and the maximum of the curve shifts slightly beyond the position of the classical Wigner molecule shell, since this central electron pushes the other electrons towards the edge of the disk. At $N=6$, a local maximum at $r=0$ appears. At $N=7$ a giant maximum arises in the disk center, with the value of $n_e(r=0)$, larger than the $n_e$-maximum at a finite $r\simeq R_s$. 

Figure \ref{fig:DensityExact}(b) shows the density of electrons in the exact ground state (${\cal L}=63$) and in the first excited state (${\cal L}=57$) for seven particles. It can be seen that the distribution of electrons in the first excited state also resembles a Wigner molecule, but with somewhat different parameters: in the excited state, the probability of finding an electron in the disk center is higher than in the ground state, and the radius of the outer shell of the electron ring is slightly smaller. Physical reasons of such a behavior of the densities of the ground and the first excited states will be clarified in Section \ref{sec:ExactSolutionNu<1}.

The exact quantum-mechanical solution thus clearly shows that the ground state of a few (up to $N=7$) two-dimensional electrons at the Landau level filling factor $\nu=1/3$ has the form resembling a floating (or sliding) Wigner crystal molecule \cite{Mikhailov02d,Yannouleas03,Yannouleas07,Lewin18,Lewin19}. This agrees with the intuitive understanding of the physics of Coulomb-interacting particles. 

\subsection{Pair correlation function in the ground state \label{sec:ExactPCF}}

\begin{figure}[ht!]
\includegraphics[width=0.49\columnwidth]{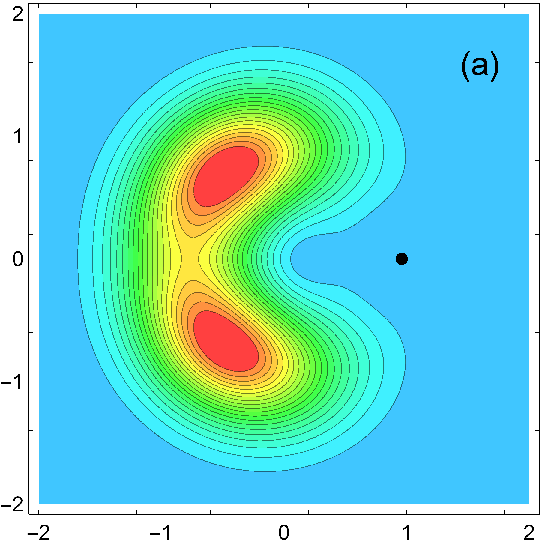}
\includegraphics[width=0.49\columnwidth]{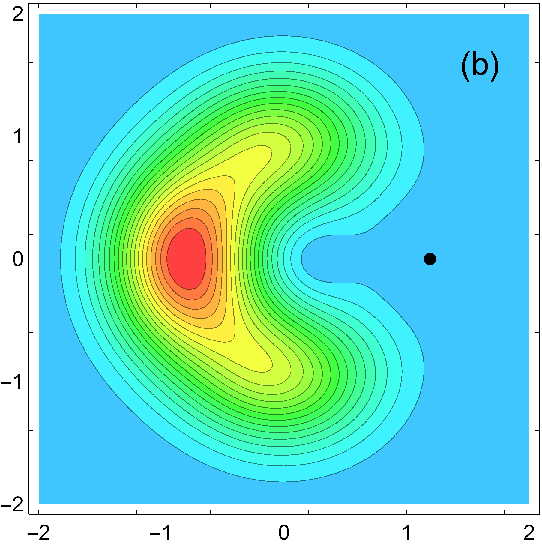}\\
\includegraphics[width=0.49\columnwidth]{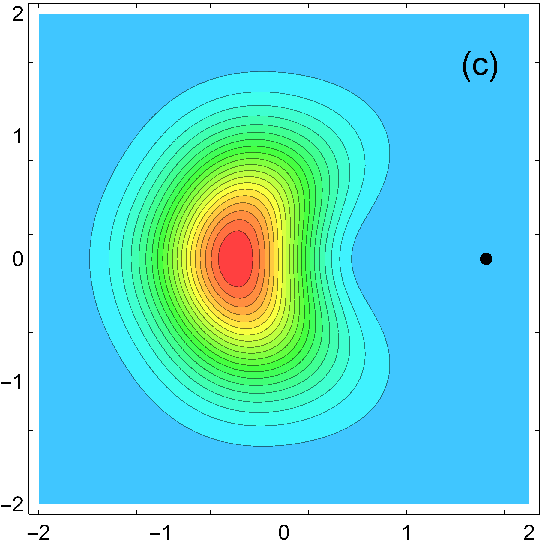}
\includegraphics[width=0.49\columnwidth]{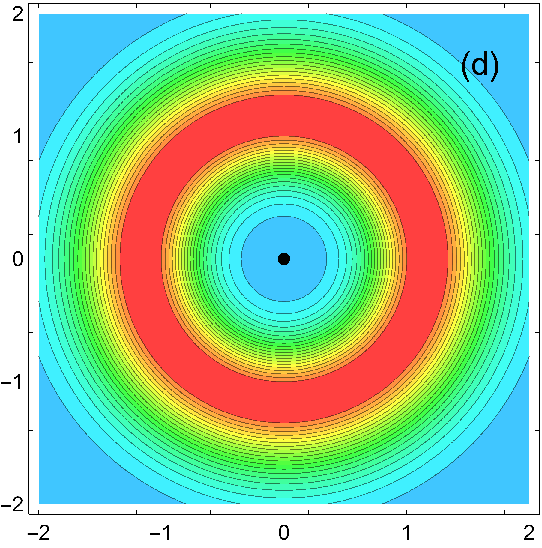}
\caption{\label{fig:PCFexact}The pair correlation function $P_{\rm GS}(\bm r,\bm r')$ in the ground state, as a function of $x/a_0$ and $y/a_0$, for different $N$ and different coordinates $\bm r'$: (a) $N=3$ and $\bm r'/a_0=(0.96,0)$, (b) $N=4$ and $\bm r'/a_0=(1.19,0)$, (c) $N=6$ and $\bm r'/a_0=(1.65,0)$, and (d) $N=7$ and $\bm r'/a_0=(0,0)$. The positions of the $\bm r'$ points are shown by small black circles; in (a),(b), and (c) they correspond to the maxima of the electron density, see Figure \ref{fig:DensityExact}.}
\end{figure}

The pair correlation function in the ground state can be calculated using the formula
\be 
P_{\rm GS}(\bm r,\bm r')=\sum_{s=1}^{N_{mbs}} A_s^{\rm GS}\sum_{s'=1}^{N_{mbs}} A_{s'}^{\rm GS}\langle\Psi_s|\hat P(\bm r,\bm r')|\Psi_{s'}\rangle,
\label{PCF}
\ee
with the matrix elements determined by Eqs. (\ref{Prron})--(\ref{Prroff}). Examples of the calculated function $P_{\rm GS}(\bm r,\bm r')$ are shown in Figure \ref{fig:PCFexact}. Here, the colored distribution shows the probability to find an electron at a point $\bm r$, under the condition that another electron is fixed at the point $\bm r'$ shown by small black circles on the panels. The panels (a), (b), and (c) illustrate the cases when the points $\bm r'$ are outside the disk center at the distance corresponding to the maxima of the electron density, see Figure \ref{fig:DensityExact}. Figure \ref{fig:PCFexact}(a) illustrates the case of $N=3$ particles. One sees a triangular structure which resembles the classical Wigner molecule configuration shown in Figure \ref{fig:WCconfigs}(b). Starting from  $N=4$, crystalline electron-electron correlations are significantly weakened: they are still weakly seen in Figure \ref{fig:PCFexact}(b) corresponding to $N=4$, but becomes invisible at larger $N$, see Figure \ref{fig:PCFexact}(c) for $N=6$, where only one maximum is seen opposite to the point $\bm r'$. This is a consequence of the significant overlap of single-particle wave functions and of the averaging of their positions over the angular coordinate.  

Figure \ref{fig:PCFexact}(d) shows the pair correlation function for the case $N=7$ and $\bm r'=\bm 0$. Now the function $P_{\rm GS}(\bm r,\bm 0)$ is circularly symmetric and has a maximum at $r\approx 1.16a_0$. The behavior of $P(\bm r,\bm 0)$ at small $r$ is quadratic, $P_{\rm GS}(\bm r,\bm 0)\propto r^2$ at $r\to 0$.

Thus, although the ground state of the system at $\nu = 1/3$ resembles a Wigner molecule, its internal structure is not rigid like that of a real crystal. The analogy with the Wigner crystal appears only in the fact that the maxima of the electron density are located where they are expected according to the distribution of classical particles. 

\section{Maximum density droplet state, $\nu=1$\label{sec:MDD}}

Using the mathematical apparatus developed in Section \ref{sec:formulation}, one can now calculate properties of the states (\ref{LaughlinWF}) for few-electron systems and compare them with those of the exact ground state. But before proceeding to the discussion of the case $m = 3$ ($\nu = 1/3$), it will be useful to briefly overview physical properties of the MDD configuration \cite{Bychkov81} which is a special case of the function (\ref{LaughlinWF}) at $m = 1$. Only results for the step-like profile of the uniform background will be shown in this Section.

\subsection{Wave function}

The MDD configuration 
\be
\Psi_{\rm mdd}=|0,1,2,\dots,N-1\rangle=
\frac{1}{\sqrt{N!}}\det
\left |\begin{array}{cccc}
\psi_{0}(\bm r_1) & 
\psi_{0}(\bm r_2) & 
\dots &
\psi_{0}(\bm r_N) \\
\psi_{1}(\bm r_1) & 
\psi_{1}(\bm r_2) &
\dots & 
\psi_{1}(\bm r_N) \\
\dots &\dots &\dots &\dots \\
\psi_{N-1}(\bm r_1) & 
\psi_{N-1}(\bm r_2) & 
\dots & 
\psi_{N-1}(\bm r_N) \\
\end{array}
\right |,
\label{manybodyMDD}
\ee
describes the ground state of the system at $\nu=1$, if the influence of higher Landau levels is neglected, Ref. \cite{Bychkov81}. In this configuration, $N$ electrons occupy the lowest Landau level single-particle states with the smallest possible individual angular momenta from $L=0$ up to $L=N-1$. Since $\psi_{L}(\bm r)$ is proportional to $z^L$, the matrix in (\ref{manybodyMDD}) has the form of the Vandermonde matrix, and its determinant can be presented in the form (\ref{LaughlinWF}) with $m=1$. The total angular momentum in the MDD state is ${\cal L}={\cal L}_{\min}=N(N-1)/2$, Eq. (\ref{Lmin}). The number of many-body configurations for a given $N$ and ${\cal L}={\cal L}_{\min}$ equals one, $N_{mbs}(N,{\cal L}_{\min})=1$. 

\subsection{Energy of an $N$-particle system}

\begin{figure}[ht!]
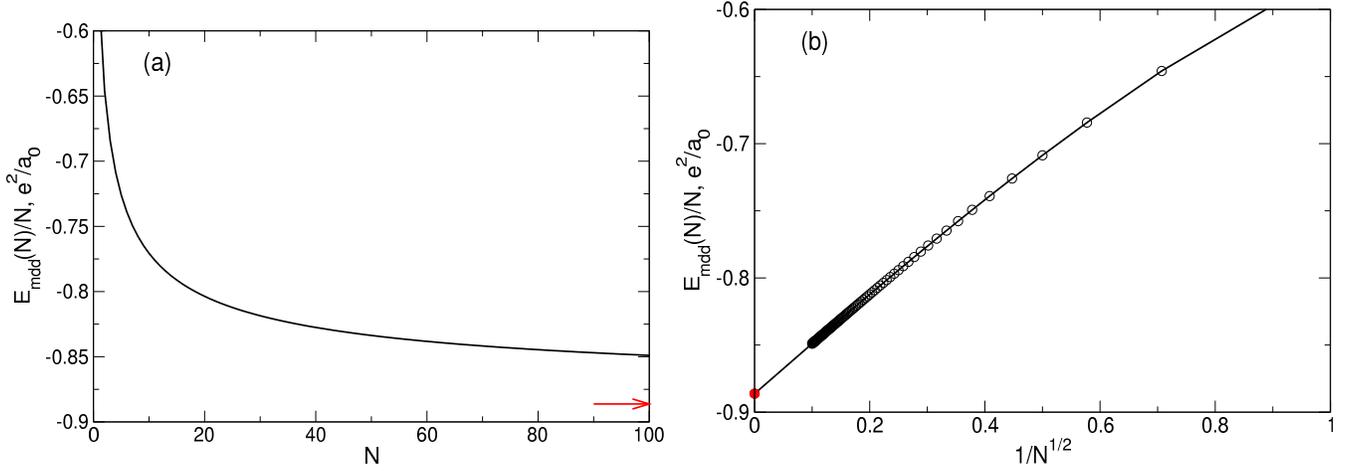

\includegraphics[width=0.49\columnwidth]{MDD_Energy_St.eps}
\includegraphics[width=0.49\columnwidth]{MDD_Energy_St_Vs1SqrtN.eps}
\caption{\label{fig:mdd-energy}The energy per particle of the MDD state (\ref{manybodyMDD}), measured in units $e^2/a_0$, (a) as a function of the number of electrons $N$ and (b) as a function of $1/\sqrt{N}$. The red arrow in (a) and red point in (b) show the asymptotic value (\ref{EmddPpLargeN}), reported in Ref. \cite{Laughlin83}.
}
\end{figure}

The energy of the MDD state 
\be 
E_{\rm mdd}(N)=\langle \Psi_{\rm mdd}|\hat V_{bb}+\hat V_{be}+\hat V_{ee}|\Psi_{\rm mdd}\rangle
\label{EmddStep}
\ee
can be calculated using the Coulomb matrix elements found in Section \ref{sec:MBmatrixelements}. The background-background interaction energy in this formula is given, for the step-like background density profile, by Eq. (\ref{BBmbStep-matrixelements}), calculation of the background-electron interaction energy in the MDD case gives 
\be 
\langle \Psi_{\rm mdd}|\hat V_{be}|\Psi_{\rm mdd}\rangle =
-N\frac{e^2}{a_0} 2\frac{\Gamma\left(N+\frac 12\right)}{\Gamma\left(N\right)}{_2F_2}\left(-\frac 12,\frac 12;2,\frac 12-N;-N\right),
\label{EMDDbeStep}
\ee
see Appendix \ref{app:mddBEmdd}, and the electron-electron (Hartree minus Fock) interaction energy is
\be
\langle \Psi_{\rm mdd}|\hat V_{ee}|\Psi_{\rm mdd}\rangle=
\frac{e^2}{a_0} \sqrt{\frac{\pi}2}
\sum_{L=0}^{N-2} 
\sum_{L'=L+1}^{N-1} 
\Big[
{\cal K}(L,L',0) -{\cal K}(L,L,L'- L)
\Big],
\label{EMDDee}
\ee
The energy (\ref{EmddStep}) (per particle) calculated for $N$ up to $N=100$ is shown in Figure \ref{fig:mdd-energy} by the black curves and symbols, as a function of $N$ in Figure \ref{fig:mdd-energy}(a), and as a function of $N^{-1/2}$ in Figure \ref{fig:mdd-energy}(b). 

\subsection{Electron density}

\begin{figure}[ht!]
\includegraphics[width=0.5\columnwidth]{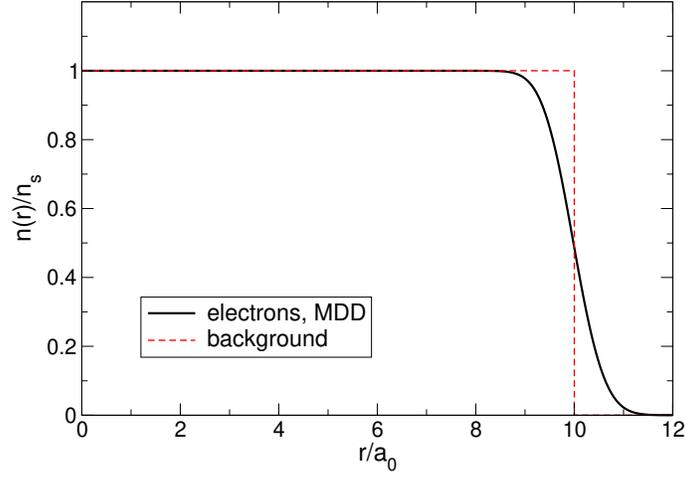}
\caption{\label{fig:MDDdens}The density of electrons in the MDD state at $\nu=1$ (black solid curve) and the density of the positive background $n_b(r)$ (red dashed curve) for $N = 100$. The density is plotted as a function of $r/a_0$ or $r/\lambda$ (at $\nu=1$ $\lambda=a_0$).}
\end{figure}

The density of electrons in the MDD state (\ref{manybodyMDD}), according to (\ref{electrondensity}), is
\be 
n_{e}^{\rm mdd}(\bm r) =
\sum_{L=0}^{N-1} |\psi_{L}(\bm r)|^2
=\frac{1}{\pi\lambda^2} 
e^{-r^2/\lambda^2}
\sum_{L=0}^{N-1} \frac{\left(r/\lambda \right)^{2L}}{L!}
=n_s Q(N,r^2/\lambda^2).
\label{electrondensityMDD}
\ee
The density $n_{e}^{\rm mdd}(r)$ is constant and equals $n_s$ up to $r\simeq R-\lambda=R-a_0$, Figure \ref{fig:MDDdens}. In the limit $N\to\infty$ the density $n_{e}^{\rm mdd}(r)$ becomes homogeneous in the whole 2D space. The state (\ref{manybodyMDD}) thus describes an ideally uniform liquid. The structure typical for a Wigner crystal does not arise in this state, since the function (\ref{manybodyMDD}) does not satisfy the many-body Schr\"odinger equation (\ref{NbodySchrEq}), (\ref{MBhamilt}), i.e. in the MDD state, the Coulomb interaction is completely ignored. In order to get a more accurate description of the ground state at $\nu=1$, the single-particle states from the higher Landau levels should be taken into account. 

\subsection{Pair correlation function}

\begin{figure}[t!]
\includegraphics[width=0.49\columnwidth]{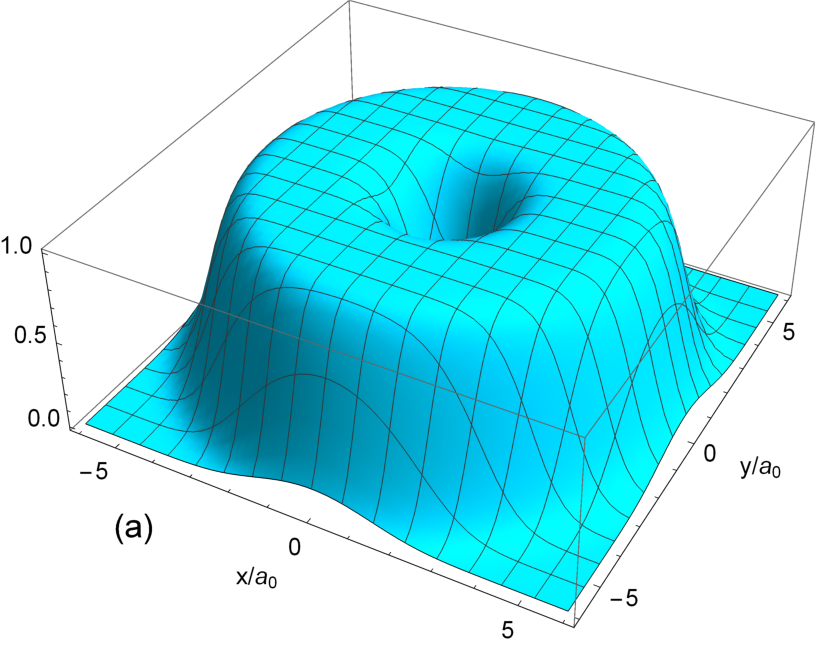}
\includegraphics[width=0.49\columnwidth]{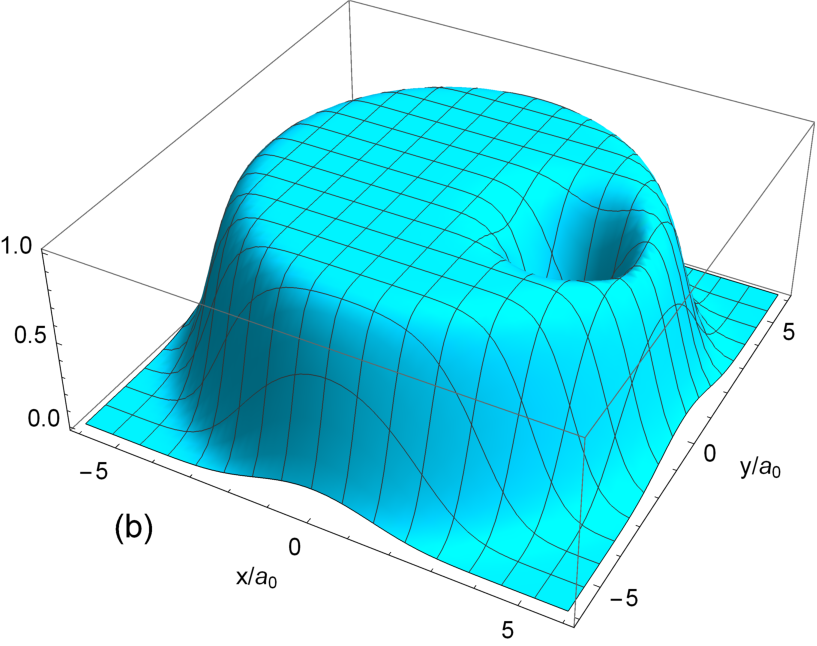}
\caption{\label{fig:pcf-mdd}The pair correlation function of the MDD state (\ref{manybodyMDD}), as a function of $\bm r/a_0$, at $N=30$ and (a) $\bm r'/a_0=(0,0)$ and (b) $\bm r'/a_0=(3,0)$.
}
\end{figure}

The pair correlation function of the MDD configuration can be found by calculating the average value of the operator (\ref{PairCorrFun}) with the wave function (\ref{manybodyMDD}). It can be presented in the form
\be 
P(\bm r,\bm r')=
n_{e}(\bm  r)n_{e}(\bm  r') + \delta P(\bm r,\bm r'),
\label{deltaPCF}
\ee
where 
\be 
\delta P_{\rm mdd}(\bm r,\bm r')=-n_s^2 e^{-|z -z'|^2}
Q(N,zz'^\star)Q(N,z^\star z'),
\label{PCF-mdd}
\ee
and $z=(x-iy)/\lambda$.
The dependence of $P_{\rm mdd}(\bm r,\bm r')$ on $x/a_0$ and $y/a_0$ at $\bm r'/a_0=(0,0)$ and $\bm r'/a_0=(3,0)$ for $N=30$ particles is shown in Figure \ref{fig:pcf-mdd}. The pair correlation function tends to unity when $|\bm r-\bm r'|/a_0\gg 1$ and both points, $\bm r$ and $\bm r'$, are far from the system boundaries. In the limit $|\bm r-\bm r'|\to 0$ the function $P_{\rm mdd}(\bm r,\bm r')$ tends to zero as $|\bm r-\bm r'|^2$, due to the exchange ``interaction''. 

\subsection{Thermodynamic limit}

If $\Psi$ is a many-body wave function of an $N$-particle system, the average value of the energy in this state $E(N)=\langle \Psi|\hat{\cal H}|\Psi\rangle=\langle \Psi|\hat V_C|\Psi\rangle$ can be presented in the form (the kinetic energy contribution $N\hbar\omega_c/2$ is omitted)
\be 
E(N)=\langle \Psi|\hat V_C|\Psi\rangle=\frac{e^2}{2}\int \frac{d\bm r d\bm r'}{|\bm r-\bm r'|}\delta n(\bm r)\delta n(\bm r')
+\frac{e^2}{2}\int \frac{d\bm r d\bm r'}{|\bm r-\bm r'|} \delta P(\bm r,\bm r'),\label{EnergyViaPCF}
\ee 
where $\delta n(\bm r)=n_e(\bm r)-n_b(\bm r)$, and $\delta P(\bm r,\bm r')$ is defined in Eq. (\ref{deltaPCF}). The first term in (\ref{EnergyViaPCF}) is the Hartree energy, which vanishes if the electron and positive background densities are identical. This condition is satisfied at $\nu=1$ in the MDD state in the limit $N\to\infty$, when the density of electrons $n_{e}^{\rm mdd}(r)$ becomes ideally uniform in the whole 2D space, $n_{e}^{\rm mdd}( r)\to n_s$. The regularized incomplete Gamma function $Q(N,x)$ tends to unity at $N\to\infty$, and the pair correlation function (\ref{PCF-mdd}) becomes a function of only the inter-particle distance $|\bm r-\bm r'|$,
\be 
\delta P_{\rm mdd}(\bm r-\bm r')=-n_s^2e^{-|\bm r-\bm r'|^2/a_0^2}.
\label{PCF-mddInfN}
\ee
For the MDD energy one then gets from (\ref{EnergyViaPCF}) 
\be 
E_{\rm mdd}(N)\approx \frac{e^2}{2}\int d\bm r \int \frac{d\bm r'}{|\bm r-\bm r'|} \delta P_{\rm mdd}(|\bm r-\bm r'|) \approx \frac{e^2}{2}\pi R^2 \int \frac{d\bm r'}{|\bm r'|} \delta P_{\rm mdd}(|\bm r'|).\label{EmddViaPCFa}
\ee 
Since $\pi R^2 n_s=N$, this gives the energy per particle in the thermodynamic limit $N\to\infty$
\be 
\frac{E_{\rm mdd}(N)}{N} =\frac{\pi e^2}{n_s}  \int_0^\infty dr \delta P_{\rm mdd}(r)=-\frac{e^2}{a_0}\frac{\sqrt{\pi}}{2}.
\label{EmddPpLargeN}
\ee
This formula was reported for $\nu=1$ in Ref. \cite{Laughlin83}. The asymptotic value (\ref{EmddPpLargeN}) is shown by the red arrow and red point in Figures \ref{fig:mdd-energy}(a) and \ref{fig:mdd-energy}(b). Notice that the formulas (\ref{EmddViaPCFa}) and (\ref{EmddPpLargeN}) are valid only if the density of electrons is uniform in the whole 2D space -- only in this case the Hartree energy in (\ref{EnergyViaPCF}) vanishes and the pair correlation function $P_{\rm mdd}(\bm r,\bm r')$ becomes a function of $|\bm r-\bm r'|$. 

\section{Laughlin state at $\nu=1/3$\label{sec:Laughlin13}}

\subsection{General remarks\label{sec:Laughlin13general}}

Now, let us calculate the energy, electron density, and other physical properties of the system in the state (\ref{LaughlinWF}) at $\nu=1/m=1/3$. To do this I proceed as follows.
 
The function (\ref{LaughlinWF}) is the eigenfunction of the angular momentum operator with the eigenvalue 
\be 
{\cal L}=m\frac{N(N-1)}{2}=3\frac{N(N-1)}{2}.
\label{L}
\ee 
For any given $N$ and ${\cal L}$ I determine $N_{mbs}$ many-body states, formed from the lowest Landau level single-particle functions (\ref{sp-wavefun}), and expand the function (\ref{LaughlinWF}) in these basis states,
\be 
\Psi_{\rm LS}^{(m=3)}=\sum_{s=1}^{N_{mbs}} A_s^{\rm LS}\Psi_s.
\label{PsiExpand}
\ee
The coefficients $A_s^{\rm LS}$ here are real numbers, and the function (\ref{PsiExpand}) is assumed to be normalized, i.e., $\sum_{s=1}^{N_{mbs}} \left(A_s^{\rm LS}\right)^2=1$. After the coefficients $A_s^{\rm LS}$ are found, any physical quantity $F$ can be calculated as the average
\be 
F=\langle \Psi_{\rm LS}^{(m=3)}|\hat F|\Psi_{\rm LS}^{(m=3)}\rangle =
\sum_{s=1}^{N_{mbs}}\sum_{s'=1}^{N_{mbs}} A_s^{\rm LS} A_{s'}^{\rm LS}\langle \Psi_{s}|\hat F|\Psi_{s'}\rangle,
\ee
where the matrix elements $\langle \Psi_{s}|\hat F|\Psi_{s'}\rangle$ are found in Section \ref{sec:MBmatrixelements} for commonly used operators. How to calculate the coefficients $A_s^{\rm LS}$?

\subsection{Expansion coefficients of the Laughlin function \label{sec:LaughExpansion}}

The coefficients $A_s^{\rm LS}$ in (\ref{PsiExpand}) can be found as follows. First, using the binomial expansion of the polynomial factors in the function (\ref{LaughlinWF}), I obtain integer binomial coefficients $C_s$, as in the following example for two particles:
\be
(z_1-z_2)^3=z_1^3-3z_1^2z_2+3z_1z_2^2-z_2^3=( -1)\det 
\left|\begin{array}{cc}
z_1^0 & z_2^0 \\
z_1^3 & z_2^3 \\
\end{array}
\right|+3\det
\left|\begin{array}{cc}
z_1^1 & z_2^1 \\
z_1^2 & z_2^2 \\
\end{array}
\right|;\label{z1z2expan}
\ee
I denote the coefficients $C_s$ as
\be 
C_{|0,3\rangle}=-1 \ \ \textrm{and }\ \ C_{|1,2\rangle}=3.
\label{coefs2part}
\ee
The determinants in (\ref{z1z2expan}) are proportional to the basis functions $|0,3\rangle$ and $|1,2\rangle$. For example 
\be 
|0,3\rangle=
\frac 1{\sqrt{2!}} \det
\left|\begin{array}{cc}
\psi_0(\bm r_1) & \psi_0(\bm r_2) \\
\psi_3(\bm r_1) & \psi_3(\bm r_2) \\
\end{array}
\right|=
\frac 1{\sqrt{2!}} \det
\left|\begin{array}{cc}
z_1^{0}  & z_2^{0}  \\
z_1^{3}  & z_2^{3}  \\
\end{array}
\right|
\frac{e^{-(|z_1|^2+|z_2|^2)/2}}{\pi \lambda^2\sqrt{0! 3!}},
\ee
so that one gets
\be 
\det
\left|\begin{array}{cc}
z_1^{0}  & z_2^{0}  \\
z_1^{3}  & z_2^{3}  \\
\end{array}
\right| \propto \sqrt{0! 3!}\ |0,3\rangle.
\ee 
Then the expansion of the function (\ref{LaughlinWF}) for two particles takes the form
\ba 
(z_1-z_2)^3 e^{-(|z_1|^2+|z_2|^2)/2}
\propto D_{|0,3\rangle}  |0,3\rangle
+D_{|1,2\rangle}  |1,2\rangle,
\label{LaughlinWFm3normN2}
\ea
where
\be 
D_{|L_1,L_2\rangle} = \sqrt{L_1! L_2!} C_{|L_1,L_2\rangle}.
\ee 
Thus, in order to calculate (for any $N$) the real coefficients $A_s^{\rm LS}$ in the expansion (\ref{PsiExpand}) one needs to find, first, the integers binomial coefficients $C_s=C_{|L_1,L_2,\dots,L_N\rangle}$ in the expansions of the polynomial factors in (\ref{LaughlinWF}), then calculate the $D_s$ factors according to the formula
\be 
D_{|L_1,L_2,\dots,L_N\rangle} = \sqrt{L_1! L_2!\dots L_N!} C_{|L_1,L_2,\dots,L_N\rangle},\label{coefsDs}
\ee 
and finally determine the real coefficients $A_s^{\rm LS}$ by using the normalization condition
\be 
A_s^{\rm LS}=\frac{D_s}{\sqrt{\sum_{p=1}^{N_{mbs}}D_p^2}}.\label{AsviaDs}
\ee
Now I apply the described algorithm to the Laughlin wave function for a few values of the particle number $N$.

\subsubsection{Two particles}

If $N=2$, then the angular momentum (\ref{L}) at $\nu=1/3$ equals ${\cal L}=3$. The number of many-body states in this case is $N_{mbs}=2$, and they are $|0,3\rangle$ and $|1,2\rangle$, see Table \ref{tab:N2configs}. Expanding the factor $(z_1-z_2)^3$ as in (\ref{z1z2expan}) I get the coefficients $C_s$, Eq. (\ref{coefs2part}), $D_s$, $A_s^{\rm LS}$, and $\left(A_s^{\rm LS}\right)^2$. The results are shown in Table \ref{tab:LaughCsN2}. 

 \begin{table}[ht!]
 \caption{Many-body configurations for $N=2$ and ${\cal L}=3$, with the
 expansion coefficients $C_s$, $A_s^{\rm LS}$, and $\left(A_s^{\rm LS}\right)^2$. The coefficients $D_s$ can be found
using Eq. (\ref{coefsDs}).\label{tab:LaughCsN2}}
 \begin{tabular}{c|c|r|c|c}
  No. & Configuration & $C_s$ & $A_s^{\rm LS}$ & $\left(A_s^{\rm LS}\right)^2$ \\
 \hline
 1 & $| 0, 3\rangle$ & $   -1$ & $    -0.5$ & $     0.25$ \\
 2 & $| 1, 2\rangle$ & $    3$ & $     0.86603$ & $     0.75$ \\
 \end{tabular}
 \end{table}

\subsubsection{Three particles}

If $N=3$, then the angular momentum (\ref{L}) equals ${\cal L}=9$. Now one has $N_{mbs}=7$ many-body states, see Table \ref{tab:N3configs}. Using again the binomial expansion
\be
(z_1-z_2)^3(z_1-z_3)^3(z_2-z_3)^3\propto -
\Delta_{|0,3,6\rangle} +3
\Delta_{|0,4,5\rangle} +3
\Delta_{|1,2,6\rangle} -6
\Delta_{|1,3,5\rangle} +15
\Delta_{|2,3,4\rangle},
\label{laughlin3}
\ee
where
\be 
\Delta_{|j,k,l\rangle}=\det
\left|\begin{array}{ccc}
z_1^{j}  & z_2^{j} & z_3^j \\
z_1^{k}  & z_2^{k} & z_3^k \\
z_1^{l}  & z_2^{l} & z_3^l \\
\end{array}
\right|,
\ee
I get the coefficients $C_s$, $A_s^{\rm LS}$, and $\left(A_s^{\rm LS}\right)^2$ shown in Table \ref{tab:LaughCsN3}. Not all possible basis functions $\Psi_s$ are represented in the LS (\ref{LaughlinWF}). Two configurations, $|0,1,8\rangle$ and $|0,2,7\rangle$, have zero weights in $\Psi_{\rm LS}^{(m=3)}$. 

\begin{table}[ht!]
 \caption{Many-body configurations for $N=3$ and ${\cal L}=9$, with the
 expansion coefficients $C_s$, $A_s^{\rm LS}$, and $\left(A_s^{\rm LS}\right)^2$. The coefficients $D_s$ can be found using Eq. (\ref{coefsDs}).\label{tab:LaughCsN3}}
 \begin{tabular}{c|c|r|r|c}
  No. & Configuration & $C_s$ & $A_s^{\rm LS}$\ \ \ \ & $\left(A_s^{\rm LS}\right)^2$ \\
 \hline
 1 & $| 0, 1, 8\rangle$ & $    0$ & $     0.0 $ & $     0.0 $ \\
 2 & $| 0, 2, 7\rangle$ & $    0$ & $     0.0 $ & $     0.0 $ \\
 3 & $| 0, 3, 6\rangle$ & $   -1$ & $    -0.17961$ & $     0.03226$ \\
 4 & $| 0, 4, 5\rangle$ & $    3$ & $     0.43994$ & $     0.19355$ \\
 5 & $| 1, 2, 6\rangle$ & $    3$ & $     0.31109$ & $     0.09677$ \\
 6 & $| 1, 3, 5\rangle$ & $   -6$ & $    -0.43994$ & $     0.19355$ \\
 7 & $| 2, 3, 4\rangle$ & $   15$ & $     0.69561$ & $     0.48387$ \\
 \end{tabular}
 \end{table}

\subsubsection{Four particles}

If $N=4$, then the total angular momentum (\ref{L}) equals ${\cal L}=18$, and the number of many-body states is $N_{mbs}=34$. Applying the binomial expansion procedure I calculate the numbers $C_s$. Only 16 of them are nonzero; these states, the corresponding integer binomial coefficients $C_s$, as well as the numbers $A_s^{\rm LS}$ and $\left(A_s^{\rm LS}\right)^2$, are shown in Table \ref{tab:LaughCsN4}. Other 18 states, namely the states
\ba 
|0,1,2,15\rangle, \ |0,1,3,14\rangle, \ |0,1,4,13\rangle, \ |0,1,5,12\rangle, \ |0,1,6,11\rangle, \ |0,1,7,10\rangle, \ |0,1,8,9\rangle, \nonumber \\
|0,2,3,13\rangle, \ |0,2,4,12\rangle, \ |0,2,5,11\rangle, \ |0,2,6,10\rangle, \ |0,2,7,9\rangle, \nonumber \\
|0,3,4,11\rangle, \ |0,3,5,10\rangle, \nonumber \\ 
|1,2,3,12\rangle, \ |1,2,4,11\rangle, \ |1,2,5,10\rangle, \nonumber \\
|1,3,4,10\rangle  \nonumber
\ea
have zero weights $C_s=0$ in the LS. 

 \begin{table}[ht!]
 \caption{Many-body configurations for $N=4$ and ${\cal L}=18$, with the
 expansion coefficients $C_s$, $A_s^{\rm LS}$, and $\left(A_s^{\rm LS}\right)^2$. The coefficients $D_s$ can be found
using Eq. (\ref{coefsDs}).\label{tab:LaughCsN4}}
 \begin{tabular}{c|c|r|r|c}
  No. & Configuration & $C_s$\ \ & $A_s^{\rm LS}$\ \ \ \ & $\left(A_s^{\rm LS}\right)^2$ \\
 \hline
15 & $| 0, 3, 6, 9\rangle$ & $    1$ & $     0.05322$ & $     0.00283$ \\
16 & $| 0, 3, 7, 8\rangle$ & $   -3$ & $    -0.14082$ & $     0.01983$ \\
17 & $| 0, 4, 5, 9\rangle$ & $   -3$ & $    -0.13037$ & $     0.01700$ \\
18 & $| 0, 4, 6, 8\rangle$ & $    6$ & $     0.21290$ & $     0.04533$ \\
19 & $| 0, 5, 6, 7\rangle$ & $  -15$ & $    -0.42078$ & $     0.17705$ \\
23 & $| 1, 2, 6, 9\rangle$ & $   -3$ & $    -0.09219$ & $     0.00850$ \\
24 & $| 1, 2, 7, 8\rangle$ & $    9$ & $     0.24391$ & $     0.05949$ \\
26 & $| 1, 3, 5, 9\rangle$ & $    6$ & $     0.13037$ & $     0.01700$ \\
27 & $| 1, 3, 6, 8\rangle$ & $  -12$ & $    -0.21290$ & $     0.04533$ \\
28 & $| 1, 4, 5, 8\rangle$ & $   -9$ & $    -0.13037$ & $     0.01700$ \\
29 & $| 1, 4, 6, 7\rangle$ & $   27$ & $     0.33872$ & $     0.11473$ \\
30 & $| 2, 3, 4, 9\rangle$ & $  -15$ & $    -0.20614$ & $     0.04249$ \\
31 & $| 2, 3, 5, 8\rangle$ & $   27$ & $     0.27656$ & $     0.07649$ \\
32 & $| 2, 3, 6, 7\rangle$ & $   -6$ & $    -0.05322$ & $     0.00283$ \\
33 & $| 2, 4, 5, 7\rangle$ & $  -45$ & $    -0.32593$ & $     0.10623$ \\
34 & $| 3, 4, 5, 6\rangle$ & $  105$ & $     0.49787$ & $     0.24788$ \\
 \end{tabular}
 \end{table}

\subsubsection{Five to eight particles}

The cases of $N=5,\dots,8$ particles can be analyzed similarly. The full tables with all many-body configurations and their weights are very large and not shown here; they can be found in Ref. \cite{datafiles}, see Appendix \ref{app:suppl}. Here, I give only a brief overview of some key features of the Laughlin function expansions that are useful for the subsequent analysis of the problem.

The number of many-body configurations which are \textit{not} used in the LS dramatically grows with the number of particles $N$. Table \ref{tab:LaughZeroStates} shows the total number of many-body configurations for given $N$ and ${\cal L}$, $N_{mbs}(N,{\cal L})$, the number of states $N_{mbs}^{=0}$ which do not contribute to the LS, the number of states which give a nonzero contribution to $\Psi_{\rm LS}^{(m=3)}$, $N_{mbs}^{\neq 0}$, and the percentage of many-body configuration \textit{not contributing} to the function (\ref{LaughlinWF}) (denoted as ``\% zero'' in the Table \ref{tab:LaughZeroStates}). One sees that, while for $N=3$ ``only'' 28.57\% of all possible many-body configurations are not used in the LS, at $N=8$ this number increases up to 90.54\%. 

\begin{table}[ht!]
\caption{The total number of many-body states $N_{mbs}$, the number of states contributing ($N_{mbs}^{\neq 0}$) and not contributing ($N_{mbs}^{= 0}$) to the Laughlin function, as well as the percentage of non-contributing many-body configurations (`\% zero'). $N$ is the number of particles and ${\cal L}$ is the total angular momentum (\ref{L}) corresponding to $\nu=1/3$ and $m=3$ in Eq. (\ref{LaughlinWF}). \label{tab:LaughZeroStates}}
\begin{tabular}{crrrrr}
\ \ $N$\ \  & \ \ ${\cal L}$ \ \ & \ \ $N_{mbs}\ $ & \ \ \ $N_{mbs}^{=0}$ \ & \ \ \ $N_{mbs}^{\neq 0}$ \ & \ \ \ \% zero\\
\hline
2 &  3 &     2 &     0 &      2 & 00.00 \\
3 &  9 &     7 &     2 &      5 & 28.57 \\
4 & 18 &    34 &    18 &     16 & 52.94 \\
5 & 30 &   192 &   133 &     59 & 69.27 \\
6 & 45 &  1206 &   959 &    247 & 79.52 \\
7 & 63 &  8033 &  6922 &   1111 & 86.17 \\
8 & 84 & 55974 & 50680 &   5294 & 90.54 \\
\end{tabular}
\end{table}

The first nonzero-weight many-body configurations for $N$ particles at $\nu=1/3$ have the form $| 0, 3,\dots,3(N-1)\rangle$, e.g., $|0,3,6,9,12,15,18\rangle$ for seven particles. The angular momenta of individual particles in these states are $L_j=3(j-1)$ for $j=1,\dots,N$. The largest angular momentum of individual particles in all possible many-body configurations which contribute to the LS equals $L_{\max}=L_N=3(N-1)$. In general, if $\nu=1/m$ in Eq. (\ref{LaughlinWF}), $m=3,5,7$, then $L_{\max}=L_N=m(N-1)$. 

Having obtained the expansion coefficients $C_s$, $D_s$ and $A_s^{\rm LS}$ for different numbers of particles I can now calculate physical properties of the LS (\ref{LaughlinWF}). 

\subsection{Energy of the Laughlin state \label{sec:Laughlin13energy}}

\begin{figure}[ht!]
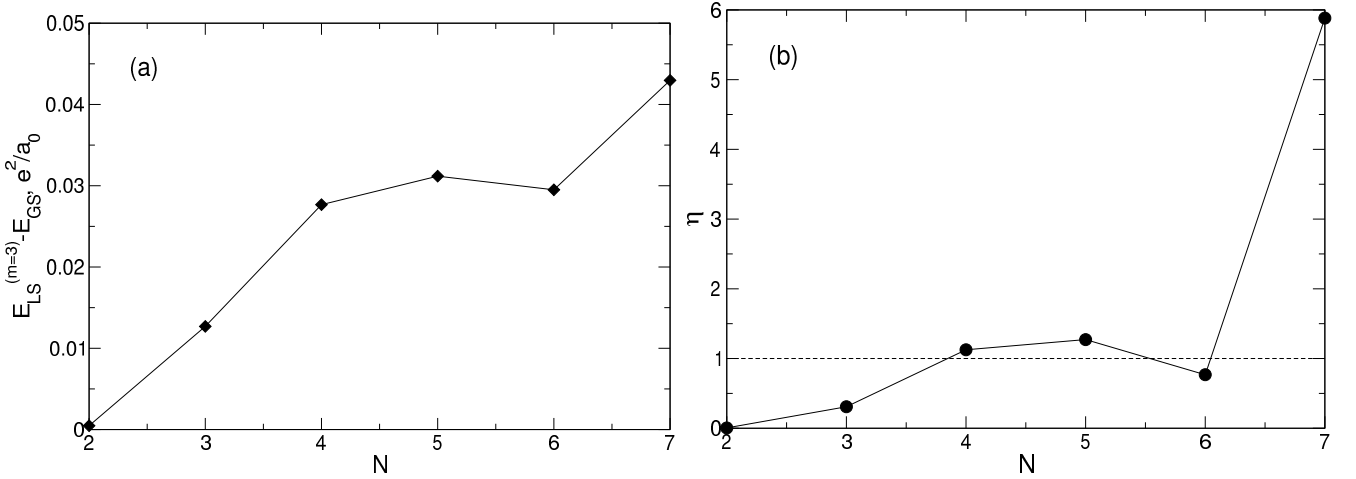

\includegraphics[width=0.5\columnwidth]{EnergyRLminusGS_NOTperParticle.eps}
\includegraphics[width=0.48\columnwidth]{RLminGSrelativeStep.eps}
\caption{\label{fig:EnergyDiffRL-GS} (a) The energy difference $E_{\rm LS}^{(m=3)}-E_{\rm GS}$ between the Laughlin state and the true ground state of a system of $N$ particles as a function of $N$; the energy unit is $e^2/a_0$. (b) The value of $\eta$ defined by Eq. (\ref{eta}), as a function of $N$. }
\end{figure}

The energy of the LS is determined by the formula (the constant kinetic energy contribution $N\hbar\omega_c/2$ is omitted)
\be 
E_{\rm LS}^{(m=3)}(N)=\langle \Psi_{\rm LS}^{(m=3)}|\hat {\cal H}|\Psi_{\rm LS}^{(m=3)}\rangle = 
\sum_{s=1}^{N_{mbs}} \sum_{s'=1}^{N_{mbs}} A_{s}^{\rm LS}A_{s'}^{\rm LS}\langle \Psi_{s}|\hat V_{bb}+\hat V_{eb}+\hat V_{ee}|\Psi_{s'}\rangle 
\label{LaughlinEnergy}
\ee
in which all required matrix elements are calculated above. The calculated values of the energy (\ref{LaughlinEnergy}), as well as the energy differences $E_{\rm LS}^{(m=3)}-E_{\rm GS}$ between the LS and the true ground state are given in Table \ref{tab:ExactGSEnergy} in the last two columns. The difference $E_{\rm LS}^{(m=3)}-E_{\rm GS}$ increases significantly with the growth of $N$ (by more than 90 times when $N$ changes from $N=2$ to $N=7$), see Figure \ref{fig:EnergyDiffRL-GS}(a). In order to quantitatively estimate how big the deviation of the Laughlin energy from the true ground state energy is, I introduce a dimensionless quantity
\be 
\eta=\frac{E_{\rm LS}^{(m=3)}-E_{\rm GS}}{E_{\rm 1st}-E_{\rm GS}}\label{eta}
\ee
which measures the difference $E_{\rm LS}^{(m=3)}-E_{\rm GS}$ in units of the energy gap $E_{\rm 1st}-E_{\rm GS}$ between the true first excited state and the true ground state at $\nu=1/3$. The value of $\eta$ is plotted as a function of $N$ in Figure \ref{fig:EnergyDiffRL-GS}(b). For $N=4$, 5 and 7 the value of $\eta$ is greater than 1, i.e. the energy of the LS is not only greater than the energy of the ground state, which is understandable for the trial wave function, but also greater than the energy of the first excited state. For $N=7$, the value of $\eta$ equals 5.8825 at $\nu=1/3$. 
 
For $N=6$, the wave functions of the LS and the true ground state are accidentally close to each other; this can be seen from the comparison of the electron densities in both states in Section \ref{sec:LaughDensity}. This is the reason why both the absolute and relative energy differences between the LS and the true ground state are smaller at $N=6$ compared to neighboring values of $N$.

\subsection{Deviation of the Laughlin wave function from the true ground state wave function \label{sec:Laughlin13projection}}

\begin{figure}[ht!]
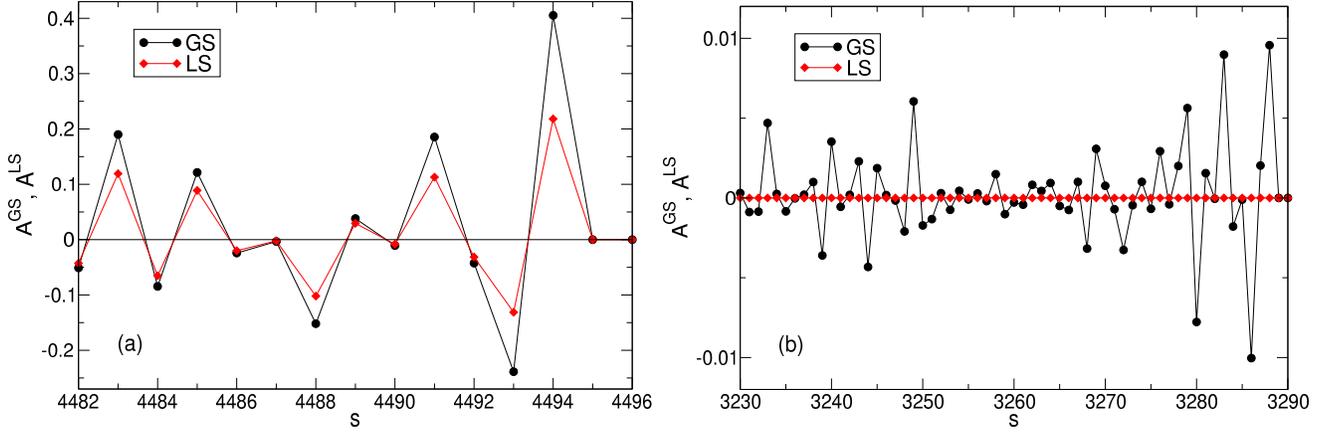

\includegraphics[width=0.48\columnwidth]{AsAsN_7a.eps}
\includegraphics[width=0.48\columnwidth]{AsAsN_7b.eps}
\caption{\label{fig:GS-RLdeviation} The expansion coefficients $A_s^{\rm GS}$ and $A_s^{\rm LS}$ of the true ground state wave function (\ref{PsiExpansion}) and of the Laughlin state (\ref{PsiExpand}) for a few selected many-body basis states: (a) $4482\le s\le 4496$ and (b) $3230\le s\le 3280$. The Landau level filling factor is $\nu=1/3$, the number of particles is $N=7$, the total angular momentum is ${\cal L}=63$, and the total number of many-body basis states is $N_{mbs}=8033$.}
\end{figure}

The energy difference between the LS and the exact ground state is very large. How big is the difference between the wave functions? Figure \ref{fig:GS-RLdeviation} shows the expansion coefficients of the true ground state wave function and the Laughlin wave function, $A_s^{\rm GS}$ and $A_s^{\rm LS}$, for several many-body basis states, in the intervals $4482\le s\le 4496$ and $3230\le s\le 3290$ selected as examples. One sees a large difference between the vectors $A_s^{\rm GS}$ and $A_s^{\rm LS}$. In the first example, Figure \ref{fig:GS-RLdeviation}a, both coefficients are finite, and $A_s^{\rm GS}$ can be almost twice as big as $A_s^{\rm LS}$. For example, for $s=4493$ and 4494 (the states $|0,7,9,10,11,12,14\rangle$ and $|0,8,9,10,11,12,13\rangle$) the ratio $A_s^{\rm GS}/A_s^{\rm LS}$ equals 1.819 and 1.858, respectively. In the second example, Figure \ref{fig:GS-RLdeviation}b, the coefficients $A_s^{\rm GS}$ of all $\sim 60$ $s$-states are on the order of $0.005-0.01$, i.e. are close to the average value $\bar A_s=N_{mbs}^{-1/2}\simeq 0.011157$, while the coefficients $A_s^{\rm LS}$ for these states are identically zero.  

In order to quantitatively characterize the overall discrepancy between the states $\Psi_{\rm GS}$ and $\Psi_{\rm LS}^{(m=3)}$, I calculate the standard deviation 
\be 
D=\sqrt{\sum_{s=1}^{N_{mbs}} \left(A_{s}^{\rm GS}-A_{s}^{\rm LS}\right)^2}.
\label{standdev}
\ee 
Table \ref{tab:projection} shows the quantity $D$ for $N$ varying from $N=2$ up to $N=7$. The second column of Table \ref{tab:projection} shows the projection of the LS onto the true ground state
\be 
P=\langle \Psi_{\rm LS}^{(m=3)}|\Psi_{\rm GS}\rangle = 
\sum_{s=1}^{N_{mbs}} A_{s}^{\rm LS}A_{s}^{\rm GS}=1-D^2/2.
\label{projection}
\ee 
For $N=2$ the deviation (\ref{standdev}) is quite small, about $3.3$\%, but as $N$ increases, it substantially grows and exceeds $\sim 40$\% for $N=7$. 

\begin{table}[ht!]
 \caption{The standard deviation (\ref{standdev}) and the projection (\ref{projection}) of the Laughlin wave function (\ref{LaughlinWF}) from/onto the ground state wave function (\ref{PsiExpansion}) at $\nu=1/3$.\label{tab:projection}}
 \begin{tabular}{c|c|c}
  $N$\ \ & $D$ & $P$ \\
 \hline
 2 & 0.0334& 0.9994  \\    
 3 & 0.1755& 0.9846  \\    
 4 & 0.2978& 0.9557  \\    
 5 & 0.3500& 0.9387  \\    
 6 & 0.2960& 0.9562  \\    
 7 & 0.4021& 0.9191  \\
 \end{tabular}
 \end{table}

In Ref. \cite{Laughlin83} Laughlin gave the following numbers for the projections of his function on the numerically calculated ground state at $\nu=1/3$: $P_3=0.99946$ for $N=3$ and $P_4=0.979$ for $N=4$. My calculations do not confirm these numbers. As seen from Table \ref{tab:projection}, the values of $P_2=0.9994$ and $P_3=0.9846$ in the first two lines are close to the numbers given in Ref. \cite{Laughlin83}, but they refer to $N=2$ and $N=3$ respectively. When $N=4$, the projection $P$ is less than 95.6\% which corresponds to the deviation $D$ of almost 30\%. 

\subsection{Electron density in the Laughlin state \label{sec:LaughDensity}}

\begin{figure}[ht!]
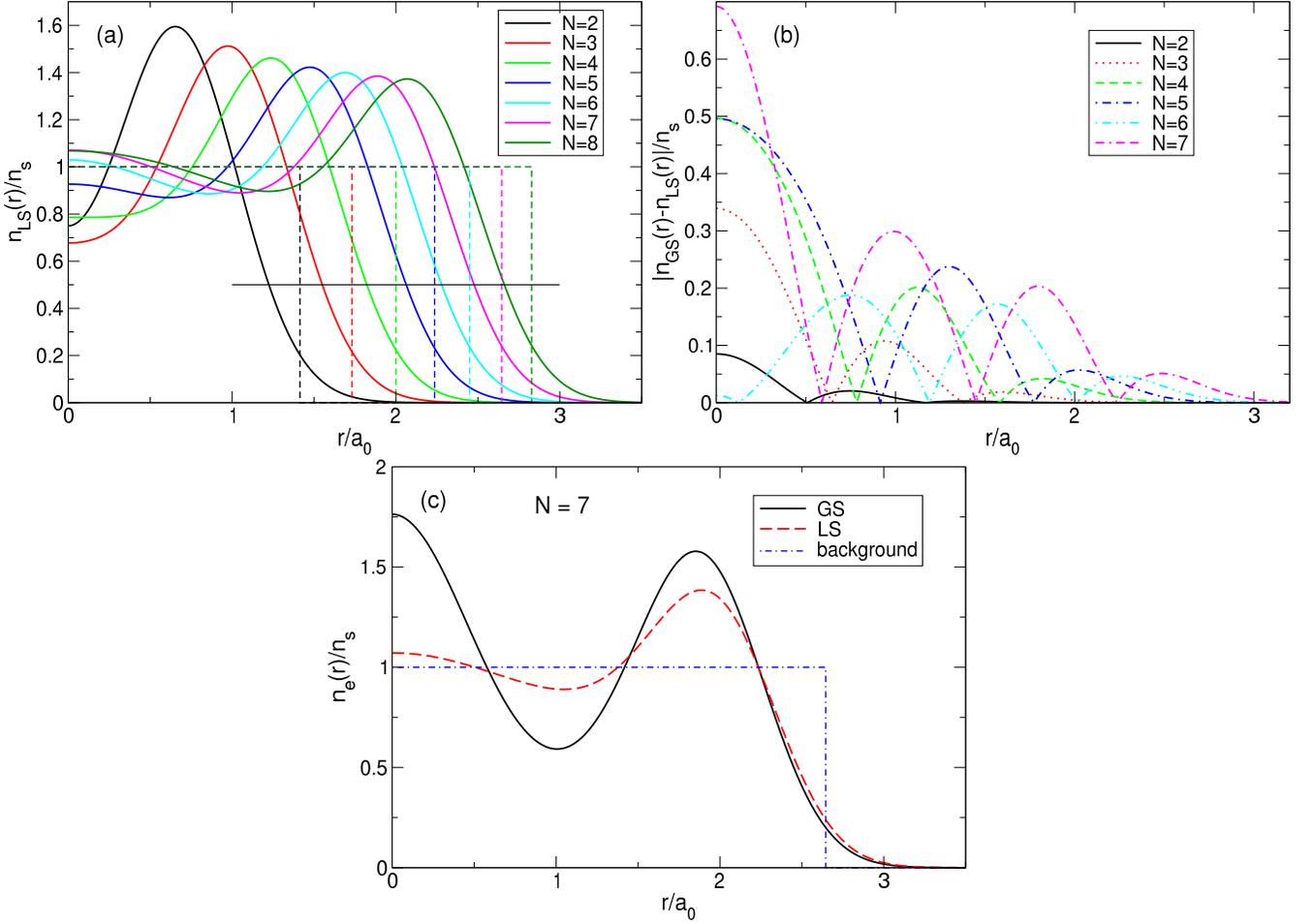

\includegraphics[width=0.49\columnwidth]{LaughlinDensity.eps}
\includegraphics[width=0.49\columnwidth]{BetragDensityDifferenceStepProfile.eps}
\includegraphics[width=0.49\columnwidth]{DensityExactLaughN7step.eps}
\caption{\label{fig:LaughDens} (a) The density of electrons $n_e^{\rm LS}(r)$ (solid curves) in the Laughlin state (\ref{LaughlinWF}) at $\nu=1/3$, as a function of the radial coordinate for $N=2$ to $8$. The dashed lines show the corresponding positive background densities; the thin horizontal black line indicates the level 0.5, which determines the radii of the electron disks. (b) The absolute value of the density difference between the true ground state and the LS for different $N$. (c) The density of electrons in the true ground state (GS) and the Laughlin state (LS) at $N=7$.
}
\end{figure}

The density of electrons in the LS (\ref{LaughlinWF}) at $\nu=1/3$ is determined by the formula similar to (\ref{DensExactGS}), 
\be 
n_e^{\rm LS}(r)=\sum_{s=1}^{N_{mbs}} \left(A_s^{\rm LS}\right)^2\sum_{j=1}^N |\psi_{L_j^{(s)}}(\bm r)|^2.
\label{DensLaughlin}
\ee
The normalized density (\ref{DensLaughlin}) is shown in Figure \ref{fig:LaughDens}(a) for $2\le N\le 8$. At $N\le 4$ my results coincide with those obtained by Ciftja et al. in Ref. \cite{Ciftja11} by a different method.

If $N\lesssim 5$, the density of the state (\ref{LaughlinWF}) behaves qualitatively similar to the exact electron density, Figure \ref{fig:DensityExact}(a): both have a maximum at a finite $r$, and this maximum shifts to a larger $r$ as $N$ increases. However, quantitatively, the density difference at $r/a_0\ll 1$ becomes very large already at $N\ge 3$, see Figure \ref{fig:LaughDens}(b): while at $N=2$ the Laughlin density differs from the exact one ``only'' by $\sim 8.5$\%, at $N=3$ the difference is already about 34\%, and at $N=4$ and 5 it approaches 50\%. At $N=6$ the densities of electrons in the both states become close to each other, but this is a coincidental result of two different trends. While for Laughlin electrons $n_{\rm LS}(0)/n_s$ tends to unity at $N\to\infty$, Ref. \cite{Ciftja03,Ciftja04}, the ratio $n_{\rm GS}(0)/n_s$ is close to 1 only because the maximum in the disk center is not yet sufficiently developed. When $N$ grows further, the local density difference becomes huge approaching $\sim 70$\% for $N=7$ at $r\ll a_0$, and the coordinate dependencies of the exact and Laughlin densities becomes \textit{qualitatively} different, see Figure \ref{fig:LaughDens}(c). While the exact density shows the formation of a structure resembling a sliding Wigner crystal, with a large density maximum arising in the disk center, the Laughlin density flattens out in the inner part of the disk.

\subsection{Pair correlation function in the Laughlin state \label{sec:LaughPCF}}

\begin{figure}[ht!]
\includegraphics[width=0.49\columnwidth]{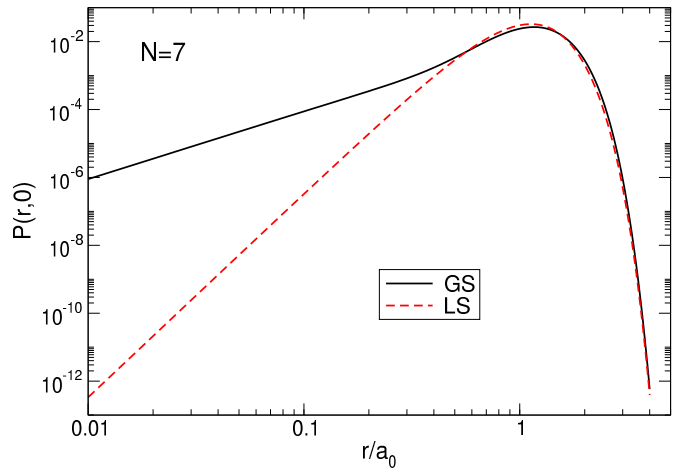}
\caption{\label{fig:LaughPCF} The pair correlation functions $P(\bm r,\bm 0)$ for the exact ground state (GS) and the Laughlin state (LS), for a system of $N=7$ particles.
}
\end{figure}

The pair correlation function $P_{\rm LS}(\bm r,\bm r')$ in the LS at $\nu=1/3$ can be calculated using a formula similar to (\ref{PCF}); only the coefficients $A_s^{\rm GS}$ should be replaced by $A_s^{\rm LS}$. In general, the function $P_{\rm LS}(\bm r,\bm r')$ looks similarly to the plots of Figure \ref{fig:PCFexact}. But there exists a very large quantitative difference at small $|\bm r-\bm r'|$, when $|\bm r-\bm r'|/a_0\lesssim 0.6$, see Figure \ref{fig:LaughPCF}. While the exact pair correlation function tends to zero at $|\bm r-\bm r'|\to 0$ as $P_{\rm GS}(\bm r,\bm r')\propto |\bm r-\bm r'|^2$, the LS pair correlation function vanishes as $P_{\rm LS}(\bm r,\bm r')\propto |\bm r-\bm r'|^6$ at $|\bm r-\bm r'|\to 0$. This is a direct consequence of the unphysical assumption in (\ref{LaughlinWF}), that the wave function contains only the polynomials $(z_j-z_k)^3$. In the real world this is not the case, see an additional discussion of this point in Sections \ref{sec:RjRk} and \ref{sec:discussionAcceptedTheory}.

\subsection{Additional remark to the case of $N=7$ particles\label{sec:remark7}}

Results obtained for the energy of the ground and the first excited states of a seven-electron system need a little more discussion. Laughlin assumed \cite{Laughlin83} that the ground state of an $N$-particle system at $\nu=1/3$ should have the total angular momentum ${\cal L}=3N(N-1)/2$. For $N=7$ this is ${\cal L}=63$. My exact results show that the ground state of the seven-particle system has the angular momentum ${\cal L}=63$ indeed, while the first excited states has the angular momentum ${\cal L}=57$, Table \ref{tab:ExactGSEnergy}. The difference between the energies of the first excited and the ground state is very small and equals  $E_{\rm 1st}-E_{\rm GS}=0.0073 e^2/a_0$. 

However, Kasner et al. \cite{Kasner94} also performed exact diagonalization calculations for $N$ electrons in the disk geometry. Their calculations showed that the ground state of the seven-particle system at $\nu=1/3$ has the angular momentum ${\cal L}=57$. This \textit{exact} result contradicted the variational theory of Laughlin, but the authors of \cite{Kasner94} could not resolve this ``dilemma'' and decided just to ``disregard this difficulty''.

Questions arise: Why do the exact results of Ref. \cite{Kasner94} contradict my exact results and the assumption of Laughlin? Were the results of Ref. \cite{Kasner94} correct?

This ``dilemma'' has a simple explanation. Results for the energy shown in Section \ref{sec:ExactEnergy} (Table \ref{tab:ExactGSEnergy}) are obtained for the positive background density having the step-like form (\ref{dens-step}). Kasner et al. \cite{Kasner94} performed their exact diagonalization calculations assuming the smooth density profile (\ref{dens-smooth}). I have also done calculations for the smooth density profile (\ref{dens-smooth}) and my results \textit{confirm} those of Ref. \cite{Kasner94}. For $N=7$ and $\nu=1/3$ my results are shown in Table \ref{tab:N7EnergyGS1st2ndLauSmooth}. One sees that, in agreement with \cite{Kasner94}, in the case of the smooth density profile the ground state has the total angular momentum ${\cal L}=57$, while the first excited state has ${\cal L}=63$. Moreover, the LS energy not only exceeds the energy of the first excited state, but is larger than the energy of the second excited state, see an additional discussion of this point in Section \ref{sec:resultsSmooth}. Notice also that for the smooth density profile the projection of the LS onto the true ground state wave function is zero, since these functions have different angular momenta ${\cal L}$ and are therefore orthogonal. 

\begin{table}[!ht]
\caption{The energies of the ground state, first and second excited states, as well as of the Laughlin state, at $\beta=1/\nu =3$, in the case of the smooth density profile. All energies are in units $e^2/a_0$. \label{tab:N7EnergyGS1st2ndLauSmooth}}
\begin{tabular}{l|c|c|c}
\hspace{3mm}State & \hspace{2mm}${\cal L}$ \hspace{2mm}  & \hspace{5mm}$E_{\rm state}$ \hspace{5mm}  & \hspace{5mm}$E_{\rm state}-E_{\rm GS}$\hspace{5mm}\\
\hline
Laughlin & 63 &	-6.6363835 &	0.0431284 \\
2nd excited &	51&	-6.6407409 & 0.0387709 \\
1st excited &	63	& -6.6613834 &	0.0181285 \\
Ground  &	57	& -6.6795119 & 0.0 \\
\end{tabular}
\end{table}

\subsection{About the variational principle in the thermodynamic limit\label{sec:doesvarprinwork}}

The results obtained above show that the trial wave function (\ref{LaughlinWF}) proposed in Ref. \cite{Laughlin83} for the ground state of the FQHE system has very little in common with the true ground state wave function. The argument of \cite{Laughlin83} about sufficiently large projections of the wave function (\ref{LaughlinWF}) onto the numerically calculated exact ground state for $N = 3$ and 4 loses its persuasiveness when the number of particles increases up to $N=7$. Another Laughlin argument, that seemed to be rather strong, was that the function (\ref{LaughlinWF}) gives the lowest energy per particle at $N\to\infty$, $E_{\rm LS}^{(m=3)}/N=-0.4156 e^2/l_B$ at $\nu=1/3$, compared to all other trial functions. How convincing is this argument? 

Let us assume that the true ground ($s=1$) and excited ($s=2,3,\dots$) states of the system are known. Their energies and the wave functions are $E_s$ and $\Psi_s$. Let an arbitrary trial wave function be a linear combination of the ground ($\Psi_1\equiv \Psi_{\rm GS}$) and several low-lying excited states,
\be 
\Psi_{\rm trial}=\alpha_1\Psi_{\rm GS}+\sum_{s>1}\alpha_s\Psi_{s}.
\ee
The energy of the trial state $E_{\rm trial}$ will then be
\be 
E_{\rm trial}=\alpha_1^2E_{\rm GS}+\sum_{s>1}\alpha_s^2E_{s}=E_{\rm GS}+\sum_{s>1}\alpha_s^2(E_{s}-E_{\rm GS}),
\ee
where the normalization condition $\sum_{s}\alpha_s^2=1$ is taken into account. 

The energy differences between the low-lying excited states and the ground state $E_{s}-E_{\rm GS}$ do not depend on $N$ and are determined by typical energies of the FQHE problem, $e^2/a_0$, $e^2/l_B$, or $\hbar\omega_c$, which are all on the meV scale. But the energy $E_{\rm GS}$ is proportional to the number of particles $N$ and tends to minus infinity in the thermodynamic limit. Under typical experimental conditions ($N\simeq 10^{11}$) it is on the GeV scale. Therefore, the energy per particle in the limit $N\to\infty$ will be the same both for the true ground state and \textit{for any} trial wave function:
\be 
\lim_{N\to\infty}\frac{E_{\rm trial}}{N}=\lim_{N\to\infty}\frac{E_{\rm GS}}N+\sum_{s>1}\alpha_s^2\lim_{N\to\infty}\frac{E_{s}-E_{\rm GS}}N=\lim_{N\to\infty}\frac{E_{\rm GS}}N.
\ee
This conclusion holds also in the case $\alpha_1=0$, when the trial and the ground state wave functions are orthogonal. Thus, while the ``thermodynamic limit'' argument may give a correct estimate for the ground state energy per particle, it completely fails to determine the correct ground state wave function. The quantum-mechanical variational principle, that perfectly works for one- or few-particle systems, is useless in the limit $N\to\infty$. Relying on this principle, one can mistakenly take any arbitrarily unreasonable wave function as a correct ground state wave function. This can be illustrated by a simple quantitative example.

In Section \ref{sec:MDD} the properties of the MDD configuration $|\Psi_{\rm mdd}\rangle=|0,1,\dots,N-1\rangle$, Eq. (\ref{manybodyMDD}), have been analyzed, in particular the energy, Figure \ref{fig:mdd-energy}, and the density, Figure \ref{fig:MDDdens}, of this state. Let us now consider two other quantum states, which I will call ``MDD-plus'' and ``MDD-shift''. The MDD-plus state $|\Psi_{\rm mdd}^+\rangle=|0,1,\dots,N-2,N\rangle$ differs from $|\Psi_{\rm mdd}\rangle$ by the angular momentum of only one, $N$-th electron: $L_N=N-1\ \to\ L_N=N$. In the MDD-shift state, $|\Psi_{\rm mdd}^\Rightarrow\rangle=|1,2,\dots,N-1,N\rangle$, individual angular momenta of all particles are increased by one as compared to their MDD $L$'s: $L_j=j-1\ \to\ L_j=j$, $j=1,\dots,N$. The total angular momenta of the plus- and shift-states are ${\cal L}=N(N-1)/2+1$ and ${\cal L}=N(N+1)/2$ respectively, and hence, these two states are orthogonal to the ground MDD state: their projections onto the ground state are zero. What are the energy and the density of these two configurations? This question can be easily answered for any $N$ with the help of the formulas obtained above. 

\begin{figure}[ht!]
\includegraphics[width=0.49\columnwidth]{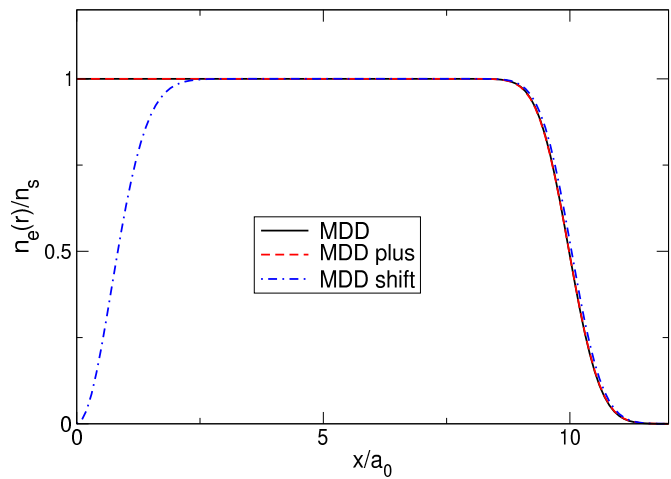}
\caption{\label{fig:MDDtypeStatesDensity} The density of electrons in the MDD, MDD-plus, and MDD-shift states as a function of $r/a_0$ for $N=100$.}
\end{figure}

Figure \ref{fig:MDDtypeStatesDensity} shows the electron density in the MDD, MDD-plus, and MDD-shift states. The densities of the first two states are very close to each other. But the density of the MDD-shift state has a deep hole in the disk center and differs significantly from both the MDD and MDD-plus states. If nothing was known about the ground state of the system, then the MDD-shift state would definitely be excluded from the list of potential candidates for the role of the ground state wave function, since it does not give a physically reasonable coordinate dependence of the electron density.

\begin{figure}[ht!]
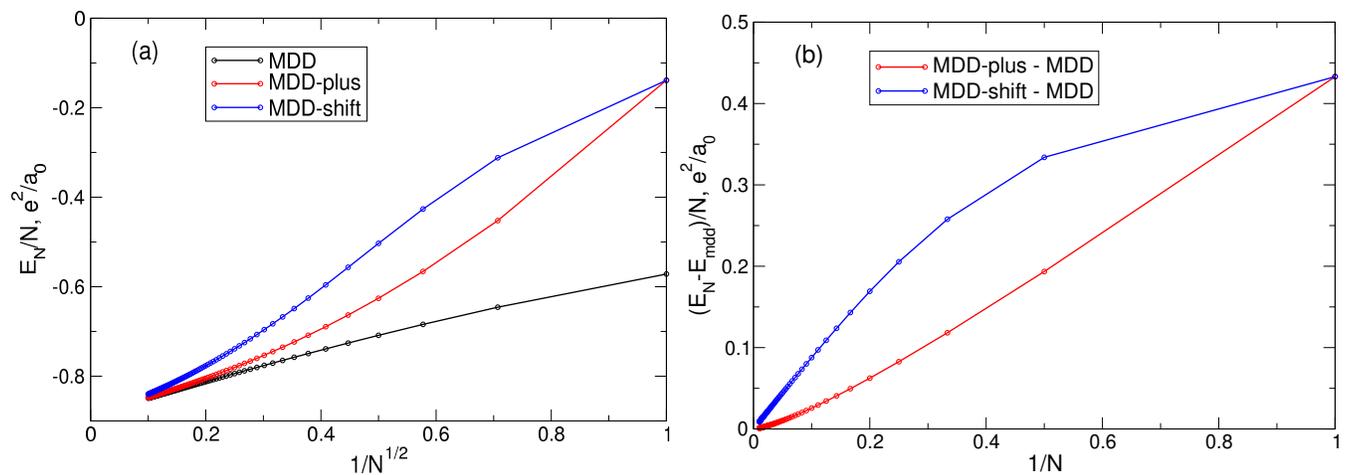

\includegraphics[width=0.49\columnwidth]{Energy3states_1sqrtN.eps}
\includegraphics[width=0.49\columnwidth]{EnergyDifference3states_1N.eps}
\caption{\label{fig:MDDtypeStates} (a) The energy of the MDD, MDD-plus, and MDD-shift states, as a function of $1/\sqrt{N}$. (b) The energy differences as a function of $1/N$.}
\end{figure}

What about the energy of these three states? Figure \ref{fig:MDDtypeStates}(a) shows the energy per particle of the MDD, MDD-plus, and MDD-shift configurations, as a function of $1/\sqrt{N}$, for $N$ varying from 1 to 100. One sees that at small $N$ the energy difference between the both excited and the ground (MDD) states is very large. But when $N$ grows, this difference quickly tends to zero. Figure \ref{fig:MDDtypeStates}(b) shows the energy differences $\delta E_{\rm mdd}^+/N$ and $\delta E_{\rm mdd}^\Rightarrow/N$ in dependence of $1/N$. In spite of the MDD-shift state has an evidently incorrect coordinate dependence of the electron density, both energy differences tend to zero linearly with $1/N$. This implies that $\delta E_{\rm mdd}^+$ and $\delta E_{\rm mdd}^\Rightarrow$ are constants at $N\to\infty$: $\delta E_{\rm mdd}^+\approx 0.118050e^2/a_0$ and $\delta E_{\rm mdd}^\Rightarrow\approx 0.8973e^2/a_0$. Obviously, there exist an infinite number of different wave functions whose energy per particle will tend to the same limit as $N\to\infty$. 

\subsection{Electron density in the Laughlin state at large $N$\label{sec:LaughDensLargeN}}

The density of electrons in the MDD-shift state, Figure \ref{fig:MDDtypeStatesDensity}, is ``physically unreasonable'', therefore, one should exclude the state $|\Psi_{\rm MDD}^\Rightarrow\rangle$ from the list of potential candidates for the ground state wave function. Could one, for similar reasons, exclude the Laughlin wave function (\ref{LaughlinWF}) from the list of potential candidates for the role of the ground state wave function at $\nu=1/m$? This question can be answered due to the studies of Ciftja and coauthors \cite{Ciftja03,Ciftja04}, who calculated, using Monte Carlo simulations, the density of electrons $n_e^{\rm LS}(r)$ in the $\nu=1/3$ Laughlin states for $N=64$, 100, 144, and 196 \cite{Ciftja03}, and in the $1/5$ and $1/7$ states for $N=196$ \cite{Ciftja04}.

\begin{figure}[ht!]
\includegraphics[width=0.49\columnwidth]{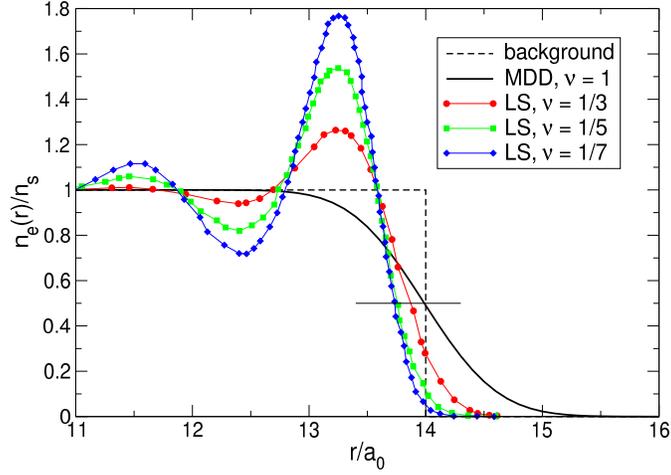}
\caption{\label{fig:LaughDensLargeN} The coordinate dependencies of the density of electrons in the Laughlin states (\ref{LaughlinWF}) at $\nu=1/3$, $1/5$, and $1/7$, calculated for $N=196$ by Ciftja et al. in Refs. \cite{Ciftja03,Ciftja04}; the discrete numerical data points are connected by lines to guide the eye. The thick black solid curve shows the MDD density at $\nu=1$. At $r/a_0\lesssim 11$ (not shown in the Figure) the normalized electron density equals 1 according to Refs. \cite{Ciftja03,Ciftja04}. Thin horizontal line at the level 0.5 visualizes the change of the electron disk radius when $\nu$ decreases from $\nu=1$ to $\nu=1/7$.}
\end{figure}

As seen in Figure \ref{fig:LaughDens}(a), the normalized LS density at small $r$ tends to unity as $N$ increases, and has a rather large peak near the edge of the disk at $r\approx R-a_0$. The value of the normalized density $n_e^{\rm LS}/n_s$ in this peak varies from $\sim 1.6$ for $N=2$ to $\sim 1.37$ for $N=8$. Do these features remain when $N$ grows further?
Figure \ref{fig:LaughDensLargeN} shows results of Refs. \cite{Ciftja03,Ciftja04}  (curves with symbols) replotted  as a function of $r/a_0$ (in Refs. \cite{Ciftja03,Ciftja04} these data were plotted in dependence of $r/l_B$). At $r/a_0\lesssim 11$ (not shown in Figure \ref{fig:LaughDensLargeN}) the normalized density $n_e^{\rm LS}/n_s$ calculated in Refs. \cite{Ciftja03,Ciftja04} is very close to 1, which agrees with the statement of \cite{Laughlin83} that the Laughlin function describes a uniform liquid. However, near the disk edge this ``liquid'' becomes strongly inhomogeneous. The ring of the high electron density, which is seen in Figure \ref{fig:LaughDens} for $N\le 8$, is also preserved for $N$ up to $N=196$, see Fig. 3 in Ref. \cite{Ciftja03} and Fig. 1 in Ref. \cite{Ciftja04}. Consider this LS feature in more detail.

\begin{figure}[ht!]
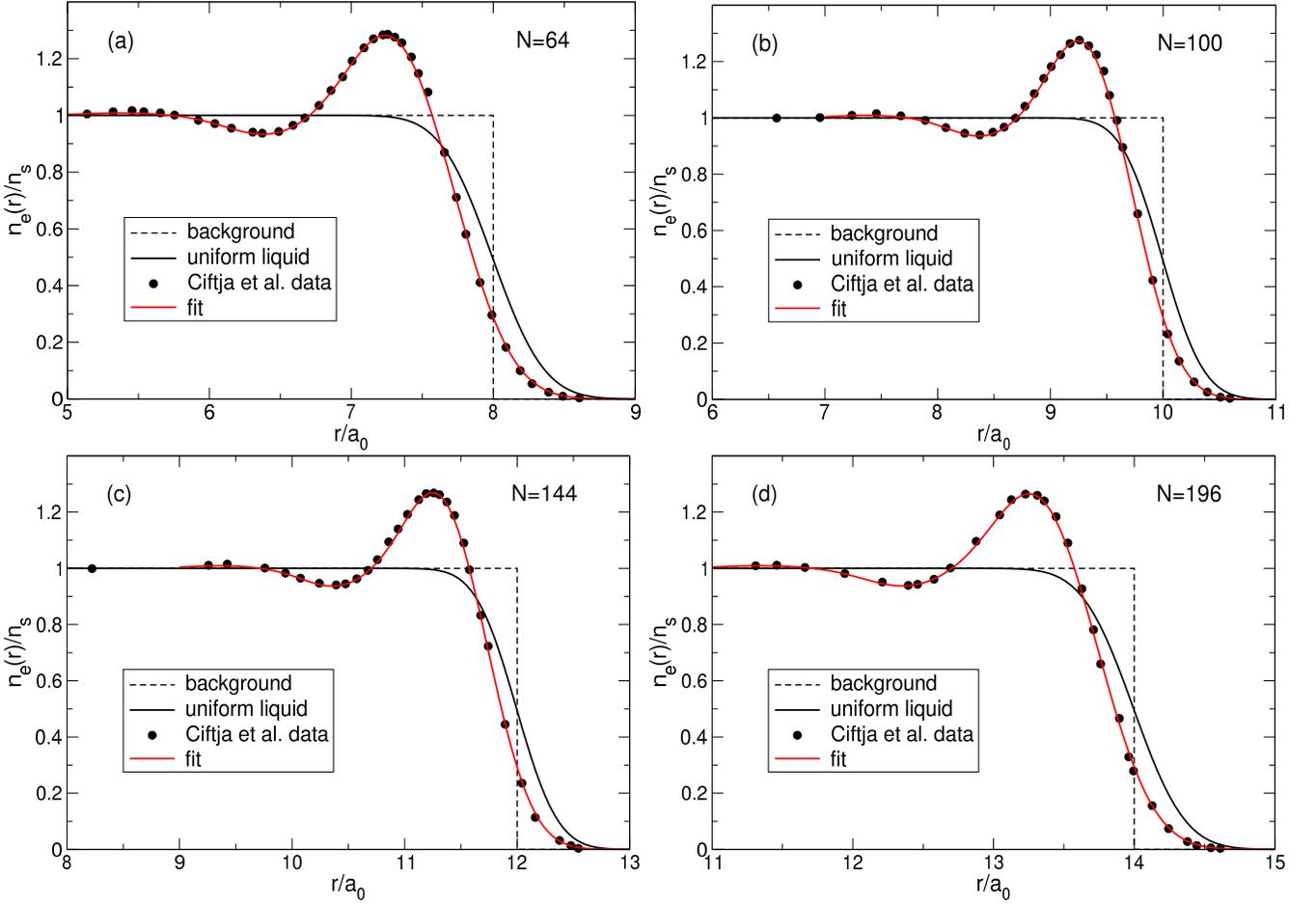

\includegraphics[width=0.49\columnwidth]{FitCiftja064nu13.eps}
\includegraphics[width=0.49\columnwidth]{FitCiftja100nu13.eps}
\includegraphics[width=0.49\columnwidth]{FitCiftja144nu13.eps}
\includegraphics[width=0.49\columnwidth]{FitCiftja196nu13.eps}
\caption{\label{fig:FitCiftjaDensity} Fitting of the numerical data \cite{Ciftja03} for the density of electrons in the $\nu=1/3$ LS by the analytical function (\ref{DensityCiftjaFit13}) for (a) $N=64$, (b) $N=100$, (c) $N=144$, and (d) $N=196$. }
\end{figure}

It is possible to fit the numerical data of Ciftja et al. \cite{Ciftja03} for $\nu=1/\beta=1/3$ by a linear combination
\be 
\frac{n_{e}^{\rm fit}(r)}{n_s} =
\sum_{L=0}^{L_{\max}(N)} \Phi_L\left(\sqrt{\beta}\frac r{a_0}\right)+
\sum_{k=1}^3(-1)^{k-1}
A_k(N)\Phi_{L_k(N)}\left(\sqrt{\beta}\frac r{a_0}\right)
\label{DensityCiftjaFit13}
\ee
of functions
\be 
\Phi_L(x)=\frac{x^{2L}}{L!}e^{-x^2},
\ee
which are related to the single-particle states (\ref{spwf_psiL}),
\be 
|\psi_L({\bm r})|^2=\frac{1}{\pi\lambda^2}\Phi_L\left(\frac{r}{\lambda}\right).
\ee
The angular momenta $L_{\max}(N)$ and $L_k(N)$, as well as the coefficients $A_k(N)$ in Eq. (\ref{DensityCiftjaFit13}) are
\be 
L_{\max}(N)=3 N-1-\sqrt{N},\ \ \ 
L_k(N)=3N-2(3k-1)\sqrt{N}-13+12k,\label{LmaxLk}
\ee
\be 
A_1(N)=1.205\sqrt{N} +0.07,\ \ 
A_2(N)=0.2525\sqrt{N} - 0.115,\ \ 
A_3(N)=\sqrt{N}-A_1(N)+A_2(N).\label{Ak}
\ee
The fitting function (\ref{DensityCiftjaFit13}) is not unique and not exact, but it very well fits the numerical data \cite{Ciftja03}, see Figure \ref{fig:FitCiftjaDensity}. Remarkably, the data for different $N$ can be equally well fitted by a single formula (\ref{DensityCiftjaFit13}), which implies that Eq. (\ref{DensityCiftjaFit13}) should also successfully work at even larger $N$, in the thermodynamic limit. 

Apart from the data of Ref. \cite{Ciftja03} and the fitting function (\ref{DensityCiftjaFit13}), I also show in Figures \ref{fig:FitCiftjaDensity}(a)-(d), by black solid curves, the density of a  ``uniform liquid'' (UL)
\be 
\frac{n_{e}^{\rm UL}(r)}{n_s} =
\sum_{L=0}^{\beta N-1} \Phi_L\left(\sqrt{\beta}\frac r{a_0}\right)=Q(\beta N,\beta r^2/a_0^2),
\label{DensityUniformLiquid}
\ee
which is similar to the density of the MDD state (\ref{electrondensityMDD}) but is defined for $\beta=1/\nu=3$. The function (\ref{DensityUniformLiquid}) satisfies the condition
\be
\int_0^{2\pi}d\phi\int_0^\infty rdr n_{e}^{\rm UL}(\bm r) =N
\label{elneutr}
\ee
and is ideally flat at $r\lesssim R-a_0$. As seen from Figures \ref{fig:FitCiftjaDensity}(a)-(d), the radius of the electron disk in the ``uniform liquid'' state (\ref{DensityUniformLiquid}), defined as the point where $n_{e}^{\rm UL}(r)/n_s=1/2$, is very close to the radius $R=a_0\sqrt{N}$ of the positively charged background disk.

If the Laughlin wave function (\ref{LaughlinWF}) really described a homogeneous liquid state, it would have to have a density close to (\ref{DensityUniformLiquid}). But, as seen from Figures \ref{fig:FitCiftjaDensity}(a)-(d), the radii of the electron disks in the LS (\ref{LaughlinWF}) is always noticeably smaller than the radii of the ``uniform liquid'' and of the positive background. Since all densities should satisfy the electroneutrality condition (\ref{elneutr}), the decrease in diameter of the ``Laughlin liquid'' disk should be compensated by a strong increase in its density in the interior of the system. This is indeed the case, that is the LS (\ref{LaughlinWF}) describes not a uniform liquid, but a highly inhomogeneous state, which is never realized in experiments.

Mathematically, one can evaluate this redistribution of the electron density by comparing the ``uniform liquid'' density (\ref{DensityUniformLiquid}) with the ``Laughlin liquid'' density, calculated for large $N$ in Ref. \cite{Ciftja03} and fitted by the formula (\ref{DensityCiftjaFit13}). The ``uniform liquid'' density consists of a sum of the functions $\Phi_L$ with $L$ running from $L=0$ up to $L=3N-1$. The LS density also contains a similar sum, but $L$ in the first term of Eq. (\ref{DensityCiftjaFit13}) runs from $L=0$ up to $L=L_{\max}=3N-1-\sqrt{N}$, Eq. (\ref{LmaxLk}). A \textit{macroscopically} large number ($\propto\sqrt{N}$) of the $L$-states are lost from the Laughlin function. To compensate this loss and to describe the oscillating behavior of the LS density near the edge of the disk, three additional terms with \textit{macroscopically} large amplitudes $A_k(N)\propto\sqrt{N}$, Eq. (\ref{Ak}), have to be added to the fitting function (\ref{DensityCiftjaFit13}). Note that the high density ring lies approximately at $R-1.2a_0\lesssim r\lesssim R-0.4a_0$, with the density maximum at $r\approx R-0.8a_0$, that is all changes in the local electron density occur inside the sample, at $r\lesssim R-a_0$, which means that they also take place in the thermodynamic limit $N\to\infty$. 

The above discussion refers to the case $\nu=1/3$. At $\nu=1/5$ and $1/7$ the Laughlin function (\ref{LaughlinWF}) demonstrates even stronger inhomogeneity of the local electron density near the edge of the system, see Figure \ref{fig:LaughDensLargeN}. The electron disk radius becomes smaller, and the maximum of the electron density near $r\approx R-0.8a_0$ gets higher. Thus, not only is the LS characterized by a non-physical feature of a strongly inhomogeneous density, but, in addition, this inhomogeneity depends significantly on the magnetic field.

\begin{figure}[ht!]
\includegraphics[width=0.49\columnwidth]{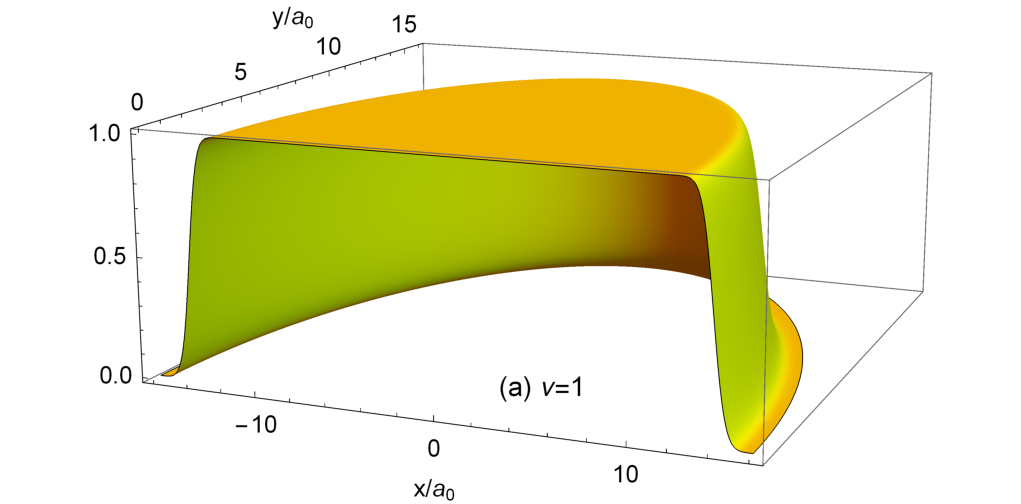}
\includegraphics[width=0.49\columnwidth]{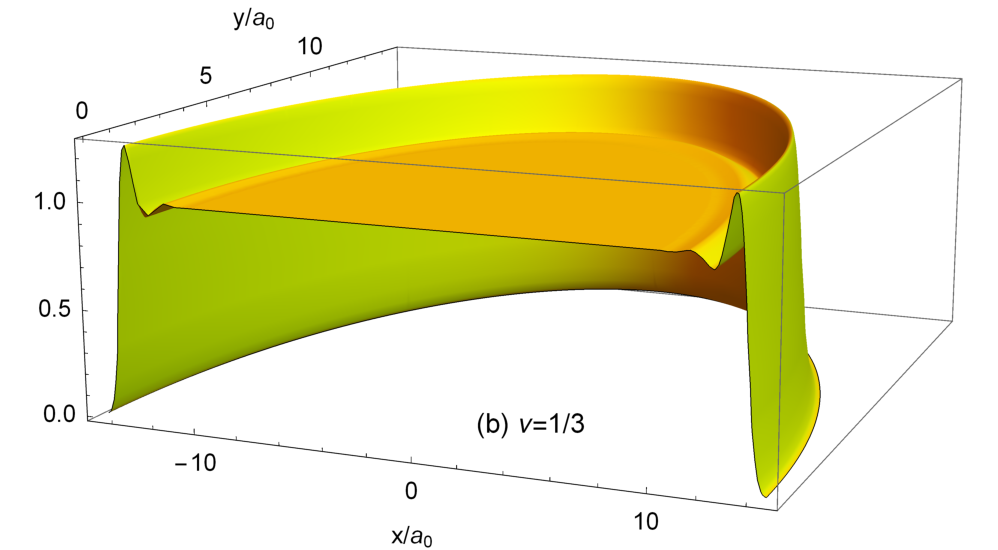}
\includegraphics[width=0.49\columnwidth]{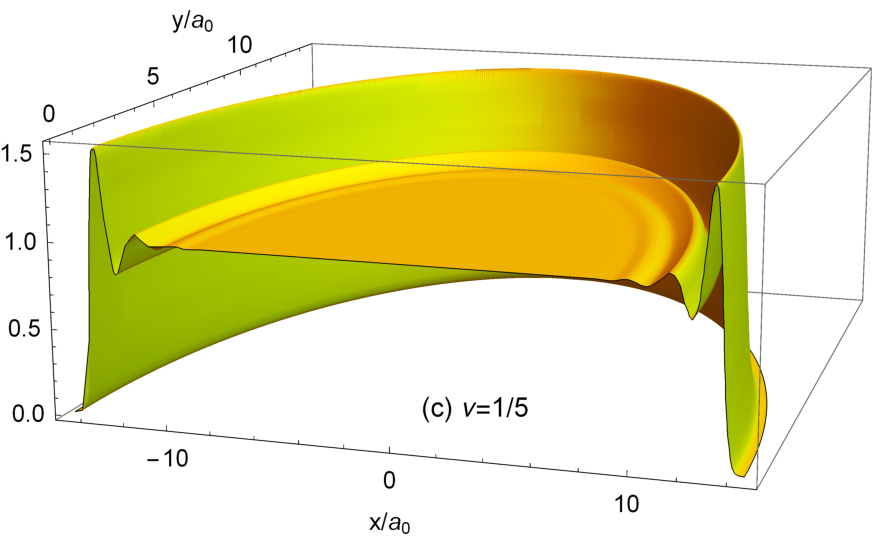}
\includegraphics[width=0.49\columnwidth]{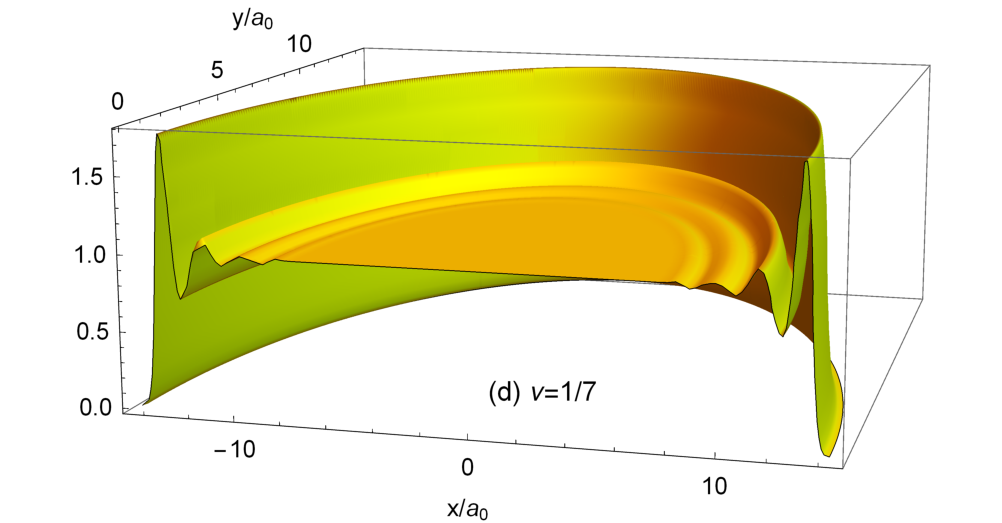}
\caption{\label{fig:3DCiftjaDens} The density of electrons in the MDD ($\nu=1$) and in the Laughlin states (\ref{LaughlinWF}) with $\nu=1/3$, $1/5$ and $1/7$. The number of particles is $N=196$. The data for the Laughlin electron densities are taken from Refs. \cite{Ciftja03,Ciftja04}.}
\end{figure}

Let us imagine a macroscopic 2DEG sample with an electron density on the order of $10^{11}$ cm$^{-2}$ \cite{Tsui82}. In equilibrium at $B=0$ the sample is locally electroneutral, i.e. the density of electrons equals the density of the positive background at any point. Now one switches the magnetic field on and increases it up to the value $\sim 50$ kG \cite{Tsui82} corresponding to the Landau level filling factor $\nu=1$. One gets the MDD ground state with a perfectly uniform density of electrons (\ref{electrondensityMDD}) at all $r\lesssim R-a_0$, Figure \ref{fig:3DCiftjaDens}(a). Then one increases the magnetic field further, up to $\sim 150$ kG \cite{Tsui82} which corresponds to the filling factor $\nu=1/3$. If the wave function (\ref{LaughlinWF}) corresponded to reality, a ring with a strongly enhanced electron density ($\sim 1.27\times 10^{11}$ cm$^{-2}$) would have to grow near the edge of the sample, Figure \ref{fig:3DCiftjaDens}(b). If after that the $B$-field is increased further, up to $\nu=1/5$ and $\nu=1/7$, then the local electron density in the edge ring should increase to $\sim 1.55\times 10^{ 11}$ cm$^{-2}$ and $\sim 1.77\times 10^{11}$ cm$^{-2}$ respectively, see Figures \ref{fig:3DCiftjaDens}(c) and (d). 

Such a strong redistribution of the electron density would require enormous energy costs, which can be estimated as follows. The density of the uncompensated charge can be modeled as
\be 
\delta \rho(\bm r)=\gamma en_s \Theta(R-r)\Theta(r-R+a_0)
\ee
where $\gamma$ is a number on the order of unity: $\gamma\approx 0.27,$ $0.55$, and $0.78$ at $\nu=1/3$, $1/5$, and $1/7$, respectively. The electrostatic energy of this uncompensated charge is
\ba 
\delta E=\int \frac{\delta \rho(\bm r)\delta \rho(\bm r')}{|\bm r-\bm r'|} d\bm r d\bm r' =
\int \frac{d\bm q}{2\pi q} |\delta \rho_{\bm q}|^2,
\label{elstaten}
\ea
where the Fourier transform of $\delta \rho(\bm r)$ equals
\be 
\delta \rho_{\bm q}=
\frac{2\pi \gamma en_s}{q^2} qR\Big( (1-\xi) J_1(qR(1-\xi))- J_1(qR)\Big),
\label{FourierDelraRho}
\ee
and 
\be 
\xi=\frac{a_0}{R}=\frac{1}{\sqrt{N}}
\ee
is a small parameter. Substituting Eq. (\ref{FourierDelraRho}) into (\ref{elstaten}) gives in the limit $\xi\ll 1$ ($N\gg 1$)
\be 
\delta E\approx\frac{e^2}{R}(2\gamma N)^2 \frac {\xi^2}{2\pi}\left(3+2\ln \frac 8\xi \right)=
\left(\frac{e^2}{a_0}\frac {6\gamma^2}{\pi }\right)\sqrt{N}\left[1+\frac 13\ln \left( 64N\right) \right].\label{elstaten-estimate}
\ee 
The electrostatic energy (\ref{elstaten-estimate}) grows as $\sqrt{N}\ln(N)$ in the thermodynamic limit. If the density of electrons assumes a typical value $n_s=3\times 10^{11}$ cm$^{-2}$ and the dielectric permittivity, which should be taken into account here, is $\epsilon=12.8$, the energy $(e^2/a_0\epsilon)(6\gamma^2/\pi)$ equals 1.52 meV for $\nu=1/3$ ($\gamma=0.27$). Then for the macroscopic number of particles $N\simeq 10^{11}$ the electrostatic energy (\ref{elstaten-estimate}) exceeds 2.5 keV at $\nu=1/3$, $\sim 10$ keV at $\nu=1/5$, and about $20$ keV at $\nu=1/7$. The ``Laughlin liquid'' requires too much energy for its existence.

\subsection{Behavior of the Laughlin function at $\bm r_j\to\bm r_k$ \label{sec:RjRk}}

Another reason why the Laughlin wave function is inappropriate for the description of the ground state of the FQHE system is as follows. The behavior of the many-body wave function at $|\bm r_j-\bm r_k|\to 0$ is governed by the Coulomb interaction terms $\sum_{jk} \Psi/|\bm r_j-\bm r_k|$ in the many-body Schr\"odinger equation. To compensate the $1/r$-singularity in this equation, the wave function should be proportional to $|\bm r_j-\bm r_k|$ at $|\bm r_j-\bm r_k|\to 0$. This is indeed the case for the MDD solution at $\nu=1$ and the true ground state solution at $\nu=1/3$. The Laughlin assumption $\Psi_{\rm LS}^{(m)}\propto|\bm r_j-\bm r_k|^m$ with $m=3,5,7$ actually suggests that electrons repel each other substantially stronger than needed in the world of Coulomb forces, as if the electron-electron interaction potential were proportional to $1/|\bm r_j-\bm r_k|^{m}$. As a result, electrons are pushed out of the disk center, with forces proportional to $1/|\bm r_j-\bm r_k|^{m+1}$, and have to accumulate near the disk edge with a local electron density growing with $m$, see Figures \ref{fig:LaughDensLargeN} and \ref{fig:3DCiftjaDens}.

In Ref. \cite{Laughlin83a} it was argued that in the case of two particles there is only one analytical function describing the motion of electrons on the lowest Landau level, and that this function is proportional to $(z_1-z_2)^m$ with $m$ odd. Then the work \cite{Laughlin83} aimed to generalize this statement to the case of $N$ particles, resulting in Eq. (\ref{LaughlinWF}). However, the statement of \cite{Laughlin83a} is not applicable to real physical systems with a neutralizing positive background and leads to physically unreasonable conclusions in such systems. Further discussion of this point can be found in Section \ref{subsec:no-backgr}.

In this Section I discussed only the properties of the wave function (\ref{LaughlinWF}). A similar analysis of the \textit{excited} Laughlin states, the fractionally charged quasiholes and quasielectrons, can also be carried out. I present results of such an analysis in Section \ref{sec:ExcStates}, after the physics of the true excited states is clarified in the next Section.

\section{Exact solutions at filling factors $\nu\le 1$\label{sec:ExactSolutionNu<1}}

As was shown in Section \ref{sec:ExactSolution}, not only the ground state at $\nu=1/3$ has the spatial particle distribution like in a floating Wigner molecule, but also the excited state, see Figure \ref{fig:DensityExact}(b). In order to better understand the nature of the ground \textit{and} excited states I now calculate the spectra and the electron densities of the system at arbitrary values of the Landau level filling factor $\nu\le 1$. The total angular momenta from ${\cal L}={\cal L}_{\min}=N(N-1)/2$ up to ${\cal L}={\cal L}_{\max}=4{\cal L}_{\min}$ will be considered, and the magnetic field parameter $\beta=1/\nu$ will be varied in the range $1\le\beta\lesssim 4$. 

As in the previous Sections, here I continue to assume that all electrons are spin-polarized and occupy only the lowest Landau level states. In general, this implies that the typical Coulomb interaction energy $e^2/a_0$ is smaller than the inter-Landau-level distance $\hbar\omega_c$, 
\be 
\frac{e^2/a_0}{\hbar\omega_c}\ll 1,\ \ \textrm{ or }\ \  \nu\ll 2\sqrt{\pi n_sa_B^2}=2\frac{a_B}{a_0};\label{strongBcondition}
\ee
here $a_B=\hbar^2/m^\star e^2$ is the effective Bohr radius and it is implicitly assumed here that the dielectric constant of the medium $\epsilon$ is included in the definition of charge, $e^2\to e^2/\epsilon$. The approximation (\ref{strongBcondition}) is very accurate for $\beta=1/\nu=3$, but begins to fail as $\beta$ approaches $1+$. Indeed, in order to get accurate results, the expansion of the $\Psi$ function (\ref{PsiExpansion}) over the basis many-body states should contain a sufficiently large number of terms, $N_{mbs}\gg 1$. But if ${\cal L}={\cal L}_{\min}$ (the MDD state) or ${\cal L}={\cal L}_{\min}+1$ (the MDD-plus state), this number is $N_{mbs}=1$. As will be seen below, under the assumption made, the MDD state is the ground state in certain intervals of the $B$-field, $\beta<\beta_0$, where $\beta_0$ depends on $N$ and for $N=7$ is about 1.5. In these magnetic fields, the results obtained provide only an upper bound on the ground state energy and should be improved in the future by taking into account states from higher Landau levels.

The positive background density profile is assumed to be step-like everywhere in this Section, except the subsections \ref{sec:resultsSmooth} and \ref{sec:density7smoo}.

\subsection{Energy spectra, energy gaps and excited states \label{sec:ExactSolutionNu<1-Energy}}

\subsubsection{Two particles}

If $N=2$, the total angular momenta considered vary from ${\cal L}_{\min}=1$ up to ${\cal L}_{\max}=4$. The many-body configurations for these ${\cal L}$ and the number of these configurations $N_{mbs}$ are given in Table \ref{tab:N2configs}. If (for any $N$) ${\cal L}\le{\cal L}_{\min}+1$, there exists only one many-body state, $N_{mbs}=1$, and there exists only one energy-vs-$B$ curve $E_{\cal L}(\beta)$. If ${\cal L}>{\cal L}_{\min}+1$, the number of many-body state is bigger than one, $N_{mbs}>1$, and the energy-vs-$B$ curves $E_{{\cal L},s}(\beta)$ will be enumerated by an additional index $s=1,2,\dots$. The goal is to determine, for each $\beta$-value, the energies of the ground $E_{\rm GS}(\beta)$ and the first excited $E_{\rm 1st}(\beta)$ states, as well as the values of the quantum numbers $({\cal L},s)$ corresponding to these states.

\begin{figure}[ht!]
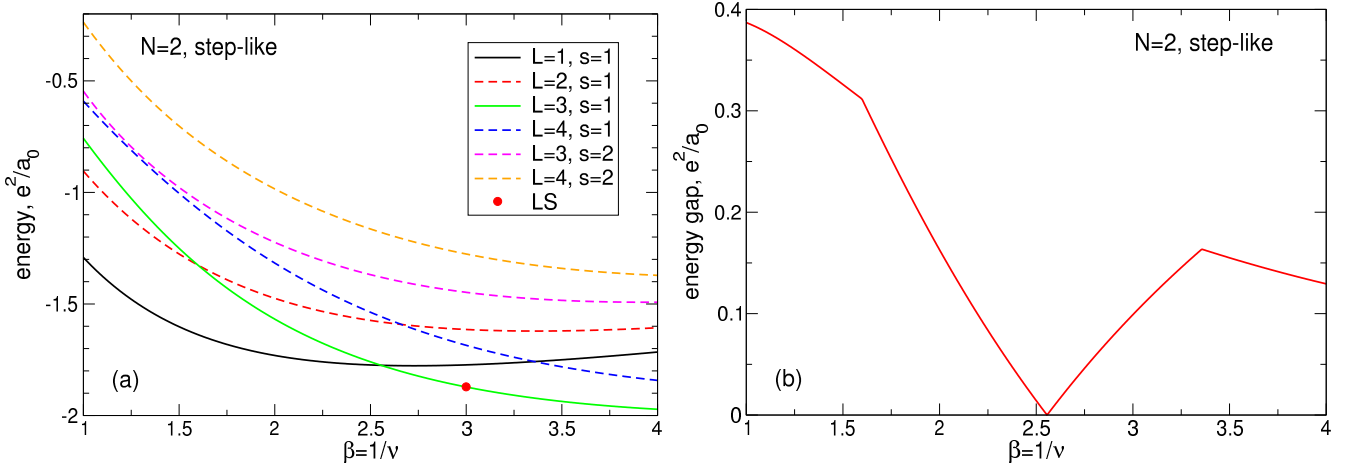

\includegraphics[width=0.49\columnwidth]{N2step.eps}
\includegraphics[width=0.49\columnwidth]{N2stepGap.eps}
\caption{\label{fig:energyN2step} (a) The energy of the many-body states with the total angular momenta from ${\cal L}={\cal L}_{\min}=1$ up to ${\cal L}={\cal L}_{\max}=4$ in the system of $N=2$ 2D electrons as a function of the magnetic field parameter $\beta=1/\nu$. The index $s$ enumerates different states with the same ${\cal L}$. The LS energy at $\nu=1/3$ is shown by a small red circle. (b) The energy gap between the ground and the first excited states as a function of $\beta$. The positive background density profile is step-like. }
\end{figure}

Figure \ref{fig:energyN2step}(a) shows the energies of all many-body states, with ${\cal L}$ varying from 1 to 4 and $s=1,2$, for the system of $N=2$ electrons. If $\beta=1/\nu=1$, the ground state is characterized by the quantum numbers $({\cal L},s)=(1,1)$ and is the MDD state $|0,1\rangle$. As $\beta$ increases, the energy of this state, shown by the black curve in Figure \ref{fig:energyN2step}(a), decreases, reaches a minimum at $\beta\approx 2.73$ and then starts to slowly grow. When $\beta$ becomes larger than $\beta_0=2.5559$, the role of the ground state $({\cal L},s)=(1,1)$ is transferred to the state $({\cal L},s)=(3,1)$, shown by the green solid curve in Figure \ref{fig:energyN2step}(a). The energies of the states with ${\cal L}=2$ and ${\cal L}=4$ are higher, at all $\beta$, than the energies of the states with ${\cal L}=1$ and ${\cal L}=3$, but the states $({\cal L},s)=(2,1)$ (the red dashed curve) and $({\cal L},s)=(4,1)$ (the blue dashed curve) may be the first excited state in certain intervals of the magnetic field, see Table \ref{tab:N2LtotGS1st}. For all ground and first excited states the value of the second quantum number is $s=1$, i.e. the states with $s>1$ ($s=2$ in the case of two particles) cannot be the first excited state at any $\beta$. This feature remains valid for any $N$. Figure \ref{fig:energyN2step}(b) shows the energy gap between the first excited and the ground state as a function of the magnetic field parameter $\beta$. The gap vanishes at $\beta=2.5559$ and reaches the value of about $0.164e^2/a_0$ at $\beta=3.3567$. The energy of the LS with $m=3$ is shown in Figure \ref{fig:energyN2step}(a) by a small red circle at $\beta=3$. 

\begin{table}[!ht]
\caption{The total angular momenta of the ground state ${\cal L}_{\rm GS}$ and of the first excited state ${\cal L}_{\rm 1st}$ assume the values shown in the last two columns in the intervals from $\beta_{\rm from}$ to $\beta_{\rm to}$ shown in the first two columns. The number of particles is  $N=2$, the density profile is step-like. \label{tab:N2LtotGS1st}}
\begin{tabular}{cccc}
$\beta_{\rm from}$ & $\beta_{\rm to}$   & ${\cal L}_{\rm GS}$   & ${\cal L}_{\rm 1st}$\\
\hline
1.0000	& 1.5986	& 1	& 2 \\
1.5986	& 2.5559	& 1	& 3 \\
2.5559	& 3.3567	& 3	& 1 \\
3.3567	& 4.0000	& 3	& 4 \\
\end{tabular}
\end{table}

\subsubsection{Three particles}

\begin{figure}[ht!]
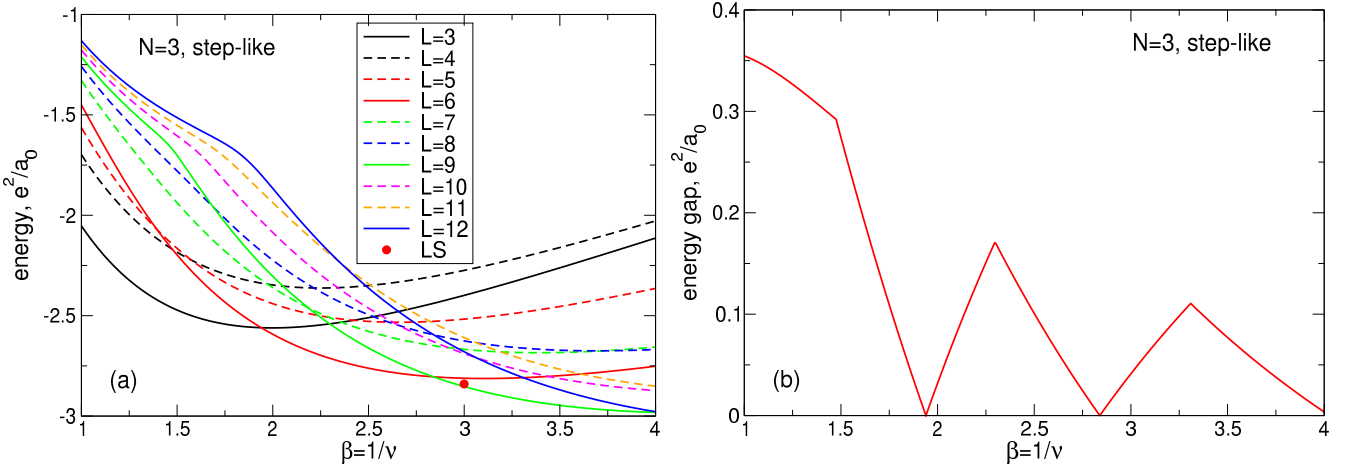

\includegraphics[width=0.49\columnwidth]{N3step.eps}
\includegraphics[width=0.49\columnwidth]{N3stepGap.eps}
\caption{\label{fig:energyN3step} (a) The energy of the many-body states with the total angular momenta from ${\cal L}={\cal L}_{\min}=3$ up to ${\cal L}={\cal L}_{\max}=12$ in the system of $N=3$ 2D electrons as a function of the magnetic field parameter $\beta=1/\nu$. For all shown states the index $s$ equals $s=1$; the states with $s>1$ are not shown. The LS energy at $\nu=1/3$ is shown by a small red circle. (b) The energy gap between the ground and the first excited states as a function of $\beta$. The positive background density profile is step-like.}
\end{figure}

If $N=3$, I consider the total angular momenta between ${\cal L}_{\min}=3$ and ${\cal L}_{\max}=12$, see Table \ref{tab:N3configs}. Figure \ref{fig:energyN3step}(a) shows the energies $E_{{\cal L},s}(\beta)$ for all many-body states with ${\cal L}$ varying from 3 to 12 and for $s=1$. The states with $s\ge 2$ are not shown, since they are neither the ground state nor the first excited state for any $\beta$. States that are the ground states in some region of the magnetic field are shown by solid curves, all others states are shown by dashed curves. 

The behavior of the curves $E_{{\cal L},1}(\beta)$ is similar to the case of $N=2$. The MDD configuration $|0,1,2\rangle $ with $({\cal L},s)=(3,1)$ is the ground state in the interval from $\beta=1$ to $\beta=\beta_0=1.9397$ (black solid curve) and then transfers its role of the ground state to the state $({\cal L},s)=(6,1)$ (red solid curve). After $\beta=2.8392$ the state $({\cal L},s)=(9,1)$ become the ground state (green solid curve), see Table \ref{tab:N3LtotGS1st}. The next state that will become the ground state at $\beta$ slightly bigger than 4 is the state $({\cal L},s)=(12,1)$ (blue solid curve). As can be seen from Figure \ref{fig:energyN3step}(a) and Table \ref{tab:N3LtotGS1st}, states with ${\cal L}$ other than 3, 6, 9 and 12 are also not the first excited states at any magnetic field. The only exception is the ${\cal L}=4$ state, which is the first excited state for $\beta$ close to 1, but in this region the problem needs to be reconsidered by including higher Landau levels as discussed above. The energy of the LS with $m=3$ is shown by a small orange circle at $\beta=3$.

\begin{table}[!ht]
\caption{The total angular momenta of the ground state ${\cal L}_{\rm GS}$ and of the first excited state ${\cal L}_{\rm 1st}$ assume the values shown in the last two columns in the intervals from $\beta_{\rm from}$ to $\beta_{\rm to}$ shown in the first two columns. The number of particles is  $N=3$, the density profile is step-like. \label{tab:N3LtotGS1st}}
\begin{tabular}{cccc}
$\beta_{\rm from}$ & $\beta_{\rm to}$   & ${\cal L}_{\rm GS}$   & ${\cal L}_{\rm 1st}$\\
\hline
1.0000	& 1.4751	& 3	& 4 \\
1.4751	& 1.9397	& 3	& 6 \\
1.9397	& 2.2971	& 6	& 3 \\
2.2971	& 2.8392	& 6	& 9 \\
2.8392	& 3.3096	& 9	& 6 \\
3.3096	& 4.0000	& 9	& 12 \\
\end{tabular}
\end{table}

Figure \ref{fig:energyN3step}(b) shows the energy gap between the first excited and the ground state as a function of the magnetic field parameter $\beta$. The gap vanishes at two $\beta$-points and is about $0.17e^2/a_0$ at $\beta=2.2971$ and $0.11e^2/a_0$ at $\beta=3.3096$. 

Note that the angular momenta values corresponding to the ground states in different $\beta$ intervals (${\cal L}=3$, 6, 9) satisfy the rule
\be 
{\cal L}_k^{\rm GS}={\cal L}_{\min}+n^{\rm st}(N)k, \label{GSrule}
\ee
where $n^{\rm st}(N)$ is given by Eq. (\ref{NsymStep}) and $k=0,1,2,\dots$ is integer. The reason for the sequence (\ref{GSrule}) is that the quantum solution should be formed from the ${\cal L}$ states that have the same circular symmetry as the classical Wigner molecule. The states from the sequence (\ref{GSrule}) satisfy this requirement, therefore they have a lower energy and serve as the ground states in different $\beta$-intervals. The rule (\ref{GSrule}) is also valid for $N=2$ and $N>3$, as shown below.

\subsubsection{Four particles}

\begin{figure}[ht!]
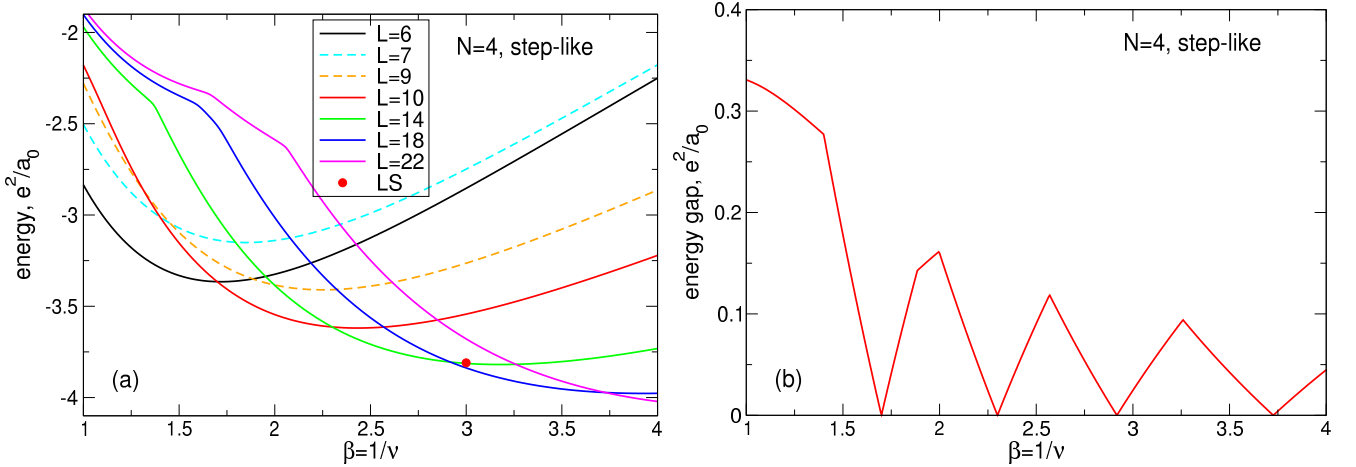

\includegraphics[width=0.49\columnwidth]{N4step.eps}
\includegraphics[width=0.49\columnwidth]{N4stepGap.eps}
\caption{\label{fig:energyN4step} (a) The energy of the many-body states with the total angular momenta from ${\cal L}={\cal L}_{\min}=6$ up to ${\cal L}=22$ in the system of $N=4$ 2D electrons as a function of the magnetic field parameter $\beta=1/\nu$. For all shown states the index $s$ equals $s=1$. Only the states which are either ground or first excited states are shown. The LS energy at $\nu=1/3$ is shown by a small red circle. (b) The energy gap between the ground and the first excited states as a function of $\beta$. The positive background density profile is step-like.}
\end{figure}

If $N=4$, the total angular momenta that I consider lie between ${\cal L}_{\min}=6$ and ${\cal L}_{\max}=24$. In order not to overload the graph with too many curves, I show in Figure \ref{fig:energyN4step}(a) only the energies of the states $({\cal L},s)$ that are either the ground or the first excited state in any interval of $\beta$. The rule (\ref{GSrule}) for the ground-states angular momenta remains valid for $N=4$. The $({\cal L},s)$-states with ${\cal L}=8$, $11\dots 13$, $15\dots 17$, $19\dots 21$, $23\dots 24$, and $s=1$, as well as all states with $s>1$, which can be only the second or higher excited state, are not shown in the Figure. The angular momenta of the ground and first excited states, with the corresponding $\beta$ intervals, are given in Table \ref{tab:N4LtotGS1st}. The energy of the LS with $\nu=1/3$ is shown by a small red circle at $\beta=3$ in Figure \ref{fig:energyN4step}(a). It lies slightly above the green curve corresponding to the first excited state ${\cal L}=14$ at $\nu=1/3$.

\begin{table}[!ht]
\caption{The total angular momenta of the ground state ${\cal L}_{\rm GS}$ and of the first excited state ${\cal L}_{\rm 1st}$ assume the values shown in the last two columns in the intervals from $\beta_{\rm from}$ to $\beta_{\rm to}$ shown in the first two columns. The number of particles is  $N=4$, the density profile is step-like. \label{tab:N4LtotGS1st}}
\begin{tabular}{cccc}
$\beta_{\rm from}$ & $\beta_{\rm to}$   & ${\cal L}_{\rm GS}$   & ${\cal L}_{\rm 1st}$\\
\hline
1.0000	& 1.4006	& 6	& 7 \\
1.4006	& 1.6991	& 6	& 10 \\
1.6991	& 1.8850	& 10	& 6 \\
1.8850	& 1.9972	& 10	& 9 \\
1.9972	& 2.2994	& 10	& 14 \\
2.2994	& 2.5689	& 14	& 10 \\
2.5689	& 2.9173	& 14	& 18 \\
2.9173	& 3.2593	& 18	& 14 \\
3.2593	& 3.7263	& 18	& 22 \\
3.7263	& 4.0000	& 22	& 18 \\
\end{tabular}
\end{table}

Figure \ref{fig:energyN4step}(b) shows the energy gap between the first excited and the ground state as a function of the magnetic field parameter $\beta$. The gap vanishes in a few $\beta$-points and has local maxima varying from $\approx 0.15e^2/a_0$ at $\beta=1.9972$ to $\approx 0.094e^2/a_0$ at $\beta=3.2593$.

\subsubsection{Five particles}

\begin{figure}[ht!]
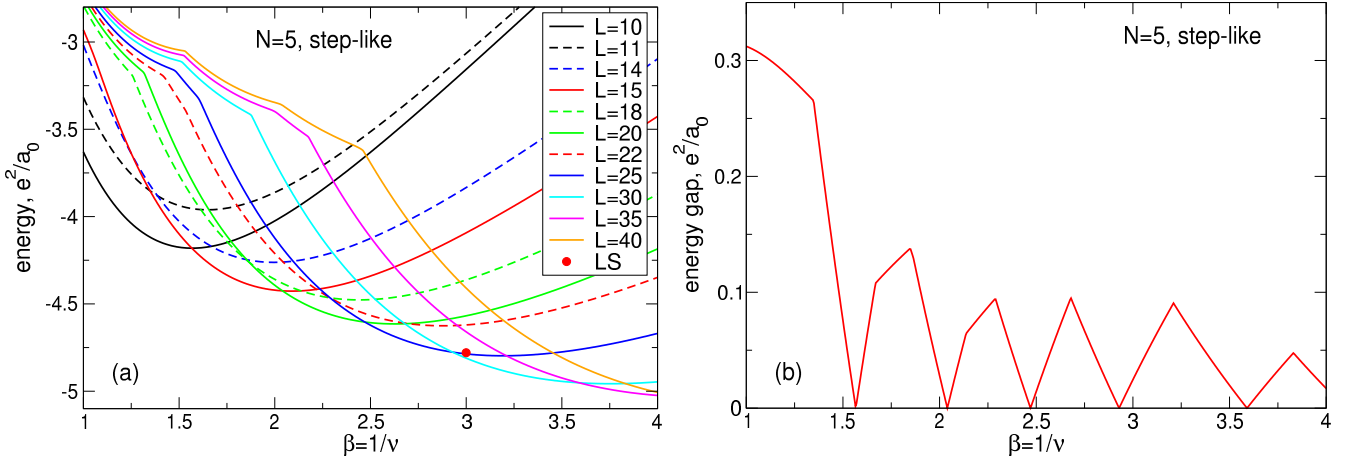

\includegraphics[width=0.49\columnwidth]{N5step.eps}
\includegraphics[width=0.49\columnwidth]{N5stepGap.eps}
\caption{\label{fig:energyN5step} (a) The energy of the many-body states with the total angular momenta from ${\cal L}={\cal L}_{\min}=10$ up to ${\cal L}={\cal L}_{\max}=40$ in the system of $N=5$ 2D electrons as a function of the magnetic field parameter $\beta=1/\nu$. For all shown states the index $s$ equals $s=1$. Only the states which are either ground or first excited states are shown. The LS energy at $\nu=1/3$ is shown by a small red circle. (b) The energy gap between the ground and the first excited states as a function of $\beta$. The positive background density profile is step-like.}
\end{figure}

As in the case of $N=4$, Figure \ref{fig:energyN5step}(a) shows the energies $E_{{\cal L},1}(\beta)$ only for those ${\cal L}$ that are either the ground state or the first excited state in some intervals of $\beta$. These intervals and the corresponding angular momenta ${\cal L}_{\rm GS}$ and ${\cal L}_{\rm 1st}$ are given in Table \ref{tab:N5LtotGS1st}. The rule (\ref{GSrule}) for ${\cal L}_j^{\rm GS}$ remains valid for $N=5$. The energy of the LS with $m=3$ is shown by a small red circle at $\beta=3$. It lies above the blue curve corresponding to the first excited state ${\cal L}=25$ at $\nu=1/3$.

\begin{table}[!ht]
\caption{The total angular momenta of the ground state ${\cal L}_{\rm GS}$ and of the first excited state ${\cal L}_{\rm 1st}$ assume the values shown in the last two columns in the intervals from $\beta_{\rm from}$ to $\beta_{\rm to}$ shown in the first two columns. The number of particles is  $N=5$, the density profile is step-like. \label{tab:N5LtotGS1st}}
\begin{tabular}{cccc}
$\beta_{\rm from}$ & $\beta_{\rm to}$   & ${\cal L}_{\rm GS}$   & ${\cal L}_{\rm 1st}$\\
\hline
1.0000	& 1.3483	& 10	& 11 \\
1.3483	& 1.5666	& 10	& 15 \\
1.5666	& 1.6688	& 15	& 10 \\
1.6688	& 1.8486	& 15	& 14 \\
1.8486	& 1.8647	& 15	& 18 \\
1.8647	& 2.0402	& 15	& 20 \\
2.0402	& 2.1372	& 20	& 15 \\
2.1372	& 2.2887	& 20	& 18 \\
2.2887	& 2.2890	& 20	& 22 \\
2.2890	& 2.4705	& 20	& 25 \\
2.4705	& 2.6791	& 25	& 20 \\
2.6791	& 2.9287	& 25	& 30 \\
2.9287	& 3.2100	& 30	& 25 \\
3.2100	& 3.5904	& 30	& 35 \\
3.5904	& 3.8305	& 35	& 30 \\
3.8305	& 4.0000	& 35	& 40 \\
\end{tabular}
\end{table}

Figure \ref{fig:energyN5step}(b) shows the energy gap between the first excited and the ground state as a function of $\beta$. Its value in local maxima varies from $\approx 0.137e^2/a_0$ at $\beta=1.8486$ to $\approx 0.048e^2/a_0$ at $\beta=3.8305$.

\subsubsection{Six particles\label{6el-energy}}

\begin{figure}[ht!]
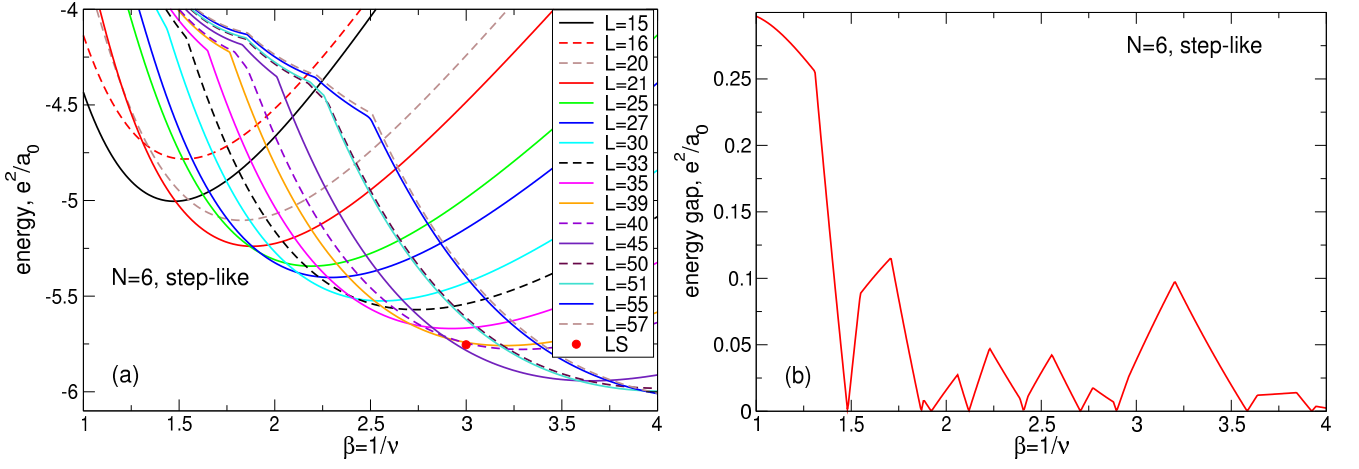

\includegraphics[width=0.49\columnwidth]{N6step.eps}
\includegraphics[width=0.49\columnwidth]{N6stepGap.eps}
\caption{\label{fig:energyN6step} (a) The energy of the many-body states with the total angular momenta from ${\cal L}={\cal L}_{\min}=15$ up to ${\cal L}=57$ in the system of $N=6$ 2D electrons as a function of the magnetic field parameter $\beta=1/\nu$. For all shown states the index $s$ equals $s=1$. Only the states which are either ground or first excited states are shown. The LS energy at $\nu=1/3$ is shown by a small red circle. (b) The energy gap between the ground and the first excited states as a function of $\beta$. The positive background density profile is step-like.}
\end{figure}

For $N=6$, Figure \ref{fig:energyN6step}(a) shows the energies $E_{{\cal L},1}(\beta)$ for those ${\cal L}$ which are either the ground state or the first excited state in some intervals of $\beta$. These intervals and the corresponding angular momenta ${\cal L}_{\rm GS}$ and ${\cal L}_{\rm 1st}$ are given in Table \ref{tab:N6LtotGS1st}. In the case of six particles, the overall picture of the energy spectra is more complex than for $N<6$. Firstly, the number of ${\cal L}$-states that take on the role of the ground state in the interval $1\le\beta\le 4$ is ten, which is substantially larger than one would expect from the evolution of plots from Figures \ref{fig:energyN2step}--\ref{fig:energyN5step}. Secondly, the length of the $\beta$ intervals in which the ground state angular momenta remain constant becomes shorter, and these intervals are less uniformly distributed along the $\beta$-axis. Thirdly, although the rule (\ref{GSrule}) is satisfied for some values of ${\cal L}$, for some other ${\cal L}$'s it is violated. 

\begin{table}[!ht]
\caption{The total angular momenta of the ground state ${\cal L}_{\rm GS}$ and of the first excited state ${\cal L}_{\rm 1st}$ assume the values shown in the last two columns in the intervals from $\beta_{\rm from}$ to $\beta_{\rm to}$ shown in the first two columns. The number of particles is  $N=6$, the density profile is step-like. \label{tab:N6LtotGS1st}}
\begin{tabular}{cccc}
$\beta_{\rm from}$ & $\beta_{\rm to}$   & ${\cal L}_{\rm GS}$   & ${\cal L}_{\rm 1st}$\\
\hline
1.0000	& 1.3094	& 15	& 16 \\
1.3094	& 1.4812	& 15	& 21 \\
1.4812	& 1.5480	& 21	& 15 \\
1.5480	& 1.7089	& 21	& 20 \\
1.7089	& 1.8685	& 21	& 25 \\
1.8685	& 1.8813	& 25	& 21 \\
1.8813	& 1.9215	& 25	& 27 \\
1.9215	& 2.0598	& 27	& 25 \\
2.0598	& 2.1190	& 27	& 30 \\
2.1190	& 2.2284	& 30	& 27 \\
2.2284	& 2.3883	& 30	& 33 \\
2.3883	& 2.4072	& 30	& 35 \\
2.4072	& 2.4303	& 35	& 30 \\
2.4303	& 2.5560	& 35	& 33 \\
2.5560	& 2.7060	& 35	& 39 \\
2.7060	& 2.7714	& 39	& 35 \\
2.7714	& 2.8798	& 39	& 40 \\
2.8798	& 2.8965	& 39	& 45 \\
2.8965	& 2.9606	& 45	& 39 \\
2.9606	& 3.2029	& 45	& 40 \\
3.2029	& 3.5837	& 45	& 51 \\
3.5837	& 3.6374	& 51	& 45 \\
3.6374	& 3.8437	& 51	& 50 \\
3.8437	& 3.9236	& 51	& 55 \\
3.9236	& 3.9456	& 55	& 51 \\
3.9456	& 4.0000	& 55	& 57 \\
\end{tabular}
\end{table}

What are the reasons for such unusual behavior of the spectra at $N=6$? 

As has been seen in Section \ref{sec:Wigner}, for $N=6$ the energies of the classical shell configurations $(1,N-1)$ and $(0,N)$ are very close to each other. Therefore, the configurations with the symmetry $C_{5}$  (five particles on the outer shell) can compete in energy with the configurations with the symmetry $C_{6}$  (six particles on the outer shell). Indeed, in the case of the $(1,5)$ shell configuration the expected ${\cal L}$ sequence (\ref{GSrule}) would be
\be {\cal L}_{\rm GS}^{(1,5)}=15,\ 20,\ 25,\  30,\  35,\  40,\  45,\  50,\  55. \label{sequence5}\ee
In the case of the $(0,6)$ shell configuration the expected (``underlined'') ${\cal L}$ sequence would be
\be {\cal L}_{\rm GS}^{(0,6)}=\underline{15},\  \underline{21},\  \underline{27},\  \underline{33},\  \underline{39},\  \underline{45},\  \underline{51},\  \underline{57}. \label{sequence6}\ee
Calculations show that for $N=6$ a mixture of the sequences (\ref{sequence5}) and (\ref{sequence6}), 
\be {\cal L}_{\rm GS}=\underline{15},\  \underline{21},\  25,\  \underline{27},\  30,\  35,\  \underline{39},\  \underline{45},\  \underline{51},\  55, \ee
is actually realized, while the states with 
\be {\cal L}_{\rm 1st}=20,\  \underline{33},\  40,\  50,\  \underline{57} \ee
serve as the first excited states. Thus the complicated structure of levels at $N=6$ is explained by the competition of one-shell and two-shells configurations which have very close energies already in the classical approach.

Figure \ref{fig:energyN6step}(b) shows the energy gap between the first excited and the ground state as a function of the magnetic field parameter $\beta$. It is noticeable that the largest energy gaps and the largest distances between the gap nodes are seen in the ranges $1.48\lesssim \beta\lesssim 1.87$ and $2.8965\lesssim \beta\lesssim 3.58$, which are close to $\nu=2/3$ and $\nu=1/3$. The gap maxima in these intervals are $\approx 0.115e^2/a_0$ at $\beta=1.6688$ and $\approx 0.097e^2/a_0$ at $\beta=3.21$, which are about twice as large as the other local gap maxima between $\beta\simeq 2$ and 3 ($\lesssim 0.05e^2/a_0$). If it were possible to show that the Hall conductivity is constant in the $\beta$-intervals, where the gaps are finite (this still needs to be done in a future theory), the widths of the plateaus would be maximal around $\nu=1/3$ and 2/3. So, some correlations with the FQHE experiment in a macroscopic sample arise already at $N=6$.

\subsubsection{Seven particles, step-like density profile}

\begin{figure}[ht!]
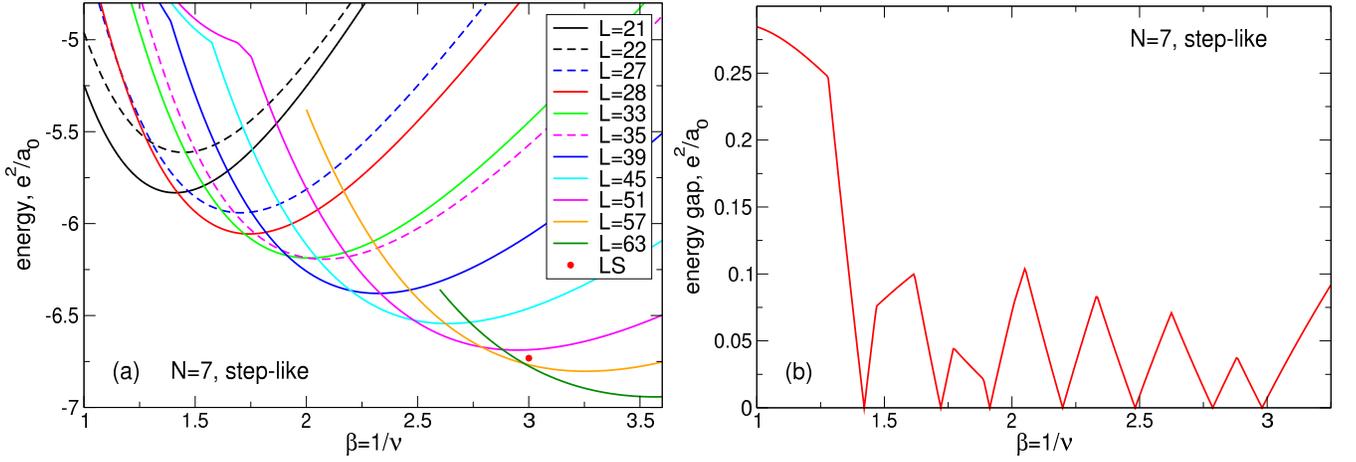

\includegraphics[width=0.49\columnwidth]{N7step.eps}
\includegraphics[width=0.49\columnwidth]{N7stepGap.eps}
\caption{\label{fig:energyN7step} (a) The energy of the many-body states with the total angular momenta from ${\cal L}={\cal L}_{\min}=21$ up to ${\cal L}=63$ in the system of $N=7$ 2D electrons as a function of the magnetic field parameter $\beta=1/\nu$. For all shown states the index $s$ equals $s=1$. Only the states which are either ground or first excited states are shown. The LS energy at $\nu=1/3$ is shown by a small red circle. (b) The energy gap between the ground and the first excited states as a function of $\beta$. The positive background density profile is step-like.}
\end{figure}

For $N=7$ the energy-vs-$\beta$ curves for the angular momenta ${\cal L}$ which are either the ground or the first excited state in some intervals of $\beta$ are shown in Figure \ref{fig:energyN7step}. The corresponding intervals and the angular momenta ${\cal L}_{\rm GS}$ and ${\cal L}_{\rm 1st}$ are given in Table \ref{tab:N7LtotGS1st}. As compared to the case of six particles the spectra look simpler again, with $N+1=8$ angular momenta, corresponding to the ground states in different intervals of $\beta$. The rule (\ref{GSrule}) is satisfied with one exception: in the expected sequence of the ground state ${\cal L}$'s ($={\cal L}_{\min}+(N-1)k$), 
\bc 21, 27, 33, 39, 45, 51, 57, 63, \ec 
the second number is replaced by 28 ($={\cal L}_{\min}+N$). The state with ${\cal L}=28$ is the ground state of the seven-particle system at $1.4211\le\beta\le 1.721$. The reason for these deviation is very interesting and is discussed in Section \ref{sec:density7step}.

\begin{table}[!ht]
\caption{The total angular momenta of the ground state ${\cal L}_{\rm GS}$ and of the first excited state ${\cal L}_{\rm 1st}$ assume the values shown in the last two columns in the intervals from $\beta_{\rm from}$ to $\beta_{\rm to}$ shown in the first two columns. The number of particles is  $N=7$, the density profile is step-like. \label{tab:N7LtotGS1st}}
\begin{tabular}{cccc}
$\beta_{\rm from}$ & $\beta_{\rm to}$   & ${\cal L}_{\rm GS}$   & ${\cal L}_{\rm 1st}$\\
\hline
1.0000	& 1.2795	& 21	& 22 \\
1.2795	& 1.4211	& 21	& 28 \\
1.4211	& 1.4690	& 28	& 21 \\
1.4690	& 1.6170	& 28	& 27 \\
1.6170	& 1.7210	& 28	& 33 \\
1.7210	& 1.7708	& 33	& 28 \\
1.7708	& 1.8886	& 33	& 35 \\
1.8886	& 1.9132	& 33	& 39 \\
1.9132	& 2.0082	& 39	& 33 \\
2.0082	& 2.0512	& 39	& 35 \\
2.0512	& 2.1988	& 39	& 45 \\
2.1988	& 2.3323	& 45	& 39 \\
2.3323	& 2.4828	& 45	& 51 \\
2.4828	& 2.6249	& 51	& 45 \\
2.6249	& 2.7866	& 51	& 57 \\
2.7866	& 2.8819	& 57	& 51 \\
2.8819	& 2.9803	& 57	& 63 \\
2.9803	& 3.2500	& 63	& 57 \\
\end{tabular}
\end{table}

The energy of the state (\ref{LaughlinWF}) with $m=3$ is shown by a small red circle at $\beta=3$ in Figure \ref{fig:energyN7step}(a). In the case $N=7$ the ground (${\cal L}=63$) and the first excited states (${\cal L}=57$) are very close in energy, while the LS energy is substantially larger than both $E_{\rm GS}$ and $E_{\rm 1st}$, see also Table \ref{tab:ExactGSEnergy}.

Figure \ref{fig:energyN7step}(b) shows the energy gap between the first excited and the ground state as a function of $\beta$. The values of the gap lie between $\sim 0.1e^2/a_0$ at $\beta=1.617$ and 2.0512 and $\sim 0.0375e^2/a_0$ at $\beta=2.8819$.

\subsubsection{Seven particles, smooth density profile\label{sec:resultsSmooth}}

As has been shown in Section \ref{sec:remark7}, at $\beta=1/\nu=3$ the angular momenta of the ground (${\cal L}=63$) and the first excited state (${\cal L}=57$) in the system with the step-like profile are reversed when the profile is smooth. It is interesting to see how the energy spectra look over a wide range of the magnetic field in a system with the smooth density profile (\ref{dens-smooth}). 

\begin{figure}[ht!]
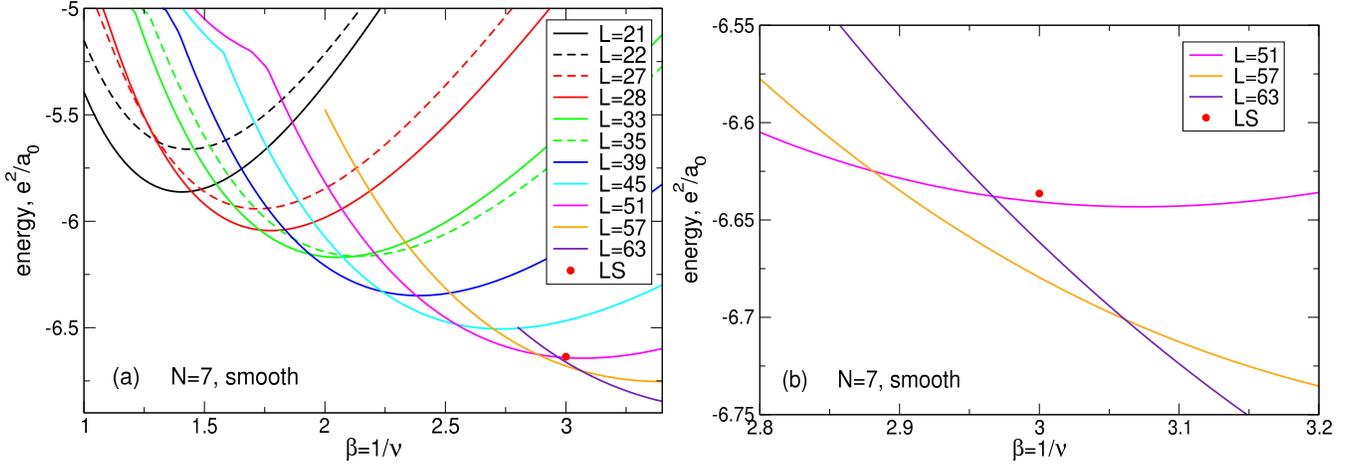

\includegraphics[width=0.49\columnwidth]{N7smoo.eps}
\includegraphics[width=0.49\columnwidth]{N7smooAtb3.eps}
\caption{\label{fig:energyN7smoo} (a) The energy of the many-body states with the total angular momenta from ${\cal L}={\cal L}_{\min}=21$ up to ${\cal L}=63$ in the system of $N=7$ 2D electrons as a function of the magnetic field parameter $\beta=1/\nu$. For all shown states the index $s$ equals $s=1$. Only the states which are either ground or first excited states are shown. The LS energy at $\nu=1/3$ is shown by a small red circle. (b) The vicinity of the point $\beta=3$ on a larger scale. The positive background density profile is smooth.}
\end{figure}

Figure \ref{fig:energyN7smoo}(a) shows these spectra. In general, the curves $E_{\rm GS}(\beta)$ and $E_{\rm 1st}(\beta)$ look qualitatively similar to the case of the step-like profile, but some details are quantitatively different. In particular, the intervals of $\beta$ corresponding to different ${\cal L}$ values differ from the case of the step-like profile, see Table \ref{tab:N7LtotGS1stSmooth}. The rule (\ref{GSrule}) is also satisfied for $N=7$ and a smooth density profile, with the exception of the region of small $\beta$, where the ground state has ${\cal L}=28$ instead of the expected ${\cal L }=$27.

\begin{table}[!ht]
\caption{The total angular momenta of the ground state ${\cal L}_{\rm GS}$ and of the first excited state ${\cal L}_{\rm 1st}$ assume the values shown in the last two columns in the intervals from $\beta_{\rm from}$ to $\beta_{\rm to}$ shown in the first two columns. The number of particles is  $N=7$, the density profile is smooth. \label{tab:N7LtotGS1stSmooth}}
\begin{tabular}{cccc}
$\beta_{\rm from}$ & $\beta_{\rm to}$   & ${\cal L}_{\rm GS}$   & ${\cal L}_{\rm 1st}$\\
\hline
   1.0000 &    1.3016 &   21 &   22 \\
   1.3016 &    1.4455 &   21 &   28 \\
   1.4455 &    1.4849 &   28 &   21 \\
   1.4849 &    1.6180 &   28 &   27 \\
   1.6180 &    1.7293 &   28 &   33 \\
   1.7293 &    1.8142 &   33 &   28 \\
   1.8142 &    1.8689 &   33 &   35 \\
   1.8689 &    1.9380 &   33 &   39 \\
   1.9380 &    2.0744 &   39 &   33 \\
   2.0744 &    2.2388 &   39 &   45 \\
   2.2388 &    2.3790 &   45 &   39 \\
   2.3790 &    2.5443 &   45 &   51 \\
   2.5443 &    2.6990 &   51 &   45 \\
   2.6990 &    2.8809 &   51 &   57 \\
   2.8809 &    2.9667 &   57 &   51 \\
   2.9667 &    3.0613 &   57 &   63 \\
   3.0613 &    3.4000 &   63 &   57 \\
\end{tabular}
\end{table}

The neighborhood of the point $\beta=3$ is shown in Figure \ref{fig:energyN7smoo}(b). One sees that around the point $\beta=3$, at $2.8809<\beta<3.0613$, the ground state has the angular momentum ${\cal L}_{\rm 1st}=57$, as was found in Ref. \cite{Kasner94} and discussed in Section \ref{sec:remark7}. The state ${\cal L}_{\rm GS}=63$ becomes the ground state at larger $\beta>3.0613$. The LS point lies above the second excited state which has ${\cal L}_{\rm 2nd}=51$.

\subsection{Electron density\label{sec:ResDensity}}

The magnetic field dependencies of the energy levels investigated in Section \ref{sec:ExactSolutionNu<1-Energy} show that the ground state energy decreases with increasing $B$, oscillating due to a stepwise increase of the total angular momentum ${\cal L}_{\rm GS}$ in the ground states. The physical reasons of such oscillations are explained below. Here I analyze the density of electrons $n_e(r)$ in a number of different ${\cal L}$-states for two cases,  $N=5$ and $N=7$, representing the one-shell and two-shells configurations. 

\subsubsection{Five particles}

The energy spectra of the system of $N=5$ electrons are shown in Figure \ref{fig:energyN5step}(a). One of the states, shown there by the blue solid curve, has the total angular momentum ${\cal L}=25$. Let us consider the density of electrons in this state at different magnetic fields.

\begin{figure}[ht!]
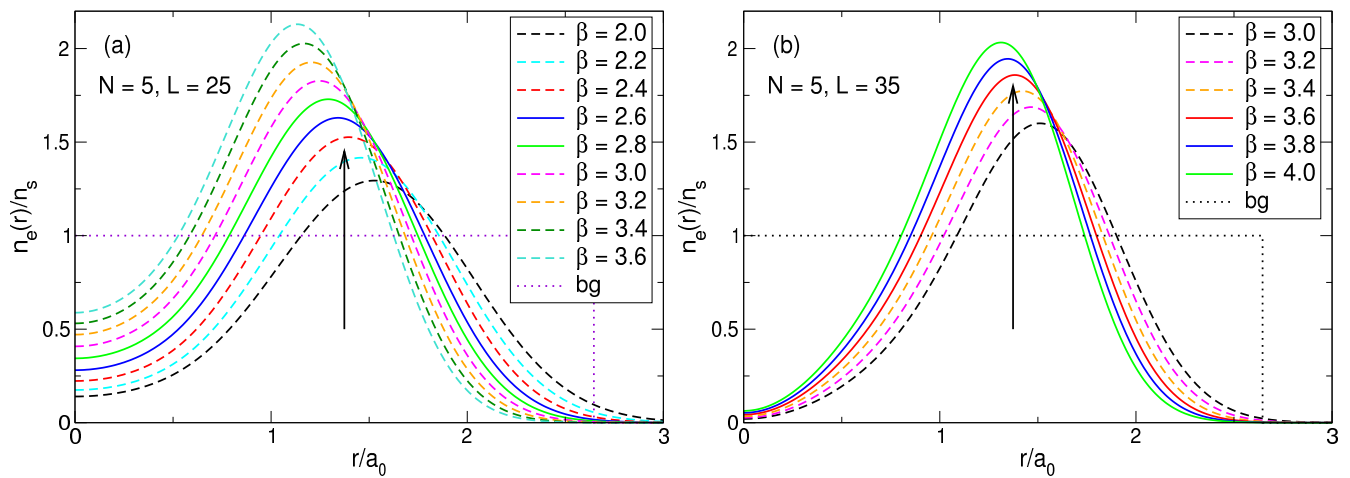

\includegraphics[width=0.49\columnwidth]{DensityN5L25.eps}
\includegraphics[width=0.49\columnwidth]{DensityN5L35.eps}
\caption{\label{fig:Density_N5L2535} The density of electrons in the system of $N=5$ particles at different magnetic fields and at (a) ${\cal L}=25$ and (b) ${\cal L}=35$. For $\beta$-values, for which the corresponding ${\cal L}$ state is the ground state (one of the excited states), the function $n_e(r)$ is shown by solid (dashed) curves. The vertical arrow at $r/a_0=R_s/a_0=1.373422$ shows the position of the shell radius in the classical Wigner molecule. The positive background density profile is step-like and shown by the black dotted curve in both panels.}
\end{figure}

Figure \ref{fig:Density_N5L2535}(a) shows the normalized function $n_e(r)$, for the state ${\cal L}=25$ and for $\beta$ varying from $\beta=2.0$ up to $\beta=3.6$ with the step 0.2. At all values of $\beta$, the function $n_e(r)$ has a maximum at a certain distance from the origin, i.e., the electron density has the shape of a ring. The position of maximum, i.e. the ring radius, as well as the width of the maximum, are large at small $\beta$ and decrease when $\beta$ grows. This is caused by a decrease of the magnetic length with increasing magnetic field. If the ${\cal L}=25$ state is the ground state of the system at a given $\beta$, the corresponding density functions are shown by solid curves; otherwise they are shown by dashed curves.

The classical shell radius at $N=5$ equals $R_s=1.3734 a_0$. It is shown by the vertical arrow in Figure \ref{fig:Density_N5L2535}. When $\beta=2.0$ (the rightmost curve in Figure \ref{fig:Density_N5L2535}(a)), the density maximum is located at $r=R_{\max}\approx 1.525 a_0$, which is noticeably larger than the classical shell radius. In this magnetic field, the state ${\cal L}=25$ is not the ground, but the fifth excited state. Then, when the magnetic field increases, the wave function shrinks, and the density maximum shifts to smaller values of $r$. At $\beta=2.2$ and $2.4$, the state ${\cal L}=25$ is the forth and first excited state respectively, and the maximum of the electron density is still located at the points larger than $R_s$. 

When the $B$-field continues to grow, the density maximum $R_{\max}$ passes through the position of the shell radius $R_s$, and the state ${\cal L}=25$ becomes the ground state. In Figure \ref{fig:Density_N5L2535}(a) this is the case for the curves labeled by $\beta=2.6$ and 2.8. The maxima of the electron density then lie at $R_{\max}\approx 1.34a_0$ and $1.29a_0$, respectively, close to the position of $R_s$. As the $\beta$-parameter increases further, the ring radius $R_{\max}$ becomes too small compared to the classical shell radius $R_s$, and the state with ${\cal L}=25$ ceases to be the ground state. 

In order to return the ring radius $R_{\max}$ back to the classical shell radius $R_s$, a state with a larger ${\cal L}$ should take over the role of the ground state. The ground state angular momentum jumps up by the value $\delta{\cal L}=n^{\rm st}(N)=5$, and the radius of the electron density ring turns out to be again close to or slightly larger than $R_s$. 

Figure \ref{fig:Density_N5L2535}(b) confirms this simple physical picture. It shows the density of electrons for the state ${\cal L}=35$ and for $\beta$ varying from $\beta=3.0$ up to $\beta=4.0$ with the step 0.2. Again, at all values of $\beta$ the density of electrons has the shape of a ring, but the ring radius $R_{\max}$ is larger (at the same values of $\beta$) then in panel (a) since the angular momentum is now bigger. At $\beta=3.0$ the state with ${\cal L}=35$ is the second excited state, and the density maximum lies well outside the shell radius $R_s$. When $\beta$ increases up to $\beta=3.2$ and 3.4 the density maxima become smaller, $R_{\max}=1.46a_0$ and $1.42a_0$, respectively, but they are still larger than $R_s$. At even larger $\beta$'s, $\beta=3.6$, 3.8 and 4.0, the ring radii become almost equal or slightly smaller than $R_s$, $R_{\max}/a_0\approx 1.38$, 1.35, and 1.31, respectively, and the state  ${\cal L}=35$ becomes the ground state. 

\subsubsection{Seven particles, step-like density profile\label{sec:density7step}}

\begin{figure}[ht!]
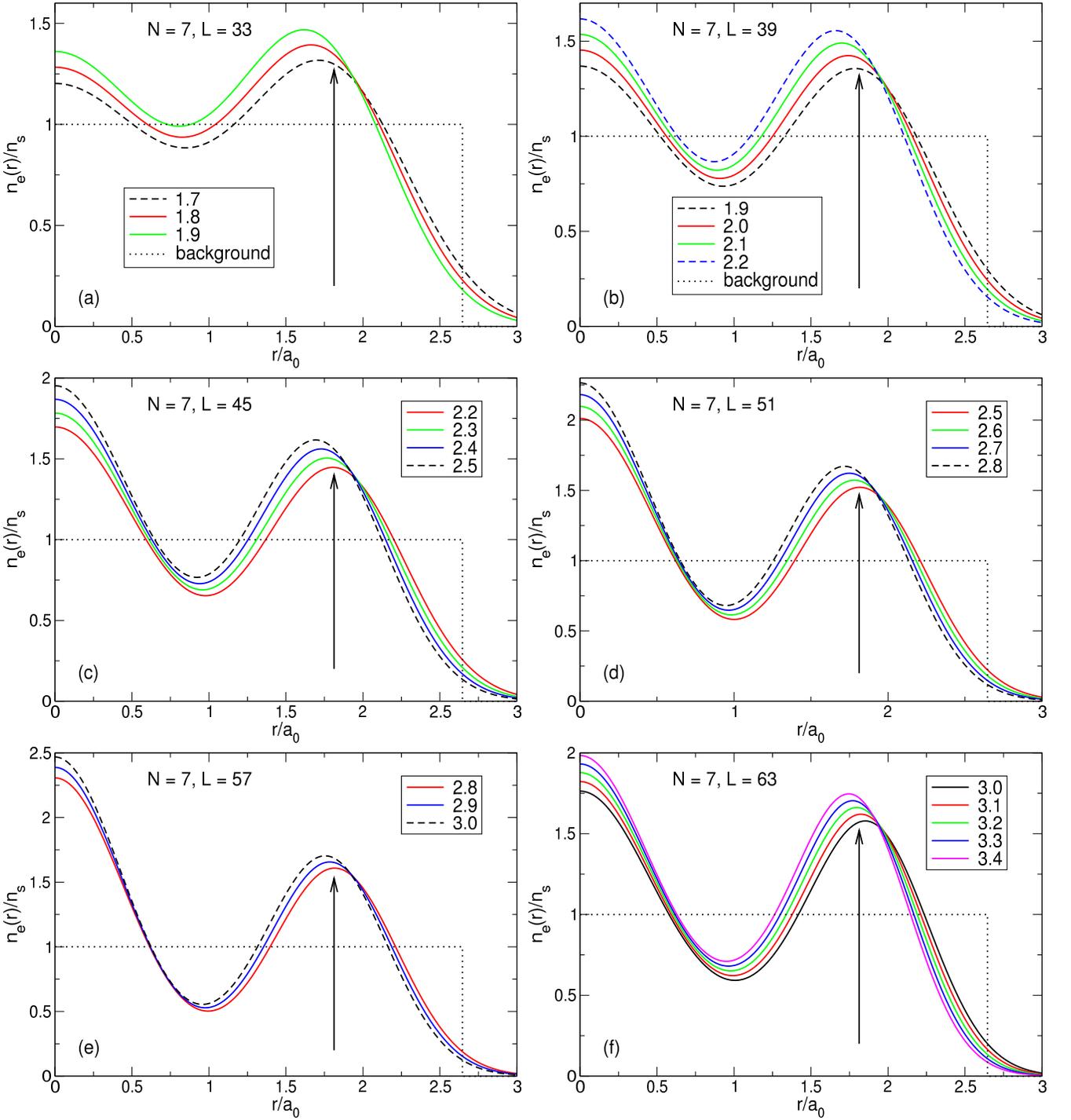

\includegraphics[width=0.49\columnwidth]{DensN7L33.eps}
\includegraphics[width=0.49\columnwidth]{DensN7L39.eps}
\includegraphics[width=0.49\columnwidth]{DensN7L45.eps}
\includegraphics[width=0.49\columnwidth]{DensN7L51.eps}
\includegraphics[width=0.49\columnwidth]{DensN7L57.eps}
\includegraphics[width=0.49\columnwidth]{DensN7L63.eps}
\caption{\label{fig:Density_N7} The density of electrons in the system of $N=7$ particles at different magnetic fields and at (a) ${\cal L}=33$, (b) ${\cal L}=39$, (c) ${\cal L}=45$, (d) ${\cal L}=51$, (e) ${\cal L}=57$, and (f) ${\cal L}=63$. For $\beta$-values, for which the corresponding ${\cal L}$ state is the ground state, the function $n_e(r)$ is shown by solid curves. The arrow at $r/a_0=R_s/a_0=1.8126$ shows the position of the shell radius in the classical Wigner molecule. The step-like positive background density profile is shown by the black dotted curve in all panels.}
\end{figure}

Let us now consider the system of $N=7$ particles which illustrates the case of the two-shells configuration. I consider first relatively large magnetic fields $\beta\ge 1.7$. Figures \ref{fig:Density_N7}(a)-(f) show the density of electrons for the ground state angular momenta from ${\cal L}=33$ to ${\cal L}=63$ and for $\beta$ from $1.7$ to 3.4 with the step 0.1. 

\begin{figure}[ht!]
\includegraphics[width=0.49\columnwidth]{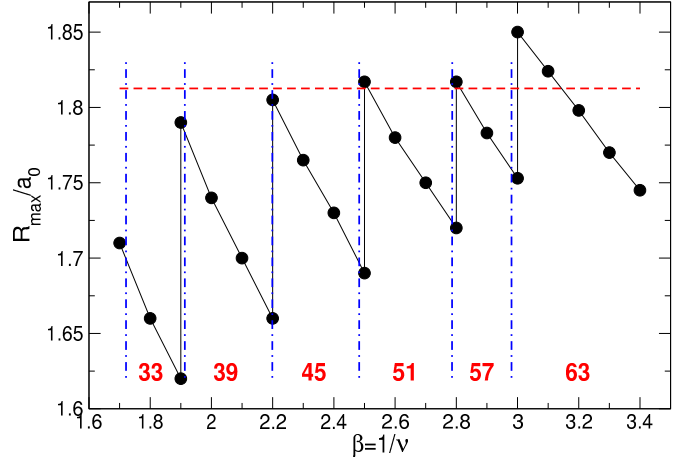}
\caption{\label{fig:DensMaxima} Positions of the maxima of the density curves shown in Figure \ref{fig:Density_N7} at different $\beta$-values and different angular momenta ${\cal L}$. The red dashed line at $R/a_0=1.8126$ shows the position of the shell radius in the classical Wigner molecule. The blue dash-dotted lines separate the areas where the ground states have different angular momenta indicated by the numbers 33, 39, $\dots$, 63 in the corresponding $\beta$-intervals.}
\end{figure}

As expected, all density curves have two maxima, one at $r=0$ and the other at a finite $r=R_{\max}$. The behavior of the density curves is similar to the case of $N=5$ particles. As the magnetic field increases, the maxima of the density curves $R_{\max}$ pass through the position of the classical shell radius $R_s$. When, for a given ${\cal L}$, $R_{\max}$ becomes close to or slightly smaller than $R_s$, this state becomes the ground state. When $R_{\max}$ becomes too small compared to $R_s$, the role of the ground state is taken over by the next state with the angular momentum increased by $\delta {\cal L}=n^{\rm st}(N)=6$. The oscillations of $R_{\max}$ around the fixed value of $R_s$ are shown in Figure \ref{fig:DensMaxima}. The quantum-mechanically calculated radii of the rings $R_{\max}$ are on average smaller than the classical shell radius $R_s$. This is because the potential energy of the attractive positively charged background is higher at $r>R_s$ and lower at $r<R_s$ compared to $V_b(r=R_s)$, see Figure \ref{fig:Ubpotential}. 

\begin{figure}[ht!]
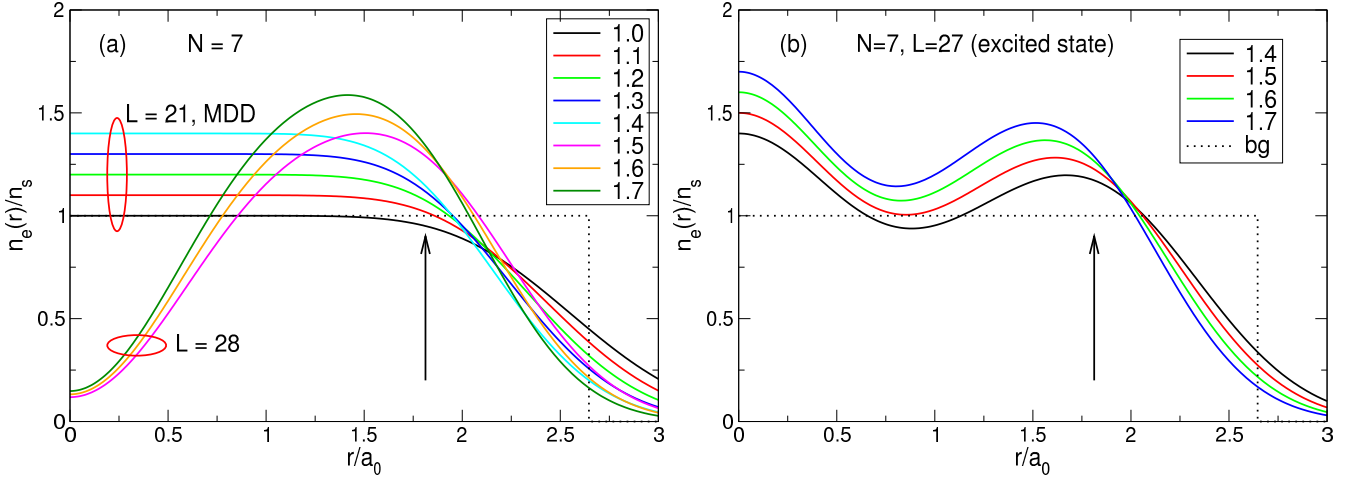

\includegraphics[width=0.49\columnwidth]{DensN7L2128.eps}
\includegraphics[width=0.49\columnwidth]{DensN7L27.eps}
\caption{\label{fig:Density_N7lowB} The density of electrons in the system of $N=7$ particles at different magnetic fields and (a) in the ground states with ${\cal L}=21$ and 28, and (b) in the excited state with ${\cal L}=27$. The arrow at $r/a_0=R_s/a_0=1.8126$ shows the position of the shell radius in the classical Wigner molecule. The step-like positive background density profile is shown by the black dotted curve in both panels.}
\end{figure}

At lower magnetic fields $\beta\lesssim 1.72$, the shape of the electron density in the ground state changes significantly, Figure \ref{fig:Density_N7lowB}(a). Here, one should consider two different situations. If $1.0\le\beta\le 1.4211$, the ground state has the total angular momentum ${\cal L}=21$, the expansion of the many-body wave function (\ref{PsiExpansion}) contains only one Slater determinant, $N_{mbs}=1$, and the wave function is the MDD state $|0,1,2,3,4,5,6\rangle$. The electron density in this state, Figure \ref{fig:Density_N7lowB}(a), does not have an internal structure typical for a Wigner molecule, but has the form of a uniform liquid, Figure \ref{fig:MDDdens}. The radius of the MDD-liquid spot decreases, while the density on the plateau increases with increasing $\beta$. As I discussed at the beginning of Section \ref{sec:ExactSolutionNu<1}, this solution is not accurate enough and should be improved by taking higher Landau level states into account. As seen from Figure \ref{fig:Density_N7lowB}(a), this should be done at least at $1.0\le\beta\lesssim 1.42$. Here I show the density $n_e(r)$ for the state ${\cal L}=21$ for completeness only.

In intermediate magnetic fields, $1.4211\le \beta< 1.721$, the total angular momentum in the ground state is ${\cal L}=28$, and the density $n_e(r)/n_s$ assumes the form of a broad ring with a \textit{single} density maximum at a finite $r$, Figure \ref{fig:Density_N7lowB}(a). That is, in this $\beta$-region, instead of the ``usual'' $(1,N-1)$ Wigner molecule configuration, the configuration $(0,N)$ is realized. The state ${\cal L}=27$, corresponding to the ``correct'' configuration $(1,N-1)$, has the density profile with two maxima, Figure \ref{fig:Density_N7lowB}(b), but is the first excited state.

There are two possible explanations for this ``unusual'' behavior of the electron density. First, it is not entirely clear whether the solution in this $\beta$-range is sufficiently accurate. If ${\cal L}=28$, the number of many-body basis states in the expansion (\ref{PsiExpansion}) is $N_{mbs}=15$. Although this number is much larger than one, it is worth checking whether this result still holds when higher Landau level states are taken into account. 

If this result is correct, it can be explained as follows. In strong magnetic fields, the size of the wave functions $\lambda\propto 1/\sqrt{\beta}$ is small, and ``thin'' electrons are more like point-like classical particles. Trying to find the configuration with the lowest energy in the field of the attractive potential, Figure \ref{fig:Ubpotential}, six electrons push one of their comrades into the center of the disk, and themselves form a ring around it. As the magnetic field decreases, the length $\lambda$ grows, and the electrons become ``fat''. A ``fat'' electron cannot fit in the center of the potential well, therefore all seven electrons are located at its edge, creating a single-shell $(0,7)$ configuration. 

To answer the question of which of the two described situations is actually the case, one should develop a more general theory that takes into account the higher Landau level states.

\subsubsection{Seven particles, smooth density profile\label{sec:density7smoo}}

\begin{figure}[ht!]
\includegraphics[width=0.49\columnwidth]{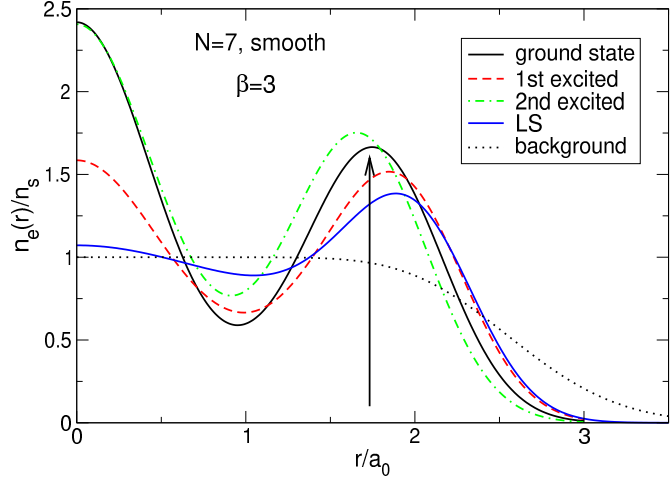}
\caption{\label{fig:densN7smooth} The density of electrons in the ground state (${\cal L}=57$), first (${\cal L}=63$) and second (${\cal L}=51$) excited states, and in the LS in a system of $N=7$ particles at $\beta=3$. The positive background density profile is smooth. The arrow at $r/a_0=R_s/a_0=1.732047$ shows the position of the outer shell of the classical Wigner molecule in the case of the smooth background density. The density of the LS at $r=0$ is 2.2576 times smaller than the density of the ground state. }
\end{figure}

In the case of the smooth background density profile (\ref{dens-smooth}) the density of electrons at $\beta=3$ is shown in Figure \ref{fig:densN7smooth} for the ground, first and second excited states, ${\cal L}=57$, 63, and 51, as well as for the Laughlin state. The density of electrons in all three exact states has the shape of the Wigner molecule, with the high maximum in the center and the second maximum at $r=R_{\max}\simeq R_s$. The density in the three exact states differs only in the position of the second maximum and the heights of both maxima. 

The difference between the electron densities in the LS and in the true ground state at $\beta=3$ in the smooth background case is even greater than in the previously considered case of the step-like profile, Figure \ref{fig:LaughDens}(c). At $r=0$, the ratio $n_e^{\rm GS}(0)/n_e^{\rm LS}(0)$ is equal to 2.2576. The density maximum in the LS is located at the point $R_{\max}^{\rm LS}/a_0=1.888$, which is significantly further from the center than the classical shell radius $R_s/a_0=1.732$. Thus, while in the exact solution one of the electrons jumps to the bottom of the deep potential well, Figure \ref{fig:Ubpotential}, lowering the energy of the system, in the LS all seven electrons accumulate at the edge of the well, increasing its energy.

\subsection{Exact solutions: Summary\label{sec:exact-summary}}

Thus, the $B$-dependencies of the energy of the ground and excited states, as well as the corresponding electron densities, give a simple and clear picture of the FQHE physics. The interplay of the Coulomb forces and the compressive action of the $B$-field leads to the alternating opening and closing of the energy gaps in the many-body spectrum of the system as the magnetic field changes, Figures \ref{fig:energyN2step} -- \ref{fig:energyN7smoo}. It is reasonable to assume that in a macroscopic 2D electron system the diagonal conductivity will tend to zero in the finite-gap regions of $B$, while the Hall conductivity will take quantized values there. It is also clear from the spectra of Figures \ref{fig:energyN2step} -- \ref{fig:energyN7smoo}, that the point $\nu=1/3$ (or any other fractional $\nu$ point), is not something special; this is just a single point in the $B$-field interval where the energy gap is finite. This fact fully agrees with the FQHE experiments in which quantized is not the filling factor but the Hall conductivity.

It is clear that the physical picture similar to that described in this Section should also occur at larger $\nu$, when the higher Landau levels are also occupied. Moreover, one sees from the described analysis that in order to explain the Hall quantization, both integer and fractional, there is no need to assume the presence of a disorder in the system. Both effects can be explained from a unified position within the framework of the many-body theory as it has been done in the present Section.

Let us now discuss some further statements of the currently accepted theory of the FQHE effect.

\section{Further comments on the currently accepted FQHE theory \label{sec:discussionAcceptedTheory}}

\subsection{Fractionally charged excitations in the Laughlin theory\label{sec:ExcStates}}

Apart from the wave function (\ref{LaughlinWF}), which was designed for the ground state of the FQHE system at $\nu=1/m$, Laughlin also ``generated'' trial many-body wave functions for the excited states. According to \cite{Laughlin83} these wave functions should have the form
\be 
\Psi_{\rm LS}^{(m),+z_0}(\bm r_1,\bm r_2,\dots,\bm r_N,\bm r_0)\propto \exp\left(-\frac 12\sum_{j=1}^N|z_j|^2\right)\left(\prod_{1\le i\le N} (z_i-z_0) \right)\left(\prod_{1\le j<k\le N} (z_j-z_k)^m \right)
\label{LaughlinHoles}
\ee
for particles that were called ``quasiholes'', and the form
\be 
\Psi_{\rm LS}^{(m),-z_0}(\bm r_1,\bm r_2,\dots,\bm r_N,\bm r_0)\propto \exp\left(-\frac 12\sum_{j=1}^N|z_j|^2\right) \left(\prod_{1\le i\le N} \left(\frac{\p}{\p z_i}-\frac{z_0}{l_B^2}\right) \right)\left(\prod_{1\le j<k\le N} (z_j-z_k)^m \right)
\label{LaughlinElectrons}
\ee
for particles that were called ``quasielectrons''. Projections of these wave functions ``onto the analogous ones computed numerically'', for four particles and $m=3$, were found to be $0.998$ for $\Psi_{\rm LS}^{(m),-0}$ and $0.982$ for $\Psi_{\rm LS}^{(m),+0}$ (at $z_0=0$). Laughlin also stated that these quasiparticles have fractional charges $e/m$. Let us explore the properties of the quasiparticles (\ref{LaughlinHoles}) and (\ref{LaughlinElectrons}) in somewhat more detail. For simplicity, I will also consider only the case $z_0=0$.

In Section \ref{sec:Laughlin13} I have expanded the Laughlin function (\ref{LaughlinWF}) in a set of many-body basis states $\Psi_s$, Eq. (\ref{PsiExpand}). A similar expansion can be also performed for the ``quasihole'' and ``quasielectron'' wave functions (\ref{LaughlinHoles}) and (\ref{LaughlinElectrons}). For example, for three particles and $m=3$, the LS function (\ref{LaughlinWF}) is expanded as
\ba
\Psi_{\rm LS}^{(m=3)}&\propto&e^{-(|z_1|^2+|z_2|^2+|z_3|^2)/2}\left[(z_1-z_2)(z_1-z_3)(z_2-z_3)\right]^3
\nonumber \\ &\propto&
-\sqrt{ 0! 3!6!}\Psi_{|0,3,6\rangle} +3\sqrt{ 0! 4!5!}
\Psi_{|0,4,5\rangle} +3\sqrt{ 1! 2!6!}
\Psi_{|1,2,6\rangle} -6\sqrt{ 1! 3!5!}
\Psi_{|1,3,5\rangle} +15\sqrt{ 2!3!4!}\Psi_{|2,3,4\rangle},
\ea
with the coefficients $C_s$ ($=-1,3,3,-6$ and $15$) from Table \ref{tab:LaughCsN3}. Similar expansions of the ``quasihole'' and ``quasielectron'' states (\ref{LaughlinHoles}) and (\ref{LaughlinElectrons}) have the form
\ba
\Psi_{\rm LS}^{(m=3),+0}&\propto&e^{-(|z_1|^2+|z_2|^2+|z_3|^2)/2}z_1z_2z_3\left[(z_1-z_2)(z_1-z_3)(z_2-z_3)\right]^3
\nonumber \\ &\propto&
-\sqrt{ 1! 4!7!}]\Psi_{|1,4,7\rangle} +3\sqrt{ 1! 5!6!}
\Psi_{|1,5,6\rangle} +3\sqrt{ 2! 3!7!}
\Psi_{|2,3,7\rangle} -6\sqrt{ 2! 4!6!}]
\Psi_{|2,4,6\rangle} +15\sqrt{ 3!4!5!}\Psi_{|3,4,5\rangle}
\label{LaughlinHolExpan}
\ea
and
\ba
\Psi_{\rm LS}^{(m=3),-0}&\propto&
e^{-(|z_1|^2+|z_2|^2+|z_3|^2)/2}\Big(\p_{z_1}\p_{z_2}\p_{z_3}\left[(z_1-z_2)(z_1-z_3)(z_2-z_3)\right]^3\Big)
\nonumber \\ &\propto&
3\cdot 1\cdot 2\cdot 6\cdot \sqrt{ 0! 1!5!}
\Psi_{|0,1,5\rangle} -6\cdot 1\cdot 3\cdot 5\sqrt{ 0! 2!4!}
\Psi_{|0,2,4\rangle} +15\cdot 2\cdot 3\cdot 4\sqrt{ 1!2!3!}\Psi_{|1,2,3\rangle}.
\label{LaughlinElecExpan} 
\ea
As seen from the definition (\ref{LaughlinHoles}) and from the example (\ref{LaughlinHolExpan}), the ``quasihole'' wave function $\Psi_{\rm LS}^{(m=3),+0}$ is an eigenfunction of the total momentum operator with 
\be 
{\cal L}_{\rm Q-hole}=\frac {3N(N-1)}2+N. \label{LtotLaughHoles}
\ee
The statement (\ref{LtotLaughHoles}) is valid for any number of particles $N$.  
Similarly, the ``quasielectron'' wave function $\Psi_{\rm LS}^{(m=3),-0}$ is an eigenfunction of the total momentum operator with 
\be 
{\cal L}_{\rm Q-elec}=\frac {3N(N-1)}2-N \label{LtotLaughElecs}
\ee
for any $N$. This obviously shows that the wave functions (\ref{LaughlinHoles}) and (\ref{LaughlinElectrons}) can in no case describe the real low lying excitations of the system. As it was seen from the exact solution of the problem, Sections \ref{sec:ExactSolution} and \ref{sec:ExactSolutionNu<1}, the angular momenta of the lowest excited states differ from the angular momenta of the ground states by a small number $\delta{\cal L}\le 6$, see Eqs. (\ref{deltaL}), (\ref{GSrule}). But according to (\ref{LtotLaughHoles}) and (\ref{LtotLaughElecs}), the angular momenta difference between the ground and the excited state is \textit{macroscopically large}. For example, in real samples with $N\simeq 10^{11}$ (in the thermodynamic limit) $\delta{\cal L}$ would be about $\delta{\cal L}=N\simeq 10^{11}$. This is evidently impossible. 

\begin{figure}[ht!]
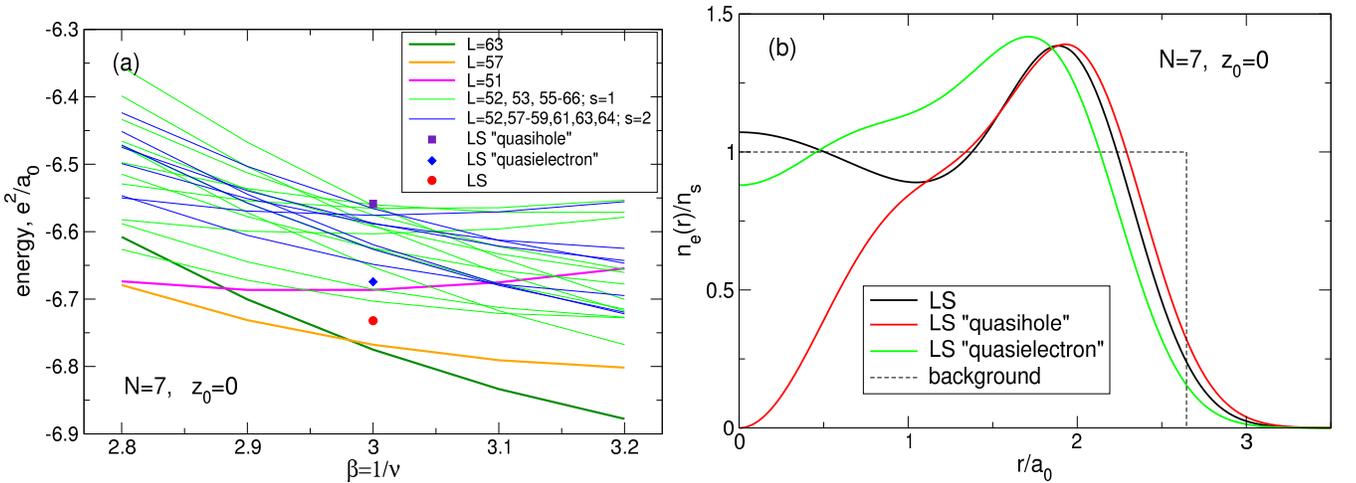

\includegraphics[width=0.49\columnwidth]{LaughlinExcitationEnergiesN7step.eps}
\includegraphics[width=0.49\columnwidth]{LaughlinExcitationDensitiesN7step.eps}
\caption{\label{fig:LaughExcitations} (a) The energy of the Laughlin state and the Laughlin excitations (``quasiholes'' and ``quasielectrons''), together with the energies of exact ground and excited states in the vicinity of the filling factor $\nu=1/3$. The green (blue) thin curves show the energy of the $s=1$ ($s=2$) exact excited states with different ${\cal L}$; not all exact excited states with energies lower than the energy of the Laughlin ``quasihole'' are shown. (b) The density of electrons in the Laughlin ``quasiholes'' and ``quasielectrons'' states (\ref{LaughlinHoles}) and (\ref{LaughlinElectrons}). }
\end{figure}

Using the expansions like (\ref{LaughlinHolExpan}) and (\ref{LaughlinElecExpan}), I calculated the energy and the density of electrons in the ``quasihole'' and ``quasielectron'' states (\ref{LaughlinHoles}) and (\ref{LaughlinElectrons}) for $N=7$. The results are shown in Figure \ref{fig:LaughExcitations}. One sees that, Figure \ref{fig:LaughExcitations}(a), while the LS energy lies between the first and the second exact excited states, see also Figure \ref{fig:energyN7step}, the energy of the Laughlin ``quasielectron'' is between the forth and fifth exact excited states. As for the energy of the ``quasihole'' state, it lies above at least 24 different exact excited states shown by thin green (the states with different ${\cal L}$ and $s=1$) and thin blue curves (the states with different ${\cal L}$ and $s=2$). Note that the number of different exact states lying below the ``quasihole'' state is definitely larger since I did not check \textit{all} possible $({\cal L},s)$ states that could have lower energy than the ``quasihole'' state (\ref{LaughlinHoles}).

Figure \ref{fig:LaughExcitations}(b) shows the density of electrons in the Laughlin ``quasihole'' and ``quasielectron'' states (\ref{LtotLaughHoles}) and (\ref{LtotLaughElecs}), together with the density of the LS (\ref{LaughlinWF}). One sees that the ``quasihole'' and ``quasielectron'' curves are qualitatively different from the densities of the exact low-lying excited states, compare with Figures \ref{fig:DensityExact}(b) and \ref{fig:densN7smooth}.

Since the total angular momenta of the ``quasihole'' and ``quasielectron'' states (\ref{LtotLaughHoles}) and (\ref{LtotLaughElecs}) for $N=7$ equal ${\cal L}_{\rm Q-hole}=70$ and ${\cal L}_{\rm Q-elec}=56$ respectively, the projections of the both states onto the first exact excited state ${\cal L}_{\rm 1st}=57$ are identically equal to zero.

Thus, the ``fractionally charged quasiparticles'' generated in Ref. \cite{Laughlin83} have no relation to physical reality.

\subsection{Can the FQHE problem be studied in models without a positively charged background?\label{subsec:no-backgr}}

Electrons repel each other by strong and long-range Coulomb forces. To hold them together requires a positively charged background, which is always present in real physical systems. However, in many publications, e.g. in Refs. \cite{Haldane83,Haldane85,Morf02,Tsiper01,Girvin04,Johri16}, the FQHE problem was considered in systems without edges, i.e., without taking into account the positively charged background. In some works \cite{Haldane83,Haldane85,Morf02} it has been assumed that electrons are on the surface of a sphere. In other papers, e.g. Refs.  \cite{Tsiper01,Girvin04,Johri16}, electrons were assumed to occupy a 2D plane, but the background-electron interaction was removed from the Hamiltonian. Are such models suitable for describing the FQHE effect in real physical systems?

\begin{figure}[ht!]
\includegraphics[width=0.5\columnwidth]{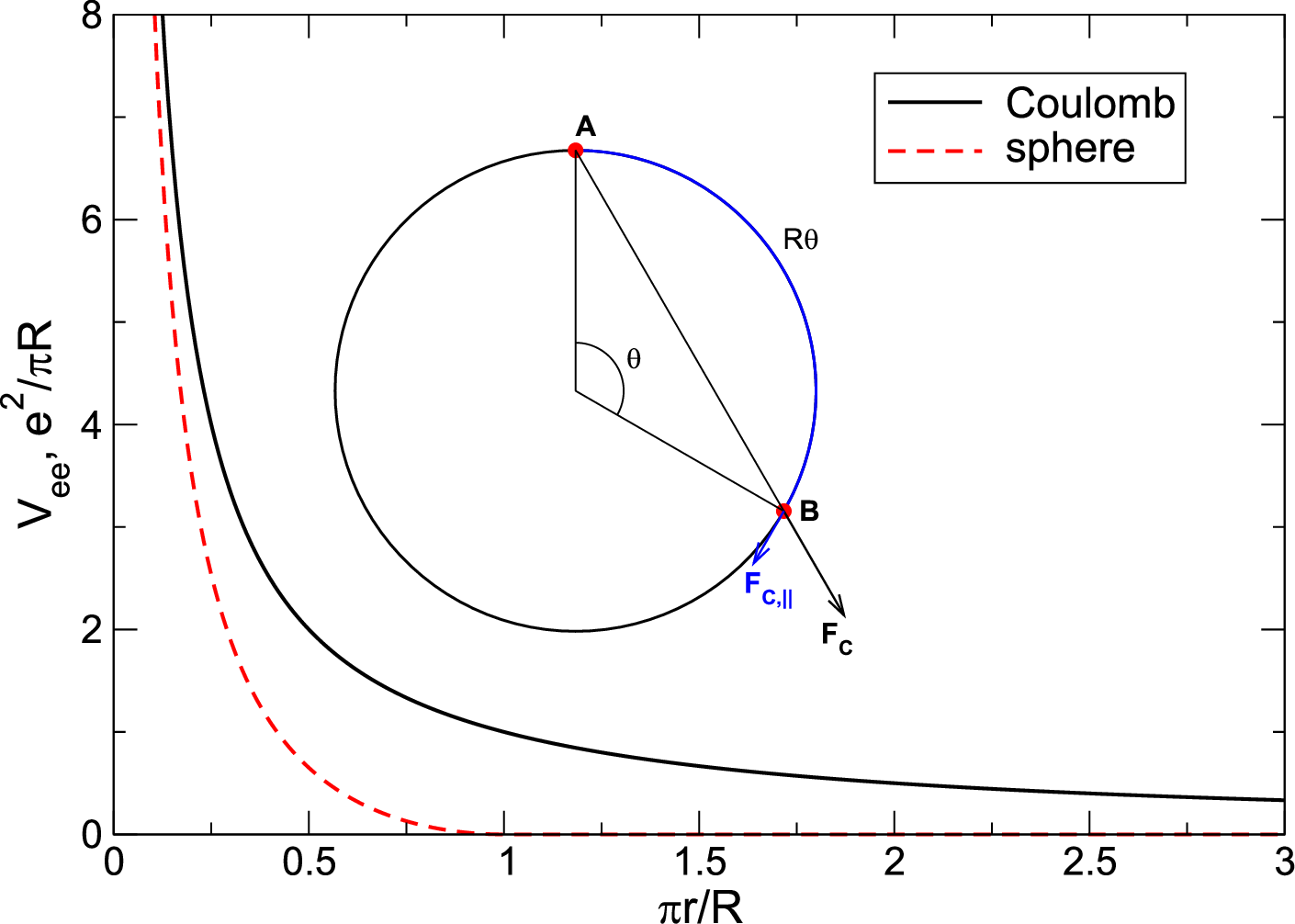}
\caption{\label{fig:haldane-sphere}The Coulomb interaction potential (black curve) and the potential (\ref{ee-int-sphere}). The energy is in units $e^2/\pi R$. Inset shows a sphere with two interacting electrons $A$ and $B$ separated by a polar angle $\theta$.}
\end{figure}

\paragraph{2D electrons on the surface of a sphere.} By considering 2D electrons on the surface of a sphere, one is actually dealing with a different problem. Let us consider two particles, $A$ and $B$, separated by a polar angle $\theta$, on the surface of a sphere, see inset to Figure \ref{fig:haldane-sphere}. The Coulomb force acting on the particle $B$ from the particle $A$ is shown as the vector ${\bm F}_C$, where
\be 
|{\bm F}_{C}|=\frac {e^2}{|AB|^2}=\frac {e^2}{(2R)^2\sin^2(\theta/2)}.
\ee
However, the projection of the force ${\bm F}_C$ on the direction perpendicular to the sphere surface does not matter for 2D electrons. The force that matters, i.e., the force that really acts on electrons in the direction parallel to the sphere surface, equals 
\be 
F_{C,\parallel}=F_C\cos\frac\theta 2=\frac {e^2}{(2R)^2}\frac{\cos(\theta/ 2)}{\sin^2(\theta/2)}.
\ee
This force equals zero at $\theta=\pi$, although the real Coulomb force is always finite and decreases very slowly with $r$. 

In addition, since the final goal is to model the behavior of the 2D electrons on a plane, the distance $r$ between them should be counted not along the chord $AB$ \cite{Haldane83,Haldane85}, but along the ark, shown by the blue curve, so that $r=R\theta$. Thus the force acting on the 2D electrons along the 2DEG surface equals
\be 
F_{C,\parallel}=\frac {e^2}{(2R)^2}\frac{\cos(r/ 2R)}{\sin^2(r/2R)},\ \ \ \textrm{ if }r<\pi R ,
\ee
and zero otherwise. The corresponding interaction potential energy of particles on the sphere,
\be 
V_{ee}^{\rm sphere}=\frac {e^2}{2R}\left(\frac{1}{\sin(r/2R)}-1\right)\theta(\pi-r/R),
\label{ee-int-sphere}
\ee
is shown in Figure \ref{fig:haldane-sphere} together with the Coulomb interaction energy $e^2/r$ of real 2D electrons. The difference is significant. The potential (\ref{ee-int-sphere}) is actually short-range, in contrast to the real Coulomb potential. Therefore, results of calculations in the spherical geometry can hardly correctly describe real systems of Coulomb interacting electrons.

\paragraph{No positively charged background for 2D electrons on a plane.} In the case of a flat geometry without the positive background the electrons are assumed to be at the lowest Landau level and are held together due to the angular momentum conservation \cite{Laughlin83a,Girvin04}.  The idea to ignore the positive background potential in this situation can lead to incorrect results and physically unreasonable conclusions. Let us consider, following \cite{Laughlin83a,Girvin04}, two particles with the relative angular momentum $m$ and the center of mass angular momentum $M$ in the plane $z=0$. The particles are at the lowest Landau level. Under these conditions, according to \cite{Laughlin83a,Girvin04}, the \textit{unique} analytic wave function that describes the two-body problem is
\be 
\psi_{mM}(z_1,z_2)\propto (z_1-z_2)^m(z_1+z_2)^Me^{-(|z_1|^2+|z_2|^2)/2}.
\ee
Further, it is stated ``Remarkably, this is the exact (neglecting Landau level mixing) solution for the Schr\"odinger equation for \textit{any} central potential $V(|z_1 - z_2|)$ acting between the two particles'' \cite{Girvin04}.

This statement is formally correct, but it obviously contradicts common sense if we think about real systems of electrons. Let us assume that the interaction of the two particles is described by the screened Coulomb potential
\be 
V_{ee}^{\rm scr}=\frac{e^2}{r}e^{-r/b}.\label{CoulScreen}
\ee
If $b\to\infty$, then $V_{ee}^{\rm scr}$ tends to the usual long-range Coulomb potential, which strongly repels electrons from each other. If $b$ tends to zero, the potential becomes short-range and the repulsive forces weaken significantly. In any real system, this will cause the distance between particles to decrease. But from the statement \cite{Laughlin83a,Girvin04} it follows that the wave function describing the motion of two particles is completely independent of the screening parameter $b$.

The contradiction arises because the above consideration does not take into account the attractive potential of a positively charged background, which is always present in real physical systems. If to add to the $V_{ee}^{\rm scr}$  the $be$ and $bb$ interactions, Eq. (\ref{CoulombEnergy}), the relative and center-of-mass angular momenta are not conserved separately. Only the total angular momentum is conserved, and the two-particle problem should be solved as described in Section \ref{sec:ExactSolution}. If $N=2$ and $m=3$, the solution should be sought as a linear combination 
\be 
\Psi(\bm r_1,\bm r_2)=A_{|0,3\rangle}|0,3\rangle+A_{|1,2\rangle}|1,2\rangle.\label{WF2part}
\ee
Having solved the problem for the screened Coulomb potential (\ref{CoulScreen}) I obtained the coefficients $A_{|0,3\rangle}$ and $A_{|1,2\rangle}$ shown in Table \ref{tab:AcoefsGS} for different values of the parameter $b/\lambda$. The wave function (\ref{WF2part}) with the coefficients from Table \ref{tab:AcoefsGS} describes a physically correct solution in which electrons come closer to each other when $b/\lambda$ tends to zero. In contrast, if to ignore the background-electron interaction, the coefficients $A_s$ equal to $A_{|0,3\rangle}=-1/2$ and $A_{|1,2\rangle}=\sqrt{3}/2$ \textit{for any value} of the screening parameter $b/\lambda$. These coefficients give the Laughlin wave function $\propto (z_1-z_2)^3$ but this does not describe physical reality.

\begin{table}[!ht]
\caption{The coefficients $A_s^{\rm GS}$ in the true ground state for the system of $N=2$ electrons with the screened Coulomb potential. \label{tab:AcoefsGS}}
\begin{tabular}{rcc}
$b/\lambda$ & $A_{|0,3\rangle}$ & $A_{|1,2\rangle}$\\
\hline
10.0 & $-0.47033$ & $0.88249$ \\
1.0 & $-0.44351$ & $0.89627$ \\
0.6 & $-0.40232$ & $0.91550$ \\
0.2 & $-0.12297$ & $0.99241$ \\
0.1 & $-0.02107$ & $0.99978$ \\
\end{tabular}
\end{table}

\paragraph{Contradiction in the thermodynamic limit.} The fact that the positively charged background cannot be neglected when solving the FQHE problem, especially in the thermodynamic limit, can be seen from the following consideration. The energies of the electron-electron ($ee$) and background-background ($bb$) interactions are positive and grow as $N^{3/2}$ in the limit $N\to\infty$. The energy of the background-electron ($be$) interaction is negative and its absolute value also grows as $N^{3/2}$ in the same limit. The ground state energy $E_{\rm GS}$ of the real physical system is determined by the sum of all three contributions. It is negative and its absolute value grows linearly with $N$ at $N\to\infty$. The energy per particle $E_{\rm GS}/N$ in the true ground state (proportional to $N^{0}$) is thus determined, in the real physical system, by the difference of two huge (proportional to $N^{1/2}$) contributions.

By replacing the actual physical problem with the problem of interacting electrons in a sample without edges and/or without the positively charged background, one minimizes only the energy of one large $ee$ contribution ($\sim N^{3/2}$) and ignores two other contributions \textit{of the same order}. It is obvious that mathematically this procedure is more than doubtful and can lead to noticeable quantitative errors and incorrect qualitative conclusions about the nature of the true ground state of FQHE systems.

\subsection{Is the Laughlin function exact for the short-range interaction potential?\label{sec:short-range}}

In a number of publications it was stated that the Laughlin wave function is an exact solution to the problem with a short-range interaction potential, see, e.g., Refs. \cite{Haldane83,Haldane85,Trugman85,Johri16}. It was first formulated in Ref. \cite{Haldane83} in the form of an \textit{observation}, i.e., without any proof, and later in Ref. \cite{Haldane85} it was mentioned that ``a formal proof (of this statement) has not yet been found''. Later, the clear statement ``Laughlin's states $\psi_m$ are shown to be exact for any number of particles in the limit in which the particles have a repulsive interaction of vanishing range'' was formulated in Ref. \cite{Trugman85}, see also \cite{Johri16}. The authors of \cite{Trugman85,Johri16} considered the FQHE problem ignoring the presence of the background potential -- the point already criticized in Section \ref{subsec:no-backgr}. Nevertheless, let us find out whether this statement is correct and how relevant it is to the FQHE problem.

The statement formulated in Ref. \cite{Trugman85} sounds very general, but raises questions and doubts. For example, does it apply to \textit{any} type of short-range interaction with vanishing range? Let us consider the well-known in physics Lennard-Jones potential 
\be 
V_{\rm LJ}(r)=4\epsilon \left[\left(\frac\sigma r\right)^{12} - \left(\frac\sigma r\right)^{6}\right] ; \label{LenJones}
\ee
here $\sigma$ and $\epsilon$ are the characteristic length and energy scales of the interaction. The parameter $\sigma$ can be chosen to be much smaller than the magnetic length, so that this is truly a short-range potential. But it is obvious that the function (\ref{LaughlinWF}) does not satisfy the many-body Schr\"odinger equation 
\be 
\frac 12\sum_{j\neq k=1}^N V_{\rm LJ}(\bm r_j-\bm r_k)\Psi_{\rm LS}^{(m)}(\bm r_1,\bm r_2,\dots,\bm r_N)=E\Psi_{\rm LS}^{(m)}(\bm r_1,\bm r_2,\dots,\bm r_N),
\ee
because at $|\bm r_j-\bm r_k|\to 0$ the left-hand side of this equation tends to infinity (as $|\bm r_j-\bm r_k|^{m-12}$), while the right-hand side tends to zero (as $|\bm r_j-\bm r_k|^{m}$, $m=3,5,7$).

Then, one can show that for any $N\ge 3$ the Laughlin function does not satisfy the Schr\"odinger equation 
\be 
\frac 12\sum_{j\neq k=1}^N V_{ee}^{\rm scr}(\bm r_j-\bm r_k)\Psi(\bm r_1,\bm r_2,\dots,\bm r_N)=E\Psi(\bm r_1,\bm r_2,\dots,\bm r_N),\label{SE-screenedCoul}
\ee
with the screened Coulomb interaction (\ref{CoulScreen}). Using the methods developed in Section \ref{sec:formulation} and recalculating the matrix elements $\langle \Psi_{s}| \hat V_{ee}|\Psi_{s'}\rangle$ for the screened Coulomb interaction potential (\ref{CoulScreen}) one can solve the problem exactly for a small number of particles. The results show, for example, that for $N=3$ the exactly calculated expansion coefficients $A_s$ do not tend to those of the LS (Table \ref{tab:LaughCsN3}) at $b/\lambda\ll 1$.

In principle, one can consider a short-range repulsive potential that is formally described by the $\delta$-function. Then the Schr\"odinger equation 
\be 
\frac 12\sum_{j\neq k=1}^N \delta(\bm r_j-\bm r_k)\Psi(\bm r_1,\bm r_2,\dots,\bm r_N)=E\Psi(\bm r_1,\bm r_2,\dots,\bm r_N)
\ee
is indeed satisfied by the function (\ref{LaughlinWF}), but it is also satisfied by \textit{any} antisymmetric wave function $\Psi$. To get a more meaningful result the authors of Ref. \cite{Trugman85} expanded some real short-range potential $V_{b}^{\rm sr}(\bm r)$ in powers of its range $b$,
\be 
V_{b}^{\rm sr}(\bm r)= \sum_{s=0}^\infty c_{s}b^{2s} \Delta^s\big[\delta(\bm r)\big]
\label{expanDelta}
\ee
and omitted all terms higher than the term $\sim b^{2}$. Then the Schr\"odinger equation is satisfied because $b^{2}\Delta\Psi_{\rm LS}^{(m=3)}=0$, that is, simply because the Laughlin wave function is by construction proportional to $|\bm r_j-\bm r_k|^3$ at $|\bm r_j-\bm r_k|\to 0$. But, if we are talking about a \textit{real} interaction potential, for example, the screened Coulomb interaction (\ref{CoulScreen}), then \textit{all} terms in the expansion (\ref{expanDelta}) must be taken into account. Then the Schr\"odinger equation is not satisfied because, for example, $b^{6} \Delta^3\Psi_{\rm LS}^{(m=3)}\neq 0$. 

Thus, the statement of Ref. \cite{Trugman85} can only be proved for the artificial strongly singular interaction potentials 
\be 
V_{pp}(\bm r-\bm r')\equiv c_{0} \delta(\bm r-\bm r')+c_{2}b^{2} \Delta\big[\delta(\bm r-\bm r')\big]
\label{expanDelta2}
\ee
that do not exist in nature. The procedure of terminating the series like (\ref{expanDelta}) on terms of order $\sim b^2$ is mathematically incorrect for realistic potentials. In addition, the corresponding eigenenergy which has been obtained in \cite{Trugman85} is $E=0$, which means that the interaction has no effect on the wave function. It can then be chosen as an arbitrary eigenfunction of the kinetic energy operator, which can hardly help in understanding the FQHE problem.

\subsection{Is there an edge reconstruction of the Laughlin state? \label{sec:edge-reconstr}}

The exact calculations by Ciftja et al. \cite{Ciftja03,Ciftja04}, performed for a macroscopically large number of electrons, showed that the density of electrons in the LS is perfectly flat in the bulk of the 2DEG, but strongly deviates from the positive background density near its edge, Figures  \ref{fig:FitCiftjaDensity}--\ref{fig:3DCiftjaDens}. Perhaps because of this problem, the currently accepted FQHE theory states that the Laughlin function (\ref{LaughlinWF}) well describes the bulk properties of the 2DEG (an unproved claim), but undergoes a ``reconstruction'' near its edge. 

The edge reconstruction of the fractional $\nu=1/3$ state has been discussed in detail in the paper \cite{Tsiper01}, where the authors performed an exact diagonalization study of the systems with up to $N=12$ electrons and interpreted their results as ``formation of an edge striped phase (ESP) with wave vector $q_{ESP}\approx \pi/2l_B$, possibly smectic liquid crystal, at the edge of a FQH system''. Do the results obtained in Ref. \cite{Tsiper01} really indicate some reconstruction of the electronic state at the sample edge?

If $N\le 7$, the density of electrons calculated in \cite{Tsiper01} (Figure 1 there) looks very similar to the electron density calculated in this work, see Figures \ref{fig:DensityExact}(b), \ref{fig:LaughDens}(c), \ref{fig:Density_N7} and \ref{fig:densN7smooth}. That is, the results of \cite{Tsiper01} for $N\le 7$ evidently suggest the formation of Wigner molecules in the 2D disks. What happens when $N>7$?

At larger $N$, the $n_e(r)$ curves in Ref. \cite{Tsiper01} begin to change qualitatively. Starting from $N=10$, a second maximum appears at a finite distance from the center. At $N=12$ the maxima are at the distances $\sim 2.5 l_B$ and $\sim 6.5 l_B$ from the disk center. Does one need to treat these changes of the electron density by formation of edge striped phases and smectic liquid crystals?

\begin{figure}[ht!]
\includegraphics[width=0.5\columnwidth]{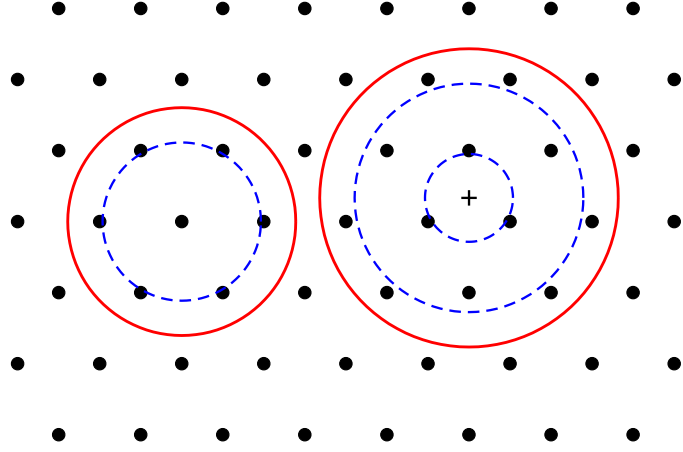}
\caption{\label{fig:edge-recon}Wigner crystal fragments of the macroscopic lattice with $N=7$ and $N=12$ electrons. The red circles show the boundaries of the disks containing $N$ electrons, the blue dashed circles correspond to the maxima of the electron density calculated in \cite{Tsiper01}. The common center of all 12 electrons is shown by a small cross in the right part of the figure.
}
\end{figure}

Obviously, these changes have a much simpler explanation. At $N\sim 7-8$, the Coulomb interaction arranges electrons into the Wigner molecule in the $(1,N-1)$ shell structure, Figure \ref{fig:WCconfigs}(g,h). At larger $N$, $N\sim 10-12$, the shell structure $(1,N-1)$ costs too much energy and the electrons are rearranged into the structure $(3,N-3)$, with three electrons dancing around the common center and other surrounding them at a larger distance, see the classical result for this case in Ref. \cite{Bedanov94}. Figure \ref{fig:edge-recon} shows the $(1,N-1)$ and $(3,N-3)$ fragments of the triangular lattice with 7 and 12 electrons respectively. The red circles show the boundaries of the disks containing this number of particles; their radii equal to $R=\sqrt{N/\pi n_s}=3^{1/4}a\sqrt{N/2\pi}$, where $a$ is the distance between the triangular lattice points. The blue dashed circles correspond to the maxima of the electron density calculated in Ref. \cite{Tsiper01}. In both cases, $N=7$ and $N=12$, they ideally coincide with the positions of classical particles in the Wigner lattice. 

In Ref. \cite{Tsiper01} it was noted that the amplitude of the density peaks noticeably decreases with distance from the edge of the disk to its center. This is completely consistent with what is expected in the case of a Wigner crystal. In a macroscopic system, it is reasonable to expect that, due to the presence of the edge, which is inevitable in any real structure, small density fluctuations with a period of the order of $a_0=1/\sqrt{\pi n_s}$ should be observed near it. As one moves away from the edge toward the sample center, weak density oscillations should decay over a distance of the order of several $a_0$, as seen from exact calculations of Ref. \cite{Tsiper01}, so that inside the sample the density becomes uniform. In the LS, on the contrary, a macroscopically large part of the electrons ($\sim\sqrt{N}$) remains near the edge, violating the principle of local electroneutrality, Figures \ref{fig:FitCiftjaDensity}--\ref{fig:3DCiftjaDens}. Thus, the exact diagonalization results of Ref. \cite{Tsiper01} perfectly confirm the formation of a Wigner-crystal-type structure in the FQHE system, which is in full agreement with the results of this paper. There is no need to invent an alternative interpretation of the results of \cite{Tsiper01}, such as an edge reconstruction.

\section{Summary and conclusions\label{sec:conclusions}}

I have presented the results of a detailed theoretical study of the ground and excited states of the system of few 2D electrons at the Landau level filling factors varying from $\nu=1$ to $\nu\simeq 1/4$. The results obtained show that
\begin{enumerate}
\item both the ground and low-lying excited states of the system have the form resembling a sliding Wigner molecule at any $\nu\lesssim 0.7$ ($\beta\gtrsim 1.42$): the maxima of the quantum-mechanically calculated electron density are at the same distance from the disk center as the radii of the shells in the classical Wigner molecules;
\item when the magnetic field changes, energy gaps between the ground and the first excited states arise and disappear as a result of the competition between repulsive Coulomb forces and compressive action of the magnetic field; the positions of the energy gaps on the magnetic field axis and their values are in reasonable agreement with experimental observations;
\item both the Laughlin wave function (\ref{LaughlinWF}) and the fractionally charged excitations (\ref{LaughlinHoles}) and (\ref{LaughlinElectrons}) do not describe the physical reality. The experimental results of Refs. \cite{Goldman95,Saminadayar97,Picciotto97} on the observation of ``fractionally charged quasiparticles'' need to be reinterpreted.
\end{enumerate}

This work represents the first step toward the development of a new, well-founded theory of the fractional quantum Hall effect.

\begin{acknowledgments}
The author thanks Elliott Lieb for useful discussions, Orion Ciftja for providing details and additional information about his works published in Refs. \cite{Ciftja03,Ciftja04,Ciftja11}, Blagoje Oblak for a useful remark, and Nadya Savostiyanova for reading the manuscript and useful comments on it.
\end{acknowledgments}

\appendix

\section{Integral ${\cal J}$\label{app:intJ}}

The matrix elements of different operators in this work can be expressed in terms of the integral ${\cal J}$ defined as
\be 
{\cal J}(n_1,n_2,l_1,l_2,k;\alpha,\beta)=\sqrt{\frac 8\pi}\int_0^\infty dx x^{2k} 
L_{n_1}^{l_1}\left(\alpha x^2\right)
L_{n_2}^{l_2}\left(\beta x^2\right)
e^{-(\alpha+\beta)x^2} 
\label{intJ},
\ee
where $n_1$, $n_2$, $l_1$, $l_2$, and $k$ are integers, $\alpha$, $\beta$ are real numbers, and $L_n^l(x)$ are Laguerre polynomials. Substituting the explicit expression for the generalized Laguerre polynomial into the definition (\ref{intJ}) one gets a finite sum of integrals of the type $\int_0^\infty x^{2n}e^{-ax^2}dx$, which can be analytically calculated. The final result can be written in the form
\be 
{\cal J}(n_1,n_2,l_1,l_2,k;\alpha,\beta)=
\sqrt{\frac{2}{\pi(\alpha+\beta)}}
\sum_{m_1=0}^{n_1}\left(\begin{array}{c}
n_1+l_1 \\ m_1+l_1 \\
\end{array}\right)\frac{(-\alpha)^{m_1}}{m_1!}
\sum_{m_2=0}^{n_2}\left(\begin{array}{c}
n_2+l_2 \\ m_2+l_2 \\
\end{array}\right)\frac{(-\beta)^{m_2}}{m_2!}
  \frac{\Gamma\left(m_1+m_2+k+\frac 12\right)}{(\alpha+\beta)^{m_1+m_2+k}}.
\label{intJsolution}
\ee
The integrals ${\cal K}(n_1,n_2,k)$ are related to ${\cal J}$ as follows
\be 
{\cal K}(n_1,n_2,k)=\sqrt{\frac{n_1!}{(n_1+k)!}\frac{n_2!}{(n_2+k)!}}{\cal J}\left(n_1,n_2,k,k, k;1,1\right).
\label{intK}
\ee

\section{Matrix elements of the background-electron interaction energy \label{app:mddBE}}

The formulas derived here are obtained for the step-like density profile (\ref{dens-step}).

\subsection{General formulas\label{app:mddBEgen}}

The matrix elements of the background-electron interaction energy are determined by Eq. (\ref{EBmatr_el}). Substituting in that formula the Fourier transforms of the background and electron densities from Eqs. (\ref{BGdensStep_Fourier}) and (\ref{ELdens_Fourier}) and replacing the variable $q\lambda/2=x$ I obtain 

\be
\langle \Psi_s|\hat V_{be}|\Psi_{s'}\rangle =
-\delta_{ss'} \frac{e^2}{R} 2N \sum_{j=1}^N \int_0^\infty \frac {dx }{x}J_1(x2R/\lambda)
e^{-x^2} 
L_{L_j^{(s)}}^{0}\left(x^2\right).
\ee 
Using the definition of the Laguerre polynomials,
\be 
L_n^{l}(x)
=\sum_{m=0}^n(-1)^m 
\left(\begin{array}{c}
n+l \\ m+l \\
\end{array}\right)\frac{x^m}{m!},
\label{Laguerre}
\ee
I get
\be 
\langle \Psi_s|\hat V_{be}|\Psi_{s'}\rangle =
-\delta_{ss'} \frac{e^2}{R} 2N \sum_{j=1}^N \sum_{m=0}^{L_j^{(s)}}\left(\begin{array}{c}
L_j^{(s)} \\ m \\
\end{array}\right)\frac{(-1)^m }{m!}
\int_0^\infty dx J_1(x2R/\lambda) e^{-x^2} x^{2m-1}.
\ee
The integral in this formula is known (Ref. \cite{PrudnikovII}, integral 2.12.9.3),
\be 
\int_0^\infty dx J_1(cx) e^{-x^2} x^{2m-1}=\frac c4\frac{\Gamma\left(m+\frac{1}{2}\right)}{\Gamma(2)} {_1F_1}\left(m+\frac{1}{2},2;-\frac{c^2}{4}\right),
\ee
so that I obtain
\be 
\langle \Psi_s|\hat V_{be}|\Psi_{s'}\rangle =
-\delta_{ss'} \frac{e^2}{\lambda}  N \sum_{j=1}^N \sum_{m=0}^{L_j^{(s)}}\left(\begin{array}{c}
L_j^{(s)} \\ m \\
\end{array}\right)\frac{(-1)^m }{m!}\Gamma\left(m+\frac{1}{2}\right)
{_1F_1}\left(m+\frac{1}{2},2;-\frac{R^2}{\lambda^2}\right).\label{BEmb-matrixelementsStepApp}
\ee
Taking into account that
\be 
\frac{R^2}{\lambda^2}=\frac{\pi n_sR^2}{\pi n_s\lambda^2}=\frac{N}{\nu }=N\beta
\ee
and
\be 
\frac{a_0}{\lambda}=\sqrt{\frac{\pi n_sa_0^2}{\pi n_s\lambda^2}}=\sqrt{\frac{1}{\nu}}=\sqrt{\beta}
\ee
I get the formula (\ref{BEmb-matrixelementsStep}) in the main text.

\subsection{Maximum density droplet state\label{app:mddBEmdd}}

In the case of the MDD state the matrix elements (\ref{BEmb-matrixelementsStep}) of the background-electron interaction energy can be transformed further. In this case $\beta= 1/\nu=1$, and one has only one state $|s\rangle=|\Psi_{\rm mdd}\rangle=|0,1,\dots,N-1\rangle$, so that $L_j^{\rm mdd}=j-1$. Then Eq. (\ref{BEmb-matrixelementsStepApp}) gives

\be 
\frac{\langle \Psi_{\rm mdd}|\hat V_{be}|\Psi_{\rm mdd}\rangle}{Ne^2/a_0 } =
-\sum_{j=1}^N \sum_{m=0}^{j-1}\left(\begin{array}{c}
j-1 \\ m \\
\end{array}\right)\frac{(-1)^m }{m!}\Gamma\left(m+\frac{1}{2}\right)
{_1F_1}\left(m+\frac{1}{2},2;-N\right).
\ee
First I change the order of summation over $j$ and $m$,
\be 
\frac{\langle \Psi_{\rm mdd}|\hat V_{be}|\Psi_{\rm mdd}\rangle}{Ne^2/a_0 } =
-\sum_{m=0}^{N-1}\frac{(-1)^m }{m!}\Gamma\left(m+\frac{1}{2}\right)
{_1F_1}\left(m+\frac{1}{2},2;-N\right)\sum_{j=m+1}^N \left(\begin{array}{c}
j-1 \\ m \\
\end{array}\right).\label{A9}
\ee
The sum of binomial coefficients here gives
\be 
\sum_{j=m+1}^N \left(\begin{array}{c}
j-1 \\ m \\
\end{array}\right)=\left(\begin{array}{c}
N \\ m +1\\
\end{array}\right).
\ee
Second, I substitute into (\ref{A9}) the definition of the hypergeometric function
\be 
{_1F_1}\left(a,b;z\right)=\sum_{k=0}^\infty \frac{(a)_k}{(b)_k}\frac{z^k}{k!},
\label{DegenHyperGfun}
\ee 
where $(a)_k=\Gamma(a+k)/\Gamma(a)$ is the Pochhammer symbol, and use the fact that $\Gamma(k+2)=(k+1)!$ After some simplifications I obtain
\be 
\frac{\langle \Psi_{\rm mdd}|\hat V_{be}|\Psi_{\rm mdd}\rangle}{Ne^2/a_0 } =
-\sum_{m=0}^{N-1}\frac{(-1)^m }{m!}
\left(\begin{array}{c}
N \\ m +1\\
\end{array}\right)
\sum_{k=0}^\infty \frac{(-N)^k}{k!(k+1)!}\Gamma\left(m+k+\frac{1}{2}\right).
\ee
Third, I change the order of summation over $k$ and $m$,
\be 
\frac{\langle \Psi_{\rm mdd}|\hat V_{be}|\Psi_{\rm mdd}\rangle}{Ne^2/a_0 } =
-\sum_{k=0}^\infty \frac{(-N)^k}{k!(k+1)!}\sum_{m=0}^{N-1}\frac{(-1)^m }{m!}
\left(\begin{array}{c}
N \\ m +1\\
\end{array}\right)
\Gamma\left(m+k+\frac{1}{2}\right)
\ee
and take the sum over $m$,
\be 
\sum_{m=0}^{N-1}\frac{(-1)^m }{m!}
\left(\begin{array}{c}
N \\ m +1\\
\end{array}\right)
\Gamma\left(m+k+\frac{1}{2}\right)=\frac{\Gamma\left(k+\frac{1}{2}\right) \Gamma\left(N-k+\frac{1}{2}\right)}{\Gamma\left(\frac{3}{2}-k\right)\Gamma\left(N\right)}.
\ee
Then, using the properties of the $\Gamma$ function,
\be 
\Gamma(z+1)=z\Gamma(z),\ \ \Gamma\left(\frac{1}{2}+k\right)\Gamma\left(\frac{1}{2}-k\right)=\frac{\pi}{\cos\pi k}
\ee
I obtain
\be 
\frac{\langle \Psi_{\rm mdd}|\hat V_{be}|\Psi_{\rm mdd}\rangle}{Ne^2/a_0 } =
\frac 2{\pi\Gamma\left(N\right)} \sum_{k=0}^\infty \frac{N^k}{k!(k+1)!}\frac{\Gamma^2\left(k+\frac{1}{2}\right) \Gamma\left(N-k+\frac{1}{2}\right)}{\left(2k-1\right)}.
\ee
The series here can be presented in the form
\be 
\sum_{k=0}^\infty \frac{N^k}{k!(k+1)!}\frac{\Gamma^2\left(k+\frac{1}{2}\right) \Gamma\left(N-k+\frac{1}{2}\right)}{\left(2k-1\right)}=-\pi \Gamma\left(N+\frac{1}{2}\right){_2F_2}\left(-\frac{1}{2},\frac{1}{2};2,\frac{1}{2}-N;-N\right),
\ee
so that finally I obtain 
\be 
\frac{\langle \Psi_{\rm mdd}|\hat V_{be}|\Psi_{\rm mdd}\rangle}{Ne^2/a_0 } =
-2\frac {\Gamma\left(N+\frac{1}{2}\right)}{\Gamma\left(N\right)} {_2F_2}\left(-\frac{1}{2},\frac{1}{2};2,\frac{1}{2}-N;-N\right).
\ee
This formula is used in the main text, Eq. (\ref{EMDDbeStep}). Calculations in this Section have been done with the help of Wolfram Mathematica, https://www.wolfram.com/mathematica.

\section{Data files\label{app:suppl}}

The .dat files containing numerical data on the many-body states and expansion coefficients for different number of particles $N$ can be found in Ref. \cite{datafiles}. In this reference there are seven directories named \textsf{Nx}, where \textsf{x} denotes the number of particles $N=2,\dots,8$. Each directory has five files (except the directory \textsf{N8} that has four files). These files contain the following information: 

\begin{enumerate}
\item The file \textsf{Nmbs.Nx.Ltotyy.dat} contains a single number, which is the number of many-body states $N_{mbs}$ for given $N$ and ${\cal L}=3N(N-1)/2$. \textsf{yy} is a two-digit number corresponding to ${\cal L}$. Example: the file \textsf{Nmbs.N7.Ltot63.dat} contains the number  $N_{mbs}=8033$.
\item The file \textsf{MBstates.Nx.Ltotyy.dat} contains $N_{mbs}$ many-body states $|s\rangle$. Example: the file \textsf{MBstates.N7.Ltot63.dat} has 8033 lines containing the many-body states running from the state number 1, $|0,1,2,3,4,5,48\rangle$, up to the state number 8033, $|6,7,8,9,10,11,12\rangle$.
\item The file \textsf{LaughlinCs.Nx.Ltotyy.dat} contains the integer numbers $C_s^{\rm LS}$ obtained using the binomial expansion of the polynomial factors in the Laughlin function (\ref{LaughlinWF}) with $m=3$, see Section \ref{sec:LaughExpansion}. Example: the first and the last lines in the file \textsf{LaughlinCs.N7.Ltot63.dat} contain the numbers  $C_{|0,1,2,3,4,5,48\rangle}=0$ and $C_{|6,7,8,9,10,11,12\rangle}=135135$.
\item The file \textsf{LaughlinAs.Nx.Ltotyy.dat} contains the real expansion coefficients $A_s^{\rm LS}$, Eq. (\ref{AsviaDs}), of the Laughlin function over the basis many-body states, see Section \ref{sec:LaughExpansion}.
\item The file \textsf{GrStateAs.Nx.Ltotyy.dat} contains the real expansion coefficients $A_s^{\rm GS}$ of the true ground state function over the basis many-body states, i.e. the eigenvector of the Hamiltonian corresponding to the lowest energy eigenvalue. The subdirectory \textsf{N8} does not contain this file.
\end{enumerate}

\bibliography{}
\end{document}